\documentclass[12pt,a4paper]{article}            
 \usepackage[skins,theorems]{tcolorbox}
\tcbset{highlight math style={enhanced,
  colframe=red,colback=white,arc=0pt,boxrule=1pt}}
  \usepackage[bookmarksopen, bookmarksnumbered, bookmarksopenlevel=2]{hyperref}
  \usepackage{tikz}
  \usepackage{tikz-3dplot}
 \usetikzlibrary{calc}
 \usetikzlibrary{decorations} %
 \usepackage[UKenglish]{babel}
 \usepackage[toc,page]{appendix}
 \usepackage{amsmath}
 \usepackage{amssymb}
 \usepackage{graphicx}
 \usepackage{hhline}
 \usepackage[bf]{caption}
\usepackage{cite}
\usepackage[vcentermath]{youngtab}
\usepackage{geometry}
\usepackage{slashed}
\usepackage{color}
\usepackage{stackrel}
\usepackage{tikz-cd} 
\usepackage{cancel} 
\usepackage{multirow} 
\usepackage{longtable}
\usepackage[all]{xy}

\usepackage{tikz}
\usetikzlibrary{shadows} 
\usepackage[framemethod=tikz]{mdframed}

\global\mdfdefinestyle{myboxstyle}{%
  shadow=true,
  linecolor=black,
  shadowcolor=black,
  shadowsize=6pt,
  nobreak=false,
  innertopmargin=10pt,
  innerbottommargin=10pt,
  leftmargin=5pt,
  rightmargin=5pt,
  needspace=1cm,
  skipabove=10pt,
  skipbelow=15pt,
  middlelinewidth=1pt,
  afterlastframe={\vspace{5pt}},
  aftersingleframe={\vspace{5pt}},
  tikzsetting={%
draw=black,
very thick} }

\newmdenv[style=myboxstyle]{whitebox} \newmdenv[style=myboxstyle,backgroundcolor=black!20]{graybox}

\newmdenv[style=myboxstyle,nobreak=true]{blockwhitebox}
\newmdenv[style=myboxstyle,backgroundcolor=black!20,nobreak=true]{blockgraybox}
\newmdenv[nobreak=true,hidealllines=true]{blockbox}

\usepackage{empheq}
\usepackage{arydshln}

\newcommand{\bqa}{\begin{eqnarray}}
\newcommand{\eqa}{\end{eqnarray}}

\newcommand{\nn}{\nonumber}

\def\et24{\eta^{24}}
\def\oet24{\frac1{\eta^{24}}}
\def\IH{\mathbb{H}}

\def\IC{\mathbb{C}}
\def\IQ{\mathbb{Q}}

\def\zh{{\hat z}}

\hypersetup{
    pdftitle={},
    pdfauthor={},
    pdfsubject={}
}
\numberwithin{equation}{section}
\numberwithin{table}{section}

\makeatletter

\newcommand{\be}{\begin{equation}}
\newcommand{\ee}{\end{equation}}
\newcommand{\beq}{\begin{equation}}
\newcommand{\eeq}{\end{equation}}
\newcommand{\ba}{\begin{aligned}}
\newcommand{\ea}{\end{aligned}}

\newcommand{\bea}{\begin{eqnarray}}
\newcommand{\eea}{\end{eqnarray}}

\newcommand{\cL}{\mathcal{L}}

\newcommand{\cF}{\mathcal{F}}

\newcommand{\cR}{\mathcal{R}}

\newcommand\bi{\begin{itemize}}
\newcommand\ei{\end{itemize}}

\def\Tr{\mathop{\mathrm{Tr}}\nolimits}

\def\unit{{1\kern-.65ex {\rm l}}}
\def\1{{1\kern-.65ex {\rm l}}}

\def\IZ{\mathbb{Z}}
\def\IP{\mathbb{P}}

\newcount\hour \newcount\minute
\hour=\time \divide \hour by 60
\minute=\time
\def\now{%
\ifnum \hour<13
  \ifnum \hour=0 \advance \hour by 12 \number\hour:\else \number\hour:\fi%
     \ifnum \minute<10 0\fi%
     \number\minute%
\ A.M.%
\else \advance \hour by -12 \number\hour:%
  \ifnum \minute<10 0\fi%
  \number\minute%
  \ P.M.%
\fi%
}

\def\fnote#1#2{\begingroup\def\thefootnote{#1}\footnote{#2}
     \addtocounter{footnote}{-1}\endgroup}

\makeatother

\begin{document}

\begin{flushright}
{\tt\normalsize {CERN-TH-2020-169}}\\
{\tt\normalsize CTPU-PTC-20-23}\\
{\tt\normalsize ZMP-HH/20-23}
\end{flushright}

\vskip 40 pt
\begin{center}
{\large \bf 
Holomorphic Anomalies, Fourfolds and Fluxes
} 

\vskip 11 mm

Seung-Joo Lee${}^{1}$, Wolfgang Lerche${}^{2}$, Guglielmo Lockhart${}^{2}$, 
and Timo Weigand${}^{3}$

\vskip 11 mm
\small ${}^{1}${\it Center for Theoretical Physics of the Universe, \\ Institute for Basic Science, Daejeon 34051, South Korea} \\[3 mm]
\small ${}^{2}${\it CERN, Theory Department, \\ 1 Esplande des Particules, Geneva 23, CH-1211, Switzerland} \\[3 mm]
\small ${}^{3}${\it II. Institut f\"ur Theoretische Physik, Universit\"at Hamburg, \\  Luruper Chaussee 149, 22607 Hamburg, Germany } \\[3 mm]
\phantom{\small ${}^{3}$}{\it Zentrum f\"ur Mathematische Physik, Universit\"at Hamburg, \\ Bundesstrasse 55, 20146 Hamburg, Germany  }   \\[3 mm]

\fnote{}{Email: seungjoolee at ibs.re.kr, wolfgang.lerche, guglielmo.lockhart at cern.ch, 
timo.weigand at desy.de}

\end{center}

\vskip 7mm

\begin{abstract}

We investigate holomorphic anomalies of partition functions
underlying string compactifications on Calabi-Yau fourfolds
with background fluxes. For elliptic fourfolds the partition functions
have an alternative interpretation as elliptic genera of $N=1$
{\nobreak supersymmetric} string theories in four dimensions, or as generating functions
for relative, genus zero Gromov-Witten invariants of fourfolds with fluxes. We
derive the holomorphic anomaly equations
by starting from the BCOV formalism of topological strings, and translating
them into geometrical terms. The result can be recast into modular and
elliptic anomaly equations. As a new feature, as compared to
threefolds, we find an extra contribution 
which is given by a gravitational descendant invariant. 
This leads to linear terms in the anomaly equations, which support
an algebra of derivatives mapping between partition functions of
the various flux sectors. These geometric features are mirrored by
certain properties of quasi-Jacobi forms. We also offer an
interpretation of the physics from the viewpoint of the
worldsheet theory, and comment on holomorphic anomalies at genus one.

\end{abstract}

\vfill

\thispagestyle{empty}
\setcounter{page}{0}
\newpage

\tableofcontents

\setcounter{page}{1}
\newpage

\section{Introduction}

\subsection{Overview and Summary}

The computation of non-perturbatively exact partition functions of supersymmetric
string theories, such as elliptic genera and various pre- and superpotentials, has attracted a lot of attention over the years. 
Some of the most spectacular results in this context rely on powerful geometrical methods like mirror symmetry in combination with string dualities, or localization techniques. 
Especially fruitful
has been the use of symmetries such as modular invariance, which allows one
to obtain exact results from a finite amount of geometrical data via modular completion.

Most works in this direction are concerned with theories with eight or 16 supercharges, translating
to $N=1,2$ supersymmetries in six dimensions or to $N=2,4$ supersymmetries in four.
Many important physical results have been obtained especially concerning massless particles
or tensionless strings that arise at singularities in the moduli space. 
Nearly tensionless non-critical strings
decoupled from gravity are known to arise at finite distances in moduli space \cite{Witten:1996qb,Klemm:1996hh}. 
The modular behaviour of their partition function, or elliptic genus \cite{Witten:1986bf}, was crucial in understanding the physics of the associated superconformal theories
\cite{Haghighat:2013gba,Haghighat:2013tka,Hohenegger:2013ala,Kim:2016usy,DelZotto:2016pvm,Gu:2017ccq,Kim:2017zyo,DelZotto:2017mee,Kim:2018gak,DelZotto:2018tcj,Gu:2018gmy,Gu:2019dan,Gu:2019pqj,Gu:2020fem,Lee:2020rns,Apruzzi:2020zot,Duan:2020cta},
or other non-perturbative phenomena such as the formation of bound states of non-critical strings to yield the heterotic string~\cite{Haghighat:2014pva,Cai:2014vka}.
Recently \cite{Lee:2018urn,Lee:2019xtm,Lee:2019wij,Baume:2019sry,Lanza:2020qmt}, 
the role of nearly tensionless critical strings at infinite distance points has been clarified in the context of quantum gravity conjectures such as the Weak Gravity Conjecture \cite{ArkaniHamed:2006dz} or the Swampland Distance Conjecture \cite{Ooguri:2006in}; the modularity of the partition function of these strings lies at the heart of the 
proof of the Weak Gravity Conjecture in such theories \cite{Heidenreich:2016aqi,Montero:2016tif,Lee:2018urn,Lee:2018spm}.\footnote{For proofs in other regimes in moduli space, see e.g. \cite{Palti:2019pca,Grimm:2018ohb,Gendler:2020dfp,Bastian:2020egp} and references therein.} 

Considerably fewer works deal with four supercharges, i.e.
 $N=1$ supersymmetry in four or  $N=(2,2)$ in two dimensions.
The initial work \cite{Lee:2019tst} on the Weak Gravity Conjecture for such theories considered compactifications of F-theory on elliptically  fibered Calabi-Yau fourfolds in flux backgrounds. Solitonic critical or non-critical strings arise on the worldvolume of D3-branes wrapping curves within the fourfold \cite{Mayr:1996sh}. It was observed that the contributions to the elliptic genus of such strings do not necessarily
exhibit the naively expected modular properties for certain flux backgrounds.
In subsequent work \cite{Lee:2020gvu} an intriguing mathematical structure was noticed, according to which
partition functions induced by certain fluxes are given by derivatives of other ones, thereby explaining the apparent lack of modularity. In fact, such partition functions are in general  what are called quasi-Jacobi forms \cite{weil1976elliptic,libgober2009elliptic,oberdieck2012serre,Oberdieck_2018,Oberdieck:2017pqm}. The derivative structure in turn played a crucial role in~\cite{Klaewer:2020lfg}, where the Weak Gravity Conjecture was verified in full generality for $N=1$ supersymmetric compactifications of F-theory to four dimensions, extending the initial results of~\cite{Lee:2019tst}. 

In this paper, we study the (anomalous lack of) modularity of { elliptic genera} in situations with four supercharges.
 It has been well known since long \cite{Schellekens:1986xh,Minahan:1997ct,Minahan:1997if,Minahan:1998vr} that quasi-modular properties of partition functions are intimately tied to holomorphic anomalies, via the substitution of the quasi-modular function $E_2(\tau)$ by a mildly non-holomorphic, but modular covariant\footnote{We will instead use the word ``invariant'' throughout the paper, meaning the absence of modular anomalies.} version denoted by 
$\hat E_2(\tau)\equiv E_2(\tau)-\frac{3}{\pi{\rm Im}\tau}$. 
In this way {\it modular} anomalies can be equivalently described in terms
of {\it holomorphic} anomalies, although the latter have a different  (albeit complementary) physical origin. They can
arise from the non-decoupling of anti-chiral operators in correlation- or partition functions of topological strings due to contact terms \cite{Bershadsky:1993ta,Bershadsky:1993cx}, or from zero modes associated with non-compact directions in field space, and generally from degenerating geometries. In fact, as is common for anomalies, one and the same holomorphic anomaly can have different physical manifestations depending on the duality frame we choose to describe a given model. The important point is that they always come  packaged together with modular anomalies which they cancel,  which is why
we will use in the following the notions of holomorphic and modular anomalies interchangeably.

In the present context of  $N=1$ supersymmetry in four dimensions, the elliptic genus of a critical or non-critical string can  be non-zero only if the system exhibits a chiral $U(1)$ gauge symmetry.
This is because the anomaly polynomial is proportional to the charge generator, i.e., to $\Tr Q$. Hence in order to have a non-trivial elliptic genus, one needs to introduce an extra background gauge field, or refinement parameter denoted by $z$. 
In the simplest case of a single $U(1)$ symmetry in a model with $(0,2)$ world-sheet superymmetry, the elliptic genus reads
\bea  \label{ellgen-def1}
{\cal Z}(q,\xi)\ =\ {\rm Tr}_{RR}(-1)^{F} F_R \,q^{H_L}   \bar q^{H_R}  \xi^Q   \,,
\eea
where $q= {\rm exp}(2 \pi i \tau)$, $\xi =  {\rm exp}(2 \pi i z)$, $F= F_R$,\footnote{In certain situations where there is a left-moving fermion number as well,  one may also consider  $F= F_L+F_R$.} and the trace is over the sector of periodic boundary conditions.
  
The extra parameter $z$ leads to an elliptic extension of the modular group 
\cite{Kawai:1993jk,Kawai:1996te,Gritsenko:1999fk}, and 
modular and quasi-modular forms are promoted to Jacobi and quasi-Jacobi forms, respectively.
In particular, the elliptic analogs of $E_2(\tau)$, and its almost holomorphic variant $\hat E_2(\tau)$, are 
given by the meromorphic quasi-Jacobi form
$E_1(\tau,z)=\frac1{2\pi i}\partial_z\log \vartheta_1(\tau,z)$  and its almost meromorphic variant, $\hat E_1(\tau,z)\equiv E_1(\tau,z)+\frac{{\rm Im}z}{{\rm Im}\tau}$.

The purpose of the present paper is to elucidate the physical and mathematical underpinning of the associated modular and elliptic anomalies, in relation to the geometry of the underlying elliptic fourfold and background four-fluxes. This is an extension of our previous
work \cite{Lee:2020gvu} which focused at  the
anomalous modularity of elliptic genera in certain flux backgrounds and the appropriate generalization of the Green-Schwarz 
mechanism for anomaly cancellation.
Here we will zoom in on the intricate interplay
of flux geometry and modularity,
as well as on the connection to holomorphic anomalies from the { dual} viewpoint of topological strings. 

In fact the generalization of holomorphic anomaly equations 
{ for elliptic genera of strings on fourfolds} 
embraces a surprising amount of additional structure { as compared to the situation on threefolds}. 
More specifically, there are two mutually interwoven aspects of the interplay between modularity and flux backgrounds: First, it turns out that certain flux-induced partition functions are related to each other by $\tau$- and $z$-derivatives. These break the modular and elliptic transformation properties, which is reflected in the appearance of the quasi-modular/quasi-Jacobi forms $E_2$ and $E_1$ as alluded to above. Second, by taking derivatives with respect to 
$E_2$ or $E_1$, we can also map into the opposite direction. 

For example,
denoting by $[{\mathcal Z}_w]$ the set of (quasi-)modular flux-induced partition functions of given
modular weight $w$, depending on appropriate choices for the fluxes
we can have the following schematic structure:
\be
\label{deristruct}
\xymatrix{
[{\mathcal Z}_w]\ \  \ar@<0.6ex>[r]^{\partial_\tau} &\ \ [{\mathcal Z}_{w+2}]\ar@<0.6ex>[l]^{\partial_{E_2\,}}},
\qquad
\xymatrix{
[{\mathcal Z}_w]\ \  \ar@<0.6ex>[r]^{\partial_z} &\ \ [{\mathcal Z}_{w+1}]\ar@<0.6ex>[l]^{\partial_{E_1\,}}}
\,.
\ee
Here the lower map represents a modular anomaly equation with the significant feature that the partition function on the right-hand side appears just linearly, i.e., $ \partial_{E_i} {\mathcal Z}_{w+i} \sim   
{\mathcal Z}_w$.  This is in contrast to the most 
familiar modular anomaly equations where, on the right-hand side, partition functions appear quadratically. The latter behaviour is the manifestation of a generic phenomenon where
an object splits into two building blocks, e.g., when a heterotic string
 unbinds into a pair of non-critical, non-perturbative E-strings \cite{Haghighat:2014pva}.
 What we encounter for modular anomaly equations for elliptic fourfolds 
with fluxes is in general a mixture of this familiar
phenomenon with the novel feature sketched by~\eqref{deristruct}.
We will give a physical, though tentative 
interpretation in terms of degenerating geometries in Section~\ref{sec:outlook}.

As we will show, all this structure can be explained from the { dual} viewpoint of topological strings.
This rests on the observation  \cite{Lee:2019tst,Lee:2020gvu} that the elliptic genera (\ref{ellgen-def1}) of certain strings in four dimensions are encoded in the { genus-zero} prepotential
of the topological $A$-model on the same Calabi-Yau fourfold. This is analogous to the situation for strings in six dimensions \cite{Klemm:1996hh}.
The $A$-model prepotential   in turn plays the role of a partition function that
captures ``relative'' Gromov-Witten invariants on the Calabi-Yau fourfold with flux background. This has been used in the mathematics literature by Oberdieck and Pixton \cite{Oberdieck:2017pqm} to conjecture a modular anomaly equation for the generating function of relative Gromov-Witten invariants on general elliptically fibered varieties.

Our work provides a physically motivated derivation of this conjecture for elliptic Calabi-Yau fourfolds.
It  makes use of the 
fact that modular anomalies are equivalent to holomorphic ones, and the latter
naturally arise from contact terms in the  CFT that underlies topological strings.
This essentially boils down to the question of how to generalize the celebrated work of BCOV \cite{Bershadsky:1993ta,Bershadsky:1993cx} on threefolds to fourfolds with fluxes.

This question will be first addressed  in an overview manner in the next subsection. 
Due to their relevance for the elliptic genera (\ref{ellgen-def1}) we will mostly be concerned with the anomaly equations for the genus-zero invariants.\footnote{In Appendix \ref{genusone} we will briefly comment on the anomaly equations
for genus one invariants. Recall that for fourfolds only the invariants for $g=0,1$ can be non-zero \cite{Klemm:2007in,cao2019stable,Cao_2020}.} 
As the relevant novel feature we identify a contact term between an anti-chiral insertion and a flux vertex operator. 
This contact term is given by what is known as  a gravitational descendant in topological gravity. 
The purpose of the subsequent Section~\ref{sec_top}  is then to reformulate this BCOV-like derivation more thoroughly
in terms of  the geometry of elliptic fourfolds and relative Gromov-Witten invariants in flux backgrounds.
Along the way we will  carefully work out  what limits have to be taken in order to 
derive the holomorphic anomaly equation in terms of generating functions for 
relative Gromov-Witten invariants. Moreover we evaluate the 
descendant invariant (i.e., the extra contact term involving the gravitational descendant).
The main results are equations (\ref{HAEversion1}) and  (\ref{psiequation}).

In Section \ref{sec_modular} we then introduce quasi-Jacobi forms  and an algebra
of derivatives acting on them \cite{weil1976elliptic,libgober2009elliptic,oberdieck2012serre,Oberdieck_2018,Oberdieck:2017pqm},  which formalizes the derivative structure (\ref{deristruct}) as well as
its elliptic generalization. A~sketch of this structure will be presented later in Fig.~\ref{fig:anomalyfluxes}.
This allows one to switch from holomorphic anomaly equations to
{\it modular} and {\it elliptic} anomaly equations that involve derivatives with respect to
$E_2$ and $E_1$, resp. These will be presented in (\ref{MAEgen}) and~(\ref{E1equ}).

In Section \ref{Examples} we specialize to geometries where the base of the elliptic
fourfold fibration is a rational fibration by itself. Such geometries are dual to perturbative
or non-perturbative heterotic strings. For these we evaluate the modular and elliptic anomaly equations, and notably the descendant invariant, to put them in a concise form directly in terms of
partition functions. Subsequently we work out a detailed example, for which we explicitly determine
the various flux-induced partition functions in terms of quasi-Jacobi forms. These are shown
to indeed satisfy the modular and elliptic anomaly equations that we derived from geometry.

We conclude with some more speculative remarks about the underlying physical picture in Section~\ref{sec:outlook}, focussing on the origins of the modular anomalies from
the perspective of the worldsheet theories of the solitonic strings.
Some of the details on the computation of the descendant invariant are deferred to Appendix~\ref{app_desc}, and those on the derivation of anomaly equations to Appendix~\ref{details_AnomalyEqs}. Moreover, Appendix~\ref{app_jacobi} recalls some well-known facts about Jacobi and quasi-Jacobi forms. 
Explicit expressions
for partition functions of our example in terms of quasi-Jacobi forms are summarized in
Appendix~\ref{expertf}. Finally, in Appendix~\ref{genusone} we briefly comment on the anomaly equations for fourfolds
at genus one, 
which is the only other case where non-trivial Gromow-Witten invariants exist. It turns out that the
situation is a straightforward generalization of the one of threefolds, in that the relevant partition functions are
independent of the flux and the anomaly equations do not receive a contribution from a gravitational descendant invariant.

\subsection{BCOV for Calabi-Yau Fourfolds}\label{sec_BCOV}   \label{BCOVon4folds}

Before we delve into the intricate mathematical details of the
holomorphic (or modular) anomaly equations for Calabi-Yau fourfolds, we 
briefly review the original work \cite{Bershadsky:1993ta,Bershadsky:1993cx} of BCOV, which was primarily 
aimed at threefolds, and outline how it extends to fourfolds at genus zero.
As we will see, the main difference is an extra term that is linear in a certain prepotential (we will briefly cover genus one in Appendix~\ref{genusone},
which is like for threefolds in that  this term does not occur).
The appearence of such a linear term was, to our knowledge, first noticed
in the work \cite{Cota:2017aal} where a special property of a particular fourfold was used.

To be precise,
we consider the  topologically twisted CFT
which describes the $N=2$ supersymmetric worldsheet theory of the topological $A$-model on a Calabi-Yau fourfold with fluxes.
We will outline the generic structure of the holomorphic anomaly equations for 
correlation functions in this CFT, and note the appearence of a contact term given by a gravitational 
descendant.  This structure will be translated later,  in Section~\ref{sec_top}, into the language
of the algebraic geometry of elliptically  fibered fourfolds.
As we will show there, the aforementioned linear term arises generically from the gravitational 
descendant term and reflects an intrinsic derivative structure which links together various different flux partition functions.

Before getting to Calabi-Yau fourfolds, however, let us first briefly
review the analogous problem for the topological string on some Calabi-Yau threefold  $Y_3$ \cite{Bershadsky:1993ta,Bershadsky:1993cx}.
We will focus on the structure of the 
 genus $g=0$ correlation functions in the topological $N=2$ worldsheet theory
 with, at first, $n=4$ operator insertions.
These correlators can be written as 
\bea \label{Fi1i4}
\cF_{i_1i_2i_3i_4} \ =\ \partial_{{i_4}} \cF_{i_1i_2i_3}(t)&=&
\big\langle\phi_{i_1}\phi_{i_2}\phi_{i_3}\int\phi_{i_4}^{(2)}\big\rangle\,,\qquad {\rm where} \\
 \cF_{i_1i_2i_3}(t) & \equiv&  \big\langle\phi_{i_1}\phi_{i_2}\phi_{i_3}\big\rangle(t)\ =\  \partial_{{i_3}} \partial_{{i_2}} \partial_{{i_1}}
\cF(t)\,.\nn
\eea
Here $\phi_i$ denote chiral primary 
vertex operators with worldsheet $U(1)_R$-charges $(1,1)$ that
represent elements in the threefold cohomology; 
since we are working in the topological $A$-model on $Y_3$, we will loosely
write $\phi_i\in H^{1,1}(Y_3)$. Moreover 
\be
\cF(t)\equiv \cF^{(g=0)}(t)
\ee
denotes the free energy at genus zero
and $\partial_i$  denotes the derivative
with respect to the (complexified) flat K\"ahler coordinate $t_i$. 
Three of these operators are inserted at random points on the $g=0$ Riemann surface, while the fourth one is integrated over the worldsheet
as indicated in (\ref{Fi1i4}).
The superscript ``$(2)$''denotes as usual the
two-form descendant version of the primary operator, 
\be
\int\phi_{ i}^{(2)} = \int Q^-\bar Q^-  \phi_{i}\,,
\ee
where ${Q^\pm}$ and ${\bar Q^\pm}$ refer to the two left- and, respectively, right-moving supersymmetry generators.
By well-known Ward identities it is irrelevant which of the operators is integrated over the worldsheet.

We can now test for holomorphic anomalies of $\cF_{i_1i_2i_3i_4}$ by inserting
an extra (integrated) anti-chiral field, $\bar\phi_{\bar i}^{(2)}$, which is BRST trivial
and thus naively decouples. However, as is well known from \cite{Bershadsky:1993ta,Bershadsky:1993cx}, 
a complete decoupling fails due to contact terms which appear at the boundary of moduli space. This leads to the BCOV equation at genus zero:
\bea\label{BCOV}
\overline{\partial}_{\bar i} \cF_{i_1i_2i_3i_4}
&=&
 {1 \over 8} {\overline C_{\bar i}}^{jk}
    \sum_{ \sigma \in S_4}
   \cF_{j i_{\sigma(1)} i_{\sigma(2)}} \cF_{k i_{\sigma(3)} i_{\sigma(4)}}
- \sum_{s=1}^4 G_{\bar{i} i_s}
         \cF_{i_1 \cdots i_{s-1} i_{s+1} \cdots i_4} \,. 
\eea         
Here  
\be\label{barcijk}
{\overline C_{\bar i}}^{jk}=\overline{C}_{\bar{i}\bar{j}\bar{k}} e^{2K} G^{j \bar j}G^{k \bar k} \,,
\ee 
$G_{\bar ij}$ is the Zamolodchikov metric on moduli space,
$K$ the K\"ahler potential, and the sum over $\sigma$ runs over all
permutations.
Recall that these entities are defined as follows:
The object $C_{ijk}\equiv \cF_{ijk}(t) $ denotes the purely holomorphic three-point correlator (\ref{Fi1i4})
of chiral primary operators,
and $\overline C_{\bar i \bar j \bar k}\equiv \overline\cF_{\bar i \bar j \bar k}(\bar t)$ its anti-holomorphic counterpart for the anti-chiral fields. 
If one defines the overlap between the chiral states $|i\rangle$ and their anti-chiral counterparts $| \bar j \rangle$ by
\bea
g_{i \bar j} = \langle \bar j | i \rangle   \,,
\eea
then the Zamolodchikov metric is given by the normalised pairing
\bea   \label{Gibarjdef}
G_{i \bar j}  =  \frac{\langle \bar j | i \rangle}{\langle \bar 0 | 0 \rangle}  = e^{K}  g_{i \bar j}    \,.
\eea
Here the K\"ahler potential is defined in terms of the overlap
between the chiral and anti-chiral ground states as
\bea
 \langle \bar 0 | 0 \rangle = e^{-K}   \,.
\eea
For later purposes note that in addition to this non-holomorphic, moduli dependent pairing between the chiral and anti-chiral sectors, one furthermore introduces the topological pairings
\bea   \label{toppairingdef} 
I_{ij}   = \langle  j | i \rangle       \,,   \qquad I_{\bar i \bar j }   = \langle  \bar j | \bar i \rangle   \,.
\eea
These constant matrices can be used to raise and lower the holomorphic and anti-holomorphic indices, respectively. For instance
\bea
C_{ij}^l = I^{lm} C_{ijm}\,,
\eea
where $I^{lm}$ is defined as the inverse of $I_{lm}$ in the sense that $I^{lm} I_{mk} = \delta^{l}_k$.

To  come back to (\ref{BCOV}), the first term on the right hand side arises from the contact terms that appear
when the inserted anti-chiral operator,
\be
\int\bar\phi_{\bar i}^{(2)} = \int Q^+\bar Q^+ \bar \phi_{\bar i}\,,
\ee
 collides  with a node 
that forms as the genus zero Riemann surface degenerates into two (i.e., when the other four operators meet pairwise),  while
the second term arises when it approaches any one of the other operators. Our primary interest will be in this latter term.

Let us have a closer look at the contact term which underlies this second type of contributions~\cite{Bershadsky:1993cx}.
In the neighborhood in moduli space where the anti-holomorphic
operator comes close to a holomorphic one, say $\phi_{i_s}$, the local geometry is
described by the state (in the $-1$ picture)
\be
 \int Q^+\bar Q^+ \bar \phi_{\bar i} \cdot \phi_{i_s}\vert0\rangle\,.
\ee
This can be evaluated with the help of the operator product
\be
\bar\phi_{\bar i}(z)\cdot\phi_{i_s}(0)\ 
\sim\  \frac {1}{|z|^2}G_{\bar i i_s}{\bf 1}+ {\rm regular}\,.
\ee
While the singular leading term can be subtracted, the
subleading term produces a contribution proportional
to the two-dimensional curvature, $d\omega= \partial\bar\partial \varphi$, where $\varphi$ is the two-dimensional dilaton. 
This coincides with the two-form version of the gravitational descendant operator, i.e.,
\be
\sigma_1^{(2)} \ =\ d\omega.
\ee
Mapping back while remembering that there is a  $d^2z$-integration left,
 one arrives at a contribution of the form
\be
-G_{\bar i i_s} \big\langle \phi_{i_1}.... \int \sigma_1^{(2)}({\bf 1})... \phi_{i_4}   
\big\rangle\,,
\ee
where $\phi_{i_s}$, as shown, has been replaced by the identity operator.
 The curvature $d\omega$ can be taken to provide
delta-function support at the locations of all the other operator insertions.
In effect, one can invoke the dilaton equation \cite{Verlinde:1990ku}
\be
\big\langle \phi_{i_1}.... \int \sigma_1^{(2)}({\bf 1})... \phi_{i_n}
\big\rangle
=
(2g-2+n-1) \cF_{i_1 \cdots i_{s-1} i_{s+1} \cdots i_n}\,,
\ee
which then, for $g=0$ and $n=4$, reproduces  the second, linear term in 
(\ref{BCOV}).

We now adapt this computation to fourfolds and consider the topological CFT describing the topological $A$-model on a Calabi-Yau fourfold with a flux configuration on top.
The starting point is quite different:
The basic building block, namely the three-point function, 
\be
\label{threept}
\cF_{a; i_1i_2}(t)\ =\  \big\langle\phi_{i_1}\phi_{i_2}
\gamma_a\big\rangle(t)\,,
\ee
contains two two-form operators
$\phi_i\in H^{1,1}(Y_4)$ plus an extra four-form operator, 
$\gamma_a\in H^{2,2}(Y_4)$, 
which will correspond to the four-flux background in geometry. Therefore we consider the 
following four-point function
\be
\cF_{a; i_1i_2i_3} \ =\ \partial_{i_3} \cF_{a; i_1i_2}(t)\ = \
 \big\langle\phi_{i_1}\phi_{i_2}\int\phi_{i_3}^{(2)}
\gamma_a\big\rangle\,.
\ee
The first, quadratic term of the holomorphic anomaly equation for $\cF_{a; i_1i_2i_3}$ arises in analogy
to the one in  (\ref{BCOV}) and leads to a
direct generalization of the threefold quantities. We will write it down below.

The linear term is more interesting as something new happens. Namely there are two contributions: The first contribution arises if $\bar\phi_{\bar i}$ hits another two-form operator, and this yields a sum of three-point functions as before. The second, novel term arises when $\bar\phi_{\bar i}$
hits the four-form operator. This contact term is governed by the operator product
\be
\bar\phi_{\bar i}(z)\cdot\gamma_a(0)\ 
\sim\  \frac {1}{|z|^2}{C_{\bar i a}}^j \, \phi_j+ {\rm regular}\,,
\ee
where $\phi_j$ denotes
 another two-form operator.

\begin{figure}[t!]
\centering
\includegraphics[width=15cm]{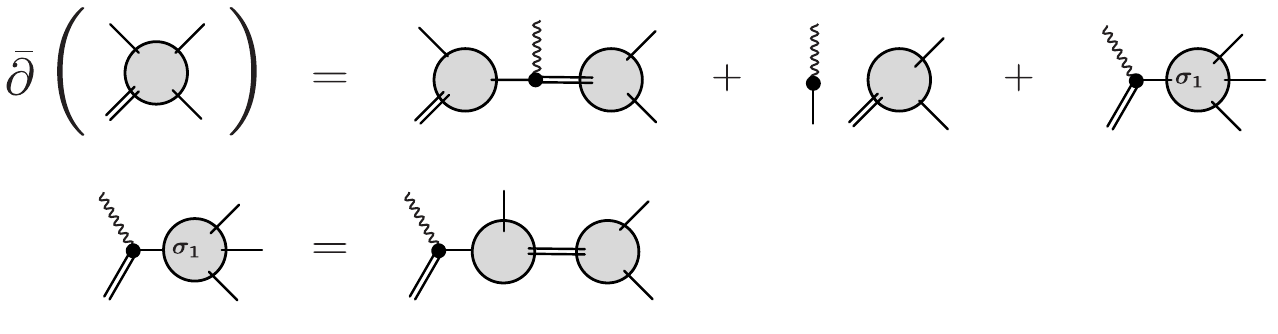}
\caption{Graphical representation of the holomorphic anomaly equation (\ref{BCOV4}) for correlation functions on
fourfolds with flux background.  Single lines denote
$(1,1)$-form fields, double lines $(2,2)$-form fields, wavy lines the antichiral $(-1,-1)$-charged field, and solid bullets correspond to classical couplings (in the limit (\ref{t-limit}) we are considering). The second line shows the factorization of the gravitational descendant term in terms of stable degenerations.
}
\label{fig:BCOV1}
\end{figure}

Using analogous arguments as above then
yields a contribution given by the following four-point function\footnote{Note that on fourfolds the naive four-point function
$\langle \phi_{i_1} \phi_{i_2} \phi_{i_3} \int \phi_j^{(2)}\rangle$ vanishes due to charge conservation.}
\be
\cF_{i_1i_2i_3j}= \big\langle \phi_{i_1} \phi_{i_2} \phi_{i_3} \int \sigma_1^{(2)}(\phi_j)
\big\rangle\,.
\ee
We thus obtain
\bea\label{BCOV4}
\overline{\partial}_{\bar i} \cF_{a;i_1i_2i_3}
&=&
 {1 \over 2}
 {\overline C_{\bar i}}^{jb}
  \sum_{ \sigma \in S_3}
  \cF_{a;j i_{\sigma(1)}} \cF_{b; i_{\sigma(2)} i_{ \sigma(3)}}
  \\
&&\qquad\qquad
- \sum_{s=1}^3 G_{\bar{i} i_s}
         \cF_{a;i_1 \cdots i_{s-1} i_{s+1} \cdots i_3} 
-      {C_{\bar ia}}^{j}   \cF_{i_1i_2i_3j}\,,\nn
\eea         
where 
\be
{\overline C_{\bar i}}^{jb}=\overline{\cF}_{\bar{b};\bar{i}\bar{j}} e^{2K} G^{j \bar j}G^{b \bar b}   \,, \qquad
{C_{\bar ia}}^{j}=I_{ab} {\overline C_{\bar i}}^{jb} \,.
\ee
The novel extra 
term can be further simplified by employing the
topological recursion formula~\cite{Hori:2003ic}
\be
\big\langle \phi_{i_1} \phi_{i_2} \phi_{i_3} \int \sigma_1^{(2)}(\phi_j)
\big\rangle
\ =\ 
\big\langle \phi_{i_1} \phi_{i_2}\gamma_b\big\rangle
I^{bc}
\big\langle \gamma_c\phi_{i_3}\phi_j\big\rangle+{\rm permutations}\,.
\ee
Here $I^{bc}$ is the inverse of the inner product
$I_{ab}=\langle\gamma_a\gamma_b\rangle$, which corresponds
to the intersection form on $H^{2,2}(Y_4)$ in geometry.
This translates to the familiar factorization of the four-point function into
three-point functions~\cite{Mayr:1996sh}:
\be\label{fourptfact}
\cF_{i_1i_2i_3j}\ =\  \cF_{b;i_1i_2}I^{bc}\cF_{c;i_3j}
+{\rm permutations}\,.
\ee
Thus  this ``linear'' term gives rise to terms quadratic in three-point functions as well,  similar to the
first term, but contracted differently corresponding to the different combinatorics of the contact terms. 
For a visualisation, see Fig.~\ref{fig:BCOV1}.
However, we will see in Section \ref{FromBCOV} that under certain circumstances, namely when we consider anomaly equations for
generating functions of {\it relative} Gromow-Witten invariants, it turns partially or completely into terms that are linear in $\cF_a$.

\subsection{Nomenclature}\label{sec_Nomenclature}

Unless stated otherwise, we will adhere to the following notation throughout the paper: 
\begin{itemize}
\item  {Geometry of the internal manifolds}
\renewcommand{\arraystretch}{1.2}
\begin{longtable}[l]{cl}
$\pi: Y_4 \to B_3$ & Elliptic fibration of a Calabi-Yau fourfold $Y_4$ over a base threefold $B_3$  \\ 
~~~~~~~~$S_0$~~~~~~~~ & Zero section to the elliptic fibration $\pi$  \\
$\bar K_{B_3}$ & Anticanonical class of $B_3$  \\
$D_{i}$ & Basis of divisor classes in $H^{1,1}(Y_4)$ \\ 

$D_\tau$ & $S_0 + \frac12 \pi^* \bar K_{B_3}$\\ 
$D_z$ & The Shioda-map image of the section generating the Mordell-Weil lattice of $\pi$ \\ 
$D_{\alpha}^{\rm b}$ & Basis of divisor classes in $H^{1,1}(B_3)$   \\ 
$D_{\alpha}$ & Pull-back divisor classes $\pi^*(D^{\rm b}_\alpha)$ in $H^{1,1}(Y_4)$ \\
$C^{i}$ & Basis of curve classes in $H_2(Y_4)$ with $D_i \cdot C^j = \delta_i^j$ \\  

$C^\tau=\mathbb E_\tau$ & Elliptic fiber of $\pi$ \\
$C^z$ & Additional fibral curve with $C^z \cdot (D_\tau, D_z, D_\alpha) = (0, 1, 0, \dots, 0)$\\
${t}^i := b^i + i v^i $ & Complexification of the volume $v^i$ of a generic curve $C^i$ \\
$(\tau, z)$ & Complexification of the volumes $({\rm Im}(\tau)=:\tau_2,\, {\rm Im}(z))$ of the curves $(C^\tau, C^z)$  \\ 
$p: B_3 \to B_2$ & Rational fibration of a base threefold $B_3$ over its own base twofold $B_2$   \\  
$C_A$ & Basis of divisor classes in $H^{1,1}(B_2)$ \\
$D^{\rm b}_A$ & Pull-back divisor classes $p^*(C_A)$ in $H^{1,1}(B_3)$ \\
$\Sigma^{\rm b}_{\dot\alpha}$ & Basis of curve classes in $H_2(B_3)$ \\

$\mathbb Y^A_3$ & Induced elliptic fibration $\pi^*(D^{\rm b}_A) = Y_4 |_{D^{\rm b}_A}$ \\
$C_0$ & Rational fiber of $p$ \\
$(C_E^1, \,C_E^2)$ & The pair of rational component curves of $C_0$ over the blowup locus $\Gamma \subset B_2$ \\
\end{longtable}

\item 
{Geometry of four-fluxes $G_*\in H^{2,2}_{{\rm vert}}(Y_4)$}
\begin{longtable}[l]{cl}
~~~~~~~$G_{\alpha_\tau}$~~~~~~~ & Basis of vertical $(0)$-fluxes $G^{(0)}_{\alpha_\tau}:=D_\tau  \wedge  \pi^\ast(D_\alpha^{\rm b})$ ($\alpha_\tau = 1_\tau, 2_\tau, \cdots$) \\
$G_{\alpha_z}$ & Basis of vertical $(-1)$-fluxes $G^{(-1)}_{\alpha_z}:= D_z  \wedge  \pi^\ast(D_\alpha^{\rm b})$  ($\alpha_z = 1_z, 2_z, \cdots$) \\
$G_{\dot\alpha}$ & Basis of vertical  $(-2)$-fluxes $G^{(-2)}_{\dot\alpha} := \pi^\ast(\Sigma_{\dot\alpha}^{\rm b})$ ($\dot \alpha = \dot 1, \dot 2, \cdots$) \\
$G_a$ & Basis of all vertical fluxes, $a \in \{\alpha_\tau, \alpha_z, \dot \alpha\}$ \\
$I_{ab}=G_a \cdot G_b$ & Intersection form \\
\end{longtable}

\item {Zamolodchikov metrics $G_{*,\bar *}$}
\begin{longtable}[l]{cl}
~~~~~~~$G_{i \bar j}$~~~~~~~ & Metric on the K\"ahler moduli space of  $(1,1)$ fields \\
$G_{a \bar b}$ & Metric of four-form fluxes for the $(2,2)$ fields \\
\end{longtable}

\item {Generating functions $\cF$, $\langle\langle -  \rangle \rangle$ of  genus-zero Gromov-Witten invariants}
\begin{longtable}[l]{cl}
$\langle A_1,\ldots, A_k \rangle_{C}^Y$ & Invariant on $Y$ for the curve class $C$ with $k$ marked points on $A_{1, \cdots, k}$ \\  
$\left<\left< A_1, \ldots, A_k \right>\right>_C^Y$ & Generating function $\sum_{n,r} \left< A_1, \ldots, A_k\right>_{C+n \mathbb E_\tau + r C^z}^Y q^n \xi^r$ \\
~~~~~~~~$\psi$~~~~~~~~ & Tautological line bundle class associated with the rightmost marked point \\
$\langle G \rangle_{C}$, $N_G(C)$ & Invariant $\langle G \rangle_{C}^{Y_4}$ on $Y_4$ for the curve class $C$ with one marked point on $G$ \\
$\langle\langle G \rangle \rangle_{C}$, $\cF_{G |C}$ & Generating function $\sum_{n,r} \left< G \right>_{C+n \mathbb E_\tau + r C^z}^{Y_4} q^n \xi^r$ for relative invariants on $Y_4$ \\ 
$\cF_{a |C}$ & $\cF_{G_a |C}$ with respect to the basis elements of four-form fluxes, $G_a$ \\ 
$\cF_{C}^{D}$  & Generating function $\left<\left< ~\right>\right>_{C}^{D}$ on the divisor $D$ of $Y_4$ containing $C$ \\
\end{longtable}

\item 
{Flux-dependent partition functions $\mathcal Z=-q^{E_0}\cF$, and modular forms}

\begin{longtable}{cl}
$\mathcal Z[G, C]=-q^{E_0}\cF_{G |C} $ & Partition function with respect to flux $G$ and curve class $C$ in $Y_4$  \\  
$\mathcal Z^{\mathbb Y_3^A}[C]$ & Partition function with respect to curve class $C$ in the threefold $\mathbb Y_3^A$  \\  
~~~$\Phi_{w,m}$~~~ &Holo- or meromorphic quasi-Jacobi form of weight $w$ and index $m$ \\
$\hat \Phi_{w,m}$ & Almost holo- or meromorphic Jacobi form of weight $w$ and index $m$ \\
${\mathcal Z}_{w,m}[G^{(w)},C]$ & Partition function for $(w)$-flux of weight $w$ and elliptic index $m$\\
\end{longtable}
\ei

\goodbreak

\section{Holomorphic Anomalies for Topological Strings 
on Elliptic Fourfolds}\label{sec_top}

In Section \ref{BCOVon4folds} we considered the topological $A$-model on a generic Calabi-Yau fourfold. 
From now on, we will specialise the geometry by imposing that the fourfold $Y_4$ be elliptically fibered. 
The motivation is two-fold:
First, on such a background the four-point functions, whose holomorphic anomaly equations were given in (\ref{BCOV4}),
exhibit distinguished modular properties, as first observed in \cite{Haghighat:2015qdq,Cota:2017aal}. The role of the modular parameter is played by the K\"ahler modulus of the elliptic fiber.
Relatedly, the prepotential of the topological string, from which the correlation functions derive, can now be expanded into generating functions of the {\it relative} Gromov-Witten invariants
on the fourfold with fluxes. 
Second, if we invoke the duality between Type IIA string theory on an elliptic $Y_4$ and F-theory on $Y_4 \times T^2$, these generating functions
 are related to the elliptic genus of certain solitonic strings in the four-dimensional $N=1$ effective theory of F-theory.

In the sequel we translate the generic expression  (\ref{BCOV4}) for the holomorphic anomaly, as derived in conformal field theory,  into geometrical language and interpret it as an equation obeyed by the
 generating functions for relative Gromov-Witten invariants on elliptic fourfolds with flux backgrounds. Our focus will be on the derivation of the resulting holomorphic anomaly equation for genus-zero invariants
 from the BCOV formalism. { See Appendix \ref{genusone} for the anomaly equations for higher genus invariants.}

Before coming to this, we observe in the next Section \ref{sec_Fldeppre} an intriguing derivative relation for the generating functions of relative genus-zero Gromov-Witten invariants for certain flux backgrounds, which is summarised in (\ref{qderivative1}).
In Section \ref{FromBCOV}, we then turn to the actual derivation of the holomorphic anomaly equations. The main result of this section is stated in (\ref{HAEfirsttermsugg}).
Since its derivation is technical, we delegate some of the details to Appendix \ref{app_desc} and \ref{details_AnomalyEqs}.

\subsection{Flux Dependent Prepotentials on Elliptic Fourfolds}   \label{sec_Fldeppre}

Let us denote by $Y_4$ a smooth elliptically fibered Calabi-Yau fourfold that forms the background of the topological $A$-model, and by $B_3$ its K\"ahler threefold base. 
We first need to introduce some notation.
The holomorphic section of the elliptic fibration is referred to as $S_0$ and the projection as $\pi: Y_4 \to B_3$.
We assume that $Y_4$ admits an additional independent rational section $S$. This is because 
its image under the Shioda map,
\bea
\sigma(S) = S - S_0 - \pi^\ast({\cal D}_S)   \,,
\eea
is associated with an extra $U(1)$ gauge symmetry group in four dimensions,  if we compactify F-theory on $Y_4$.
As remarked before, such an extra chiral $U(1)$ gauge symmetry is required in order for the elliptic genus in
four dimensions to be non-vanishing. In M-theory language, the $U(1)$ gauge potential appears by expanding the M-theory three-form as $C_3 = A \wedge \sigma(S) + \ldots$. See, for instance, the reviews \cite{Weigand:2018rez,Cvetic:2018bni} for details and original references.\footnote{This geometry can be generalised to fourfolds that admit several independent sections, and also singularities in codimension-one of the Weierstrass model associated with $Y_4$. The latter introduce non-abelian gauge symmetries in F-theory. The resolution of the singularities leads to exceptional divisors which, for our purposes, take a role similar to $\sigma(S)$. To keep the discussion simple we will not include such extra data here.}

A basis of $H^{1,1}(Y_4)$ can be defined in terms of the divisors
\be
D_i  \,, \quad   i = 1, \ldots, h^{1,1}(Y_4)  \,,
\ee
while by
\be
C^j \,, \quad   j = 1, \ldots, h^{1,1}(Y_4)       \,,
\ee
we denote a dual basis of curve classes  in $H_2(Y_4)$ which obey\footnote{Here and in the sequel, 
we abbreviate the intersection product on $Y_4$ for two (or several) forms whose total degree sums up to $(4,4)$ as $w_a\cdot w_b\equiv
\int_{Y_4}w_a\wedge  w_b$. Similarly a dot product with extra subscript $B_3$
refers to the intersection product on the threefold base, $B_3$, for forms whose total degree sums up to $(3,3)$. A dot  between forms of total degree less
 than $(4,4)$ will be used interchangeably with the wedge product  $\wedge$ on $Y_4$ (and analogously for a dot with subscript $B_3$). \label{footnote-label}}
\be
D_i \cdot C^j = \delta^j_i   \,.
\ee
Thus, if we expand the complexified K\"ahler form  ${\mathbf J}$ in terms of the divisors $D_i$ as
\be  \label{Jexp1}
{\mathbf J} = B + i J = 
{t}^i D_i  =  (b^i + i v^i ) D_i      \,,
\ee
then the $v^i$ represent the real volume moduli of the curves $C^i$.

For the geometries under consideration, a convenient basis for the divisors can be taken as
\be
\begin{split}   \label{basish11-Y4}
D_\tau &= S_0 + \frac{1}{2} \pi^\ast(\bar K_{B_3})    \,,    \cr
D_z &=   \sigma(S) \,,   \cr  
D_\alpha  &= \pi^\ast(D_\alpha^{\rm b})   \,,  \qquad   \alpha = 1, \ldots, h^{1,1}(B_3)  \,,
\end{split}
\ee
where the $D_\alpha^{\rm b}$ form a basis of the divisors on $B_3$.
Among the dual curve classes $C^j$ on $Y_4$, of particular importance will be the class of the generic elliptic fiber, 
\be
C^\tau =: \mathbb{E}_\tau   \,,
\ee
as well as the class of an additional fibral curve, $C^z$. These have the defining properties that
\bea\label{Ctau-Cz}
D_\tau \cdot \mathbb{E}_\tau&= 1,   \qquad D_z \cdot C^{z} &= 1  \,,
\eea
while the intersection numbers with all other basis elements of $H^{1,1}(Y_4)$ vanish.
If we separate out the two distinguished divisors by writing the
complexified K\"ahler form  as
\be  \label{Jexp}
{\mathbf J} = B + i J = 
\tau   D_\tau   + z   D_z + \sum_\alpha { t}^\alpha D_\alpha  \,,
\ee
the geometric volume modulus associated with the generic elliptic fiber class $\mathbb E_\tau$ is  identified with
\be
\tau_2 = {\rm Im}(\tau)   \,.
\ee
Similarly, ${\rm Im}(z)$ represents the volume modulus of the additional fibral curve $C^{z}$.

The prepotential  of the topological $A$-model is defined with respect to a choice of background fourfold fluxes, which take values in $H^{2,2}(Y_4)$.
This space splits into three mutually orthogonal subspaces  \cite{Greene:1993vm,Braun:2014xka},
\be
H^{2,2}(Y_4) = H^{2,2}(Y_4)_{\rm hor} \oplus H^{2,2}(Y_4)_{\rm vert}   \oplus H^{2,2}(Y_4)_{\rm rem}   \,,
\ee
where the labels mean ``horizontal'', ``vertical'' or ``remainder''.
The $A$-model prepotential depends explicitly on the primary vertical part of the background flux, 
\be
G \in H^{2,2}_{\rm vert}(Y_4) \,.
\ee
Elements of  $H^{2,2}_{\rm vert}(Y_4)$ are linear combinations of products of elements in $H^{1,1}(Y_4)$. 
As a basis of  $H^{2,2}_{\rm vert}(Y_4)$ we can therefore take\footnote{See footnote \ref{footnote-label} for an explanation of the notation.} 
\bea
G_a  = E_a^{ij}   \, D_i \cdot D_j
\eea
for suitable coefficients $E_a^{ij}$,
and then expand the flux in terms of this basis as 
\bea
G = g^a   \,  G_a    \,.
\eea
Note that the flux coefficients $g^a$ must take discrete values such that the flux is properly quantised, i.e.,  $G + \frac{1}{2} c_2(Y_4) \in H^4(Y_4, \mathbb Z)$ \cite{Witten:1996md}.

Among the different types of fluxes, we distinguish so-called $(0)$-fluxes, $(-1)$-fluxes and $(-2)$-fluxes, 
which are labelled according to the modular weight of the associated partition functions (see further below).
They correspond to the following basis elements of $H^{2,2}(Y_4)_{\rm vert}$\cite{Cota:2017aal,Oberdieck:2017pqm,Lee:2020gvu}:
\be   \label{basisH22vert-Y4}
\begin{aligned}
G_{\alpha_\tau}  & \equiv  G^{(0)}_{\alpha_\tau} &=&\, \,   D_\tau  \cdot  \pi^\ast(D_\alpha^{\rm b})   \,,    \cr
G_{\alpha_z } &  \equiv G^{(-1)}_{\alpha_z }   &=& \, \,  D_z  \cdot  \pi^\ast(D_\alpha^{\rm b})   \,,   \cr
G_{\dot\alpha}  & \equiv  G^{(-2)}_{\dot\alpha}     &=& \,\,   \pi^\ast(\Sigma_{\dot\alpha}^{\rm b})    \,,  \quad  \quad \quad  \Sigma_{\dot\alpha}^{\rm b}   \in  H^4(B_3)    \,.
\end{aligned}
\ee
We sometimes explicitly signify the modular weight by a superscript, as indicated.
Moreover we will split the generic label for the fluxes $\{G_a\} \equiv  \{G _{\alpha_\tau}, G_{\alpha_z} ,G_{\dot\alpha} \}$
 to indicate the respective modular weight as follows:
\be
\{ a\} \equiv\ \{  \alpha_\tau \,, \alpha_z,  \dot\alpha  \}     \,.
\ee

The structure of the intersection form on the elliptic fourfold $Y_4$ implies that the only non-vanishing products $I_{ab}$ between these basis elements are
\be   \label{intrelG-gen}
\begin{aligned}
I_{\alpha_\tau   \dot\alpha}    &:= G_{\alpha_\tau} \cdot G_{\dot\alpha}  &=& \, \,  (D_\tau \cdot \pi^\ast(D_{\alpha}^{\rm b}))    \cdot   \pi^\ast(\Sigma_{\dot\alpha}^{\rm b})  &=&  \, \,  D_\alpha^{\rm b}   \cdot_{B_3} \Sigma_{\dot\alpha}^{\rm b}   \,, \cr
I_{\alpha_z \beta_z} &:= G_{\alpha_z} \cdot G_{\beta_z} &=& \, \,  (D_z \cdot \pi^\ast(D_{\alpha}^{\rm b}))    \cdot  (D_z \cdot \pi^\ast(D_{\beta}^{\rm b}))  &=& \, \,    D_\alpha^{\rm b} \cdot_{B_3}    D_\beta^{\rm b} \cdot_{B_3} (-b)    \,,
\end{aligned}
\ee
where 
\be   \label{heightpairingdef}
b = - \pi_\ast(D_z \cdot D_z)     \in H^2(B_3)
\ee
denotes the height-pairing associated with the rational section $S$, and we will abbreviate 
\be \label{Ialphadotalpha}
 I_{\alpha \dot \alpha}   := D_\alpha^{\rm b}   \cdot_{B_3} \Sigma_{\dot\alpha}^{\rm b}     \,.
\ee

After this preparation, consider the genus-zero\footnote{By default, since in this paper  we only discuss genus-zero invariants, 
except for Appendix~\ref{genusone}, we omit a label to indicate the genus.}  prepotential
$\cF_G(t)$ 
as computed in the topological $A$-model on~$Y_4$.
It depends linearly on the vertical flux background $G = g^a G_a$:
\be \label{Fadef}
{\cal F}_G = g^a \,  {\cal F}_a   \,.
\ee

The genus-zero prepotential serves as the generating function for the genus-zero Gromov-Witten invariants on the fourfold $Y_4$ in the flux background  $G$, i.e.,
\bea
{\cal F}_G  =   \sum_{C = c_i C^i}  N_{G}(C) \, e^{2 \pi i  { t}^i c_i}       \,,
\eea
where the sum runs over all 2-cycle classes $C \in H_2(Y_4)$.
The genus-zero invariants $ N_{G}(C)$ count stable holomorphic maps 
\be   \label{stablemapf}
f: \Sigma_{g=0, k=1}  \to  C  \in H_2(Y_4)
\ee
from a genus $g=0$ Riemann surface $\Sigma$ with $k=1$ points fixed to $C$, such that the image of the distinguished point, $f(p_i) \in C$, lies on
the four-cycle $A_G \in H_4(Y_4)$ that is Poincar\'e dual to the flux $G$.
We will oftentimes denote this invariant as\footnote{Note that the symbol $\langle \ldots \rangle$ denotes both genus-zero Gromov-Witten invariants and correlation functions in the two-dimensional CFT, as in the previous section. We trust that it will be clear from the context which of the two meanings we refer to.}
\bea
N_G(C) =:   \langle   G   \rangle_{C}   \,.
\eea
For further details on the mathematics of such invariants we refer for instance to the presentations in 
\cite{Cox:2000vi,Hori:2003ic,Klemm:2007in,cao2019stable,Cao_2020}, while some aspects
of particular relevance to us will be discussed at the end of this section.

Mirroring the expansion (\ref{Jexp}) of the K\"ahler form 
of the elliptic fibration $Y_4$, one can organise the sum over all curve classes 
by introducing
\bea   \label{Cbetanr}
C_\beta(n,r) :=  C_\beta + n {\mathbb E}_\tau + r C^z   \,,
\eea
where $C_\beta$ is some curve class on $B_3$, ${\mathbb E}_\tau$ and $C^z$ denote the fibral classes introduced above via~\eqref{Ctau-Cz} and $n, r \in \mathbb Z$.
With this notation we can expand ${\cal F}_G$ as 
\bea \label{Fgexp1}
{\cal F}_G =  \sum_{C_\beta \in H_2(B_3)} {\cal F}_{G| C_\beta} \,  Q_\beta    \,,    \quad    Q_\beta  = e^{2 \pi i (C_\beta)_\alpha{t}^\alpha}    \,.
\eea
The object  
\be  \label{Fgexp2}
{\cal F}_{G| C_\beta} = g^a \,{\cal F}_{a| C_\beta} 
\ee
then represents the generating functional for the following {\it ``relative''} genus-zero Gromov-Witten invariants which are defined with reference to the given base curve class $C_\beta$,
\be\label{relativeF}
{\cal F}_{G| C_\beta} = \sum_{n, r} N_{G} (C_\beta(n,r))   \,  q^n  \, \xi^{r}   \,.
\ee
Here we denote as usual
\bea  \label{qxi}
q = e^{2 \pi i \tau}   \,, \qquad    \xi = e^{2 \pi i z}    \,,
\eea
in terms of the complexified K\"ahler parameters of the fibral curves ${\mathbb E}_\tau$ and $C^{z}$,
respectively. To stress the relation to the Gromov-Witten invariants we sometimes employ the notation
\bea
{\cal F}_{G| C_\beta} =: \langle \langle G \rangle \rangle_{C_\beta}   \,.
\eea

Let us now point out some important aspects of these generating functions that follow directly from
general properties of Gromov-Witten invariants. Namely, for special cases of $(0)$ or $(-1)$ fluxes, the
prepotentials are {\it derivatives} of generating functions for other fluxes, which in turn encode
relative invariants on certain embedded threefolds, $\mathbb Y_3^\alpha\subset Y_4$.
More precisely, 
one finds the following structure at genus zero:
\begin{whitebox}
\bea
   \label{qderivative1}
\begin{rcases}
\hspace{-.01in}G_{\alpha_\tau}=D_\tau \cdot D_\alpha :  \quad  \cF_{G_{\alpha_\tau}|C_\beta} &= \, (q \partial_q \, +E_0) \, \cF_{C_\beta}^{{\mathbb Y}_3^\alpha}  \\
G_{\alpha_z}=D_z \cdot D_\alpha:  \quad \cF_{G_{\alpha_z}|C_\beta} &= \,  \xi \partial_\xi  \,  \cF_{C_\beta}^{{\mathbb Y}_3^\alpha}       \\
\phantom{a} \, \,\hspace{.01in}G=D_\gamma \cdot D_\alpha: \quad \quad \cF_{G|C_\beta} &= \,  (D^{\rm b}_\gamma \cdot C_\beta) \, \cF_{C_\beta}^{{\mathbb Y}_3^\alpha} 
\end{rcases}
\begin{matrix}

\ \  \text{whenever}  \, \, (\ref{cond1})  \, \,  \text{holds}\,. 
\\
\end{matrix}
\eea
\end{whitebox}
The condition is that the image of the special point on the Riemann surface $\Sigma_{g=0,k=1}$ underlying the definition of the Gromov-Witten invariants, see (\ref{stablemapf}), lies on $D_\alpha$ if and only if 
the base curve 
$C_\beta$ is contained inside the base divisor $D_\alpha^{\rm b}$ (or rather if this is the case for certain members of the family of curves in class  $C_\beta$ or of divisors  $D_\alpha^{\rm b})$.
We abbreviate this condition as
\bea   \label{cond1}
f({\rm pt}) \in D_\alpha  \Rightarrow C_\beta   \subset D_\alpha^{\rm b} 
\eea
and just refer to it as the requirement that $C_\beta$ must be contained in $D_\alpha^{\rm b}$.

Moreover in  (\ref{qderivative1}) we have defined
\be  \label{E0def}
E_0 = - \frac{1}{2} \bar K_{B_3} \cdot_{B_3} C_\beta\,,
\ee
which will play  the role of a vacuum energy (hence the notation) in Section~\ref{sec_modular}, 
and we denote by
\be
\cF_{C_\beta}^{{\mathbb Y}_3^\alpha}
\ee 
the generating function for the Gromov-Witten invariants relative to\footnote{Note that by abuse of notation we use the same symbol $C_\beta$ to also denote the corresponding curve class in $H_2(D_\alpha^{\rm b})$. This is well defined since $C_\beta \subset D_\alpha^{\rm b}$.} $C_\beta$ inside the threefold cut out by the divisor $D_\alpha\equiv \pi^\ast(D_\alpha^{\rm b}) \subset Y_4$. 
In line with our previous work \cite{Lee:2020gvu} we call these ``embedded" threefolds  and refer to them as 
\be
\label{Y3adef}
{\mathbb Y}_3^\alpha :=  \pi^\ast(D^{\rm b}_\alpha) =  Y_4 |_{D^{\rm b}_\alpha}   \,.
\ee
The objects $\cF_{C_\beta}^{{\mathbb Y}_3^\alpha}$ are intrinsically geometric because they do not depend  on any further flux background.
Note, however, that the relative fourfold prepotentials ${\cal F}_{G| C_\beta}$ coincide with such generating functions 
$\cF_{C_\beta}^{{\mathbb Y}_3^\alpha}$, or their derivatives, only  if (\ref{cond1}) is satisfied.
This is a condition on the flux background.
 A prepotential ${\cal F}_{G| C_\beta}$ for more general fluxes
receives additional contributions which are not of this simple form, and in particular cannot be written as a derivative.
See our previous works \cite{Lee:2019tst,Lee:2020gvu} for initial observations and discussions of these matters, and Section~\ref{SecExS} for an explicit example.

To understand the rationale behind both the derivative structure and the appearance of  invariants of embeeded threefolds, let us 
generalise the setting to a general $n$-fold $Y_n$ (not necessarily Calabi-Yau).  The Gromov-Witten invariants count stable holomorphic maps
\bea
f: \Sigma_{g,k} \to C   \in H_2(Y_n)\,,
\eea
subject to the condition that the images of $k$ special points $p_i$ on $\Sigma$ lie on the cycles dual to the classes $A_i \in H^{m_i,m_i}(Y_n)$. We denote these invariants by
\bea\label{GW-inv}
\langle A_1,\ldots, A_k \rangle_{g,C}^{Y_n}  \,.
\eea
For simplicity of presentation, we will suppress the genus $g$ in~\eqref{GW-inv} for the $g=0$ invariants and similarly drop the subscript $Y_n$ for invariants on an elliptic fourfold with $n=4$.   
As explained for instance in \cite{Cox:2000vi,Hori:2003ic},
the moduli space of such maps has virtual dimension
\bea  \label{virtdimexp}
{\rm dim}_{\rm vir}(\overline{\cal M}_{g,k}(Y_n, C)) =  c_1(Y_n) \cdot C + (n-3) (1-g) + k  \,.
\eea
The invariants are obtained by pulling back the classes $A_i$ via the evaluation map at the $i$-th point and intersecting the result with the fundamental class of the moduli space.
In order to obtain a {non-zero} number one needs
\be   \label{dimconstraint}
\sum_{i=1}^k m_i = {\rm dim}_{\rm vir}(\overline{\cal M}_{g,k}(Y_n, C)) \,.
\ee
Note that this relation remains satisfied if we add a further fixed point on $\Sigma$, together with an additional incidence condition that its image lies on a divisor $D$.
The resulting invariants with $(k+1)$ points fixed satisfy the  well-known~\cite{Hori:2003ic} 
\bea   \label{DivEqu}
\text{Divisor Equation:}   \qquad     \langle A_1,\ldots, A_k, D   \rangle_{C}^{Y_n} =   (C \cdot D) \langle A_1,\ldots, A_k \rangle_{C}^{Y_n}    \,.
\eea

After this review, we make the following observation which is responsible for the intricate, partly
derivative structure displayed in (\ref{qderivative1}). Suppose that
one of the classes $A_i \in H^{m_i,m_i}(Y_n)$ can be written as a product
$A_i = \hat A_i \cdot D_i$  such that $\hat A_i  \in  H^{m_i-1,m_i-1}(Y_n)$ and $D_i \in H^{1,1}(Y_n)$. Assume furthermore that 
$C$ and $D_i$ satisfy a condition analogous to (\ref{cond1}).
In this case, the invariants on $Y_n$ can be expressed as invariants within the divisor $D_i$ on $Y_n$:
\bea   \label{ReductionFormula}
 \langle A_1,\ldots,( \hat A_i \cdot D_i), \ldots A_k   \rangle_{C}^{Y_n} =  \langle A_1,\ldots,\hat A_i, \ldots A_k   \rangle_{C}^{D_i}  \,,
\eea
where all classes on the right are understood as suitable pullbacks to the embedded $(n-1)$-fold $D_i \subset Y_n$.

As a special case, we now come back to relative invariants of an elliptic Calabi-Yau fourfold $Y_4$ with $k=1$ points fixed and 
combine (\ref{DivEqu}) and (\ref{ReductionFormula}):
First, consider the relative invariants for a $(0)$-flux $G_{\alpha_\tau} = D_\tau \cdot D_\alpha$   for a pullback divisor $D_\alpha =\pi^\ast(D^{\rm b}_\alpha)$, and suppose that  $C_\beta  \subset D_\alpha^{\rm b}$ in the sense of (\ref{cond1}).
Then for each curve $C_\beta(n,r)$ as defined in (\ref{Cbetanr}) we deduce that
\bea   \label{qderivative2}
\langle  D_\tau \cdot D_\alpha  \rangle_{C_\beta(n,r)}^{Y_4} = \langle  D_\tau  \rangle_{C_\beta(n,r)}^{D_\alpha} = (D_\tau \cdot C_\beta(n,r))  N_{C_\beta(n,r)}^{D_\alpha} = (n + E_0)  N_{C_\beta(n,r)}^{D_\alpha}    \,,
\eea
where $N_{C_\beta(n,r)}^{D_\alpha}$ denote the genus-zero invariants on the threefold $D_\alpha$ and the extra term proportional to $E_0 = - \frac{1}{2} \bar K_{B_3} \cdot C_\beta$
arises from the intersection of the zero-section $S_0$ with the curve class $C_\beta$ on the base.

For the generating function for the relative invariants this yields the first line in (\ref{qderivative1}). Similar reasoning applied to $(-1)$-fluxes $G_{\alpha_z} = D_z \cdot D_\alpha$
yields the second identity.
On the other hand, for a $(-2)$-flux  $G = D_\gamma \cdot D_\alpha$ the relation analogous to (\ref{qderivative2}) 
gives the same multiplicative prefactor 
$(D^{\rm b}_\gamma \cdot C_\beta)$ for all relative invariants and hence implies the third line of (\ref{qderivative1}).

Let us close this section by stressing that the properties (\ref{qderivative1}) of the prepotentials are not only interesting by themselves, but they represent special cases of more general relations 
between partition functions with respect to various flux backgrounds. 
Indeed, for the flux backgrounds as in (\ref{qderivative1}), the special $(0)$ and $(-1)$ flux prepotentials are derivatives of the prepotentials in certain $(-2)$ flux backgrounds.
More generally, as we will explain in Section \ref{sec_modular}, the appearance of derivatives $q \partial_q$ and $\xi \partial_\xi$ reflects certain modular anomalies of the prepotentials, which in turn can be translated into holomorphic anomalies. This brings us to the topic of the next section.

\subsection{From BCOV to a Holomorphic Anomaly Equation for Relative Gromov-Witten Invariants on Fourfolds} \label{FromBCOV}

We will now translate the conformal field theoretical, ``BCOV type''
holomorphic anomaly equations of Section \ref{BCOVon4folds} into 
anomaly equations for the generating functionals of relative Gromov-Witten invariants at genus zero (see Appendix~\ref{genusone} for a
discussion of invariants at genus one).
The main result of this section is the derivation of the holomorphic anomaly equations as given below in (\ref{HAEfirsttermsugg}).

It has already been noted that
the operators $\phi_i$ in the topological $A$-model with $U(1)_R$-charges $(1,1)$
are identified with the basis element $D_i$ of $H^{1,1}(Y_4)$:
\bea
\phi_i   \longleftrightarrow D_i  \in H^{1,1}(Y_4)   \,.
\eea
The  associated massless deformation moduli hence map to the  parameters ${t}^i$  in the expansion (\ref{Jexp1}) of the complexified K\"ahler form,
\bea
{\mathbf J} =   {t}^i   D_i   \,.
\eea

Similarly, the operators $\gamma_a$ with $U(1)_R$-charges $(2,2)$ map in geometry to the flux basis $G_a$ of $H^{2,2}_{\rm vert}$:
\bea \label{gammaG}
\gamma_a   \longleftrightarrow G_a   \in H^{2,2}_{\rm vert}(Y_4)  \,.
\eea
These operators are not associated with massless deformation moduli, but rather represent irrelevant operators in the topological $A$-model. Thus they should be viewed as non-dynamical background fields that define superselection sectors in
the Hilbert space, and the corresponding parameters
 $ {\mathbf g}^a$  should be interpreted only as formal sources of these operators. They can be packaged into one object specifying the flux background:
\bea
{\mathbf G}  = {\mathbf g}^a G_a  = (c^a + i g^a) G_a   \,.
\eea
In particular we identify the imaginary parts of the (massive) fields ${\mathbf g}^a$ with the vertical background flux parameters
defined via $G = g^a G_a$. The fact that $g^a$ represent discrete parameters, rather than continuous moduli, resonates with the nature of the $(2,2)$-form fields as massive objects in the CFT.

With this understanding we now revisit the holomorphic anomaly equation (\ref{BCOV4}), which applies to four-point, genus-zero correlation functions
 $\cF_{a; i_1i_2i_3}$.
By the special geometry of fourfolds \cite{Greene:1993vm,Mayr:1996sh}, any holomorphic correlation function
can be written as derivatives of flux-dependent prepotentials, $\cF_a$,
\be   \label{Fanorm}
\cF_{a; i_1i_2\dots i_n}(t)\ =\ ({ \frac{1}{2 \pi i }} \partial_{i_1}) ({ \frac{1}{2 \pi i }} \partial_{i_2}) \dots ({ \frac{1}{2 \pi i }}\partial_{i_n}) \cF_a(t)\,,
\ee
with respect to flat cordinates ${t}^i$, where $a=1,...,{\rm dim} H^{2,2}_{\rm vert}(Y_4)$.  Via mirror symmetry, these coordinates correspond simultaneously to the natural variables of the topological $A$-model,
as well as to the flat coordinates of the  topological $B$-model on the mirror
fourfould, $\hat Y_4$.
{ Here we have included additional factors of $\frac{1}{2 \pi i }$ as compared to (\ref{Fi1i4}), which account for the normalisation of the moduli as in (\ref{Fgexp1}) and (\ref{qxi}).} Geometrically, (\ref{Fanorm}) follows iteratively
from the basic relation
\be   \label{relinvDjY4}
\frac{1}{2\pi i}\partial_i  \cF_{a|C_\beta} =  \frac{1}{2\pi i}\partial_i  \langle \langle G_a \rangle \rangle_{C_\beta} = \langle \langle G_a, D_i \rangle \rangle_{C_\beta} =: \cF_{a;i|C_\beta}   \,,
\ee
which by itself is a consequence of the divisor equation (\ref{DivEqu}).

Morally speaking, the flux-dependent prepotentials $\cF_a$ represent one-point functions for the (2,2) operators $\gamma_a$ associated with $G_a$ via (\ref{gammaG}), i.e., $\cF_a=\langle \gamma_a\rangle$. Equivalently, they are
simply the generating functions in the various flux superselection sectors labelled by $a$.

Just like for the familiar holomorphic anomaly equations for threefolds,
it is thus natural to consider an integrated version of
 the holomorphic anomaly equation that acts directly on the prepotentials $\cF_a$.
It takes the form
\be  \label{HAEFaC}
{ - \frac{1}{2 \pi i }}\overline{\partial}_{\bar i} \cF_{a} \ =\ 
 {\overline C_{\bar i}}^{jb}\left(
   \cF_{a;j}   \cF_{b}
-   I_{ab} \big\langle \sigma_1^{(2)}(\phi_j)\big\rangle
\right)\,,
\ee  
where again we have included a normalisation factor for the derivative analogous to the ones appearing in (\ref{Fanorm}).

Note that  {\it a priori}  (\ref{HAEFaC})  does not make sense for genus-zero prepotentials 
viewed as correlation functions in conformal field theory.
Recall that in conformal field theory 
 a genus-zero correlation function must contain 
three non-integrated operator insertions in order to be well-defined and non-vanishing, plus an arbitrary number of integrated operator insertions.
The analogue of this condition for prepotentials is the 
constraint (\ref{dimconstraint}), which is manifestly satisfied
by all quantities that appear in (\ref{HAEFaC}). That is,
the building blocks are the
 generating functions for the genus-zero invariants with one point fixed and 
subject to the incidence relation associated with a four-form flux $G_a$.
Addition of extra integrated vertex operators for the correlators translates into fixing additional points subject to the incidence relations
associated with extra divisor classes.
The degenerations underlying the identity (\ref{HAEFaC}) are thus the possible degenerations of stable holomorphic maps counted by the Gromov-Witten invariants.

Also note that  (\ref{HAEFaC}) is valid for general Calabi-Yau fourfolds.
For elliptic fibrations, 
one can in addition expand the prepotentials in (\ref{HAEFaC}) into the generating functionals for the relative Gromov-Witten invariants as in (\ref{Fgexp1}).
Then (\ref{HAEFaC})  translates into the following equation: 
{\be \label{HAEFaC-1}
{ - \frac{1}{2 \pi i }} \overline{\partial}_{\bar i} \cF_{a|C_\beta} \ =\  
{\overline C_{\bar i}}^{jb}\left( \sum\limits_{\substack{ C_{\beta_1} + C_{\beta_2}\\  =  C_\beta }}
  \cF_{a; j |C_{\beta_1}}  \cF_{b |C_{\beta_2}}
-   I_{ab}\, \psi \cdot \langle \langle  D_j \rangle\rangle_{C_\beta}
\right)\,.
\ee  
}
\begin{figure}[t!]
\centering
\includegraphics[width=14cm]{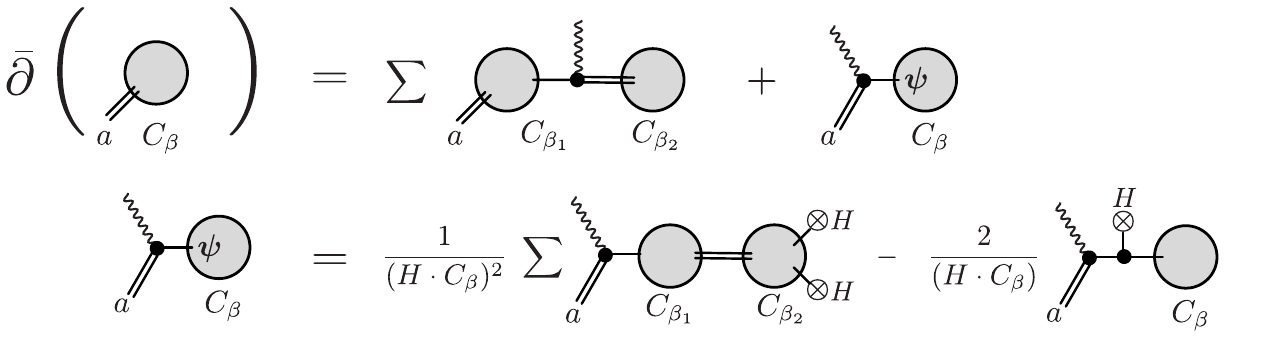}
\caption{Upper line: graphical representation of the holomorphic anomaly equation (\ref{HAEFaC-1}) for the generating function of relative Gromov-Witten invariants, $\cF_{a|C_\beta}$.
As in Fig.~\ref{fig:BCOV1}, single lines denote
$(1,1)$-form fields, double lines $(2,2)$-form fields, wavy lines the antichiral $(-1,-1)$-charged field, and solid bullets classical couplings. The second line shows the factorization of the gravitational descendant term, referring  to  Appendix ~\ref{App_A1}. The crossed circles denote insertions of an
auxiliary divisor $H$, as explained there. A noteworthy feature is that in the lower sum also
the ``trivial'' factorization
$C_\beta=C_{\beta_1}+C_{\beta_2}$, where $C_{\beta_1}=C_\beta$ and $C_{\beta_2}=$point, contributes. In this case there is only a classical contribution from $C_{\beta_2}$, which means that the other component of this factorization can contribute to the linear term in the anomaly as well.
}
\label{fig:BCOV2}
\end{figure}

The definition of $\cF_{a; j |C_{\beta_1}}$ in the first term of the bracket 
has been given in (\ref{relinvDjY4}). The quadratic first term arises whenever
the curve $C_\beta$ underlying the relative Gromov-Witten invariants is reducible into two components, 
$C_{\beta_1}$ and $C_{\beta_2}$.

The second term in the bracket of (\ref{HAEFaC-1}) denotes the generating functional for the relative gravitational descendant invariants associated with $D_j$. Here
$\psi$ denotes the class of the cotangent-line bundle on the moduli space, $\overline{\cal M}_{g=0,k=1}(Y_4, C_\beta(n,r))$, of stable holomorphic maps of genus zero with one point fixed.
This object is the Gromov-Witten-theoretic incarnation of the gravitational descendant operator  $\sigma_1^{(2)}$ in the underlying CFT \cite{Witten:1989ig,Witten:1990hr,Verlinde:1990ku,Dijkgraaf:1990qw}.\footnote{We are abusing notation here by symbolising the generating functional for the relative gravitational descendant invariants associated with $D_j$ by a dot between $\psi$ and $\langle \langle  D_j \rangle\rangle_{C_\beta}$. The more precise definition and explicit computation of this object is explained in detail in Appendix \ref{app_desc}.}
The appearance is a novel feature of the holomorphic anomaly equation for 
Calabi-Yau fourfolds, as compared to elliptic Calabi-Yau threefolds that were studied in \cite{Klemm:2012sx,Alim:2012ss}.  Notably, it leads to terms linear in prepotentials and is intimately tied to
the derivative relationships between certain flux-induced partition functions. Such terms were previously observed 
in an explicit example in ref.~\cite{Cota:2017aal}
and in our previous work \cite{Lee:2020gvu}. We will explain below how these indeed originate in the gravitational descendant term shown
in (\ref{HAEFaC-1}).

As evidenced in equation (\ref{HAEFaC-1}),  such terms can only appear in the holomorphic anomaly equations for those $\cF_a$ for which $ \overline C_{\bar i}^{jb}  I_{ab} \neq 0$. As is well known, 
the conformal field theoretic two-point function (\ref{toppairingdef}), or topological pairing, $I_{ab}$,
translates in geometry into the topological intersection numbers
\bea   \label{Iab}
I_{ab} = G_a \cdot G_b   \,.
\eea
 Its non-zero entries can be read off from (\ref{intrelG-gen}).  
 
Finally, the genus-zero gravitational descendant invariants $\psi \cdot \langle \langle D_j \rangle\rangle_{C_\beta}$ can be reduced to Gromov-Witten invariants that do not involve any powers of the contangent class $\psi$.
This reflects a general property  \cite{Witten:1989ig,Witten:1990hr,Verlinde:1990ku,Dijkgraaf:1990qw} of correlators in topological gravity, where correlators with gravitational
descendants can be expressed in terms of correlators without.  For the present geometrical
setup this is detailed in Appendix \ref{app_desc}.

Having understood the general structure of both types of terms in (\ref{HAEFaC-1}), it remains to evaluate the coefficient
 \be   \label{Cdef1}
 \overline C_{\bar i}^{jb}=\overline{\cF}_{\bar{c};\bar{i}\bar{k}} e^{2K} {G}^{j \bar k}G^{b \bar c}    
 \ee
multiplying the entire righthand side of the holomorphic anomaly
 equation (\ref{HAEFaC-1}). 
According to the general logic underlying the $tt^\ast$ formalism \cite{Bershadsky:1993cx}, the coupling (\ref{Cdef1}) 
should be evaluated in the limit where the anti-holomorphic coordinates are taken to infinity
\bea   \label{antitlimit}
\bar {t}^i   \rightarrow \infty.
\eea
Since  the three-point function $\overline{\cF}_{\bar{c};\bar{i}\bar{k}}= ({\cF}_{{c};{i}k})^\ast$ is purely anti-holomorphic, the prescription (\ref{antitlimit}) boils down to computing the latter
in the classical limit. In this regime, it reduces to the classical intersection product 
\bea   \label{tripleint}
\overline{\cF}_{{\bar c};{\bar i}\bar k}\big|_{\bar t\rightarrow\infty} 
= \int_{Y_4} G_c \wedge D_i \wedge D_k \equiv G_c \cdot D_i \cdot D_k    
 \eea
and can be easily evaluated with the help of the relations (\ref{intrelG-gen}).

Let us  next turn to the coupling matrices $G^{j \bar k}$ and $G^{b \bar c}$. From the CFT perspective, these are the inverse of the matrices
  $G_{j \bar k}$ and  $G_{b \bar c}$, which encode the overlap (\ref{Gibarjdef}) of the states associated with the respective $(1,1)$ and $(2,2)$ operators.
Here $G_{j \bar k}$ is just the familiar Zamolodchikov metric on K\"ahler moduli space for the $(1,1)$ fields,
 \be   \label{Gdef1}
 G_{j \bar k} = ({ \frac{1}{2 \pi i }} \partial_j)  ({ - \frac{1}{2 \pi i }} \partial_{\bar k}) K    \,, 
 \ee
 where $K$ denotes the K\"ahler potential. As in (\ref{Fanorm}), we have normalised the derivatives with factors of ${ \frac{1}{2 \pi i }}$ to
 properly reflect the definition of the moduli.  In the limit (\ref{antitlimit}), the metric reduces to its classical expression
 which derives from the classical K\"ahler potential
  \bea   \label{Kaehlerpot}
  K = - {\rm log}(V)    \,, \quad \quad       V = \frac{1}{4!} \int_{Y_4} J^4    \,.
 \eea

Despite the appearance of only the classical K\"ahler metric, the resulting expressions for the holomorphic anomaly equations are in general very complicated.
Luckily, as we will explain in Section \ref{sec_modularanom}, it suffices for our purposes to determine the asymptotic form of the holomorphic anomaly equations in a particular
 double scaling limit in which the K\"ahler moduli of the base, $v^\alpha$, scale to infinity much faster than the volume moduli of the fibral curves. This means that we are only interested in the limit
 \bea   \label{t-limit}
 v^i \to \infty   \qquad \text{such that} \quad \frac{\tau_2}{v^\alpha}   \to 0   \,,   \quad  \frac{{\rm Im}(z)}{v^\alpha}   \to 0    \,, \quad \frac{{\rm Im}(z)}{\tau_2} = {\cal O}(1)  \,.
 \eea

For  this limit, we will momentarily derive the following form of the holomorphic anomaly equations at genus zero:
\begin{whitebox}
\bea
 \label{HAEfirsttermsugg}
\overline{\partial}_{\bar \tau} \cF_{a|C_\beta} \ &\stackrel{(\ref{t-limit})}{=}&  \ 
{ \frac{1}{2 \pi i}} \frac{1}{ 4 \tau_2^2}    \left(  \sum_{C_{\beta_1} + C_{\beta_2} = C_\beta }     \langle\langle  G_a    \rangle \rangle_{C_{\beta_1}}  \langle\langle  \pi^\ast(C_{\beta_1}) \rangle \rangle_{C_{\beta_2}}
 -  \psi \cdot   \langle \langle \pi^\ast(\pi_\ast(G_a))  \big\rangle\rangle_{C_\beta}    \right) \nonumber \\
 \overline{\partial}_{\bar z} \cF_{a|C_\beta} \ &\stackrel{(\ref{t-limit})}{=}& 0  \label{HAEfirsttermsugg}
 \\
  \overline{\partial}_{\bar t^\alpha} \cF_{a|C_\beta} \ &\stackrel{(\ref{t-limit})}{=}& 0 \,.  \nonumber
\eea
\end{whitebox}
Here and in the sequel the symbol $\stackrel{(\ref{t-limit})}{=}$ refers to an asymptotic equality
up to terms that vanish in the limit (\ref{t-limit}).
In the first line, $\pi^\ast(C_{\beta_1})$ denotes the flux obtained by pulling back the class of the Poincar\'e dual of curve $C_{\beta_1}$ on $B_3$ to the fourfold $Y_4$.

We begin our derivation of the holomorphic anomaly equation 
with respect to the fiber parameter by setting $\bar t^{\bar i}= \bar\tau$.
The first step is to show that the coupling $\overline C_{\bar \tau}^{j b}$ appearing in (\ref{HAEFaC-1}), as  defined in (\ref{Cdef1}), 
takes the following  asymptotic form 
\bea    \label{Cbaraufinal}
\overline C_{\bar \tau}^{j b}    \stackrel{(\ref{t-limit})}{=}  { \frac{1}{(2 \pi)^2}}\frac{1}{4 \tau^2_2} \,  \delta^j_\alpha  \delta^b_{\dot\beta}  \,    I^{\alpha \dot\beta}       \,.
\eea
Here $ I^{\alpha \dot\beta} $ is the inverse of the intersection pairing, $I_{\alpha \dot\beta}$ on the base $B_3$ introduced in (\ref{Ialphadotalpha}).
The reader interested in the proof of (\ref{Cbaraufinal}) is referred to Appendix \ref{AppproofC}.

The simple structure (\ref{Cbaraufinal})  of the overall prefactor of the holomorphic anomaly
 (\ref{HAEFaC-1}) has the following consequences.
  First, recall that $\cF_{a;j|{C_{\beta_1}}}$ is the generating function for the relative
  Gromov-Witten invariants
  associated with the curve $C_{\beta_1} + n {\mathbb E}_\tau + r C^{z}$, with an additional fixed point  which must lie on $D_j$.
  By the divisor equation (\ref{DivEqu})
  these invariants equal the invariants without the additional point times the intersection number $D_j \cdot (C_{\beta_1} + n {\mathbb E}_\tau + r C^{z})$. 
  The important point is that  (\ref{Cbaraufinal}) instructs us to evaluate this intersection product only for a  
 pullback divisor $D_j = \pi^\ast(D_\alpha^{\rm b})$: In this case the intersection number is independent of the values of $n$ and $r$, and given by
  \bea
  \pi^\ast(D_\alpha^{\rm b}) \cdot (C_{\beta_1} + n {\mathbb E}_\tau + r C^{z}) = \pi^\ast(D_\alpha^{\rm b}) \cdot (C_{\beta_1})  = D_\alpha^{\rm b} \cdot_{B_3} C_{\beta_1}^{\rm b}      \,.
  \eea
Here we used that   $\pi^\ast(D_\alpha^{\rm b})$ contains the fiber and hence has vanishing intersection number with any fibral curve, and we also defined the general notation $C_{\beta_1}^{\rm b} := \pi_*(C_{\beta_1}) \in H_2 (B_3, \mathbb Z)$ for a curve class in the homology of $B_3$. 
   Therefore, with $D_j = \pi^\ast(D_\alpha^{\rm b})$ we get
   \bea
  \cF_{a;j|{C_{\beta_1}}} =  (D_\alpha^{\rm b} \cdot_{B_3} C^{\rm b}_{\beta_1})   \,  \cF_{a| C_{\beta_1}}     \,.
   \eea

Next, to evaluate the gravitational descendant term in (\ref{HAEFaC-1}), note that due to (\ref{Cbaraufinal}) the flux index $b$ must refer to a $(-2)$-flux $G_{\dot\beta}$. As a consequence of (\ref{intrelG-gen}), 
the intersection number $I_{ab}$ multiplying the second term in (\ref{HAEFaC-1}) is therefore non-zero only 
for $G_a = G_{\rho_\tau} = D_\tau  \cdot \pi^\ast(D_\rho^{\rm b})$ for some base divisor $D_\rho^{\rm b}$.
Recall that the intersection product  $I_{\rho_\tau \dot\beta} = G_{\rho_\tau} \cdot G_{\dot\beta}$ equals the intersection product $I_{\rho \dot\beta} = D_\rho^{\rm b} \cdot_{B_3} \Sigma_{\dot\beta}^{\rm b}$ on the base, i.e.,
\bea
I_{\rho_\tau \dot\beta} = I_{\rho \dot\beta} \,.
\eea
Hence
contracting $I_{a \dot\beta}$ in (\ref{HAEFaC-1})  with  $I^{\alpha \dot\beta}$ from (\ref{Cbaraufinal})  gives 
\bea
I^{\alpha \dot\beta}   I_{a \dot\beta} =   \begin{cases}  \delta^{\alpha}_\rho       &  \text{if} \quad    G_a = G_{\rho_\tau} = D_\tau  \cdot \pi^\ast(D_\rho^{\rm b}) \\
0 & \text{otherwise}   \,.
\end{cases}
\eea 
We conclude that the gravitational descendant term is present only if we compute the anomaly equation in the background of a $(0)$-flux $G_a = G_{\rho_\tau} = D_\tau  \cdot \pi^\ast(D_\rho^{\rm b})$,  and in this case the 
divisor class $D_j$ appearing in the gravitational descendant term is precisely the divisor $D_\rho$.
This fact can be compactly expressed by writing the gravitational descendant term simply as $\psi \cdot \langle \langle  \pi^\ast(\pi_{\ast} G_a) \rangle \rangle_{C_\beta}$. Here we used that 
 the pushforward formula in cohomology, applied to the basis (\ref{basisH22vert-Y4}) of fluxes, evaluates to
 \be
 \begin{split}
 \pi_{\ast} G_{\alpha_\tau} &= D_\alpha^{\rm b} \,, \qquad 
  \pi_{\ast} G_{\alpha_z} = 0   \,, \qquad
   \pi_{\ast} G_{\dot\alpha} = 0    \,.
 \end{split}
 \ee

Putting everything together we find that the holomorphic anomaly equation at genus zero with respect to $\bar\tau$, in the regime (\ref{t-limit}), takes the following form:
\be  \boxed{ \label{HAEversion1}
 { - \frac{1}{2 \pi i}}    \overline{\partial}_{\bar \tau} \cF_{a|C_\beta} \  \stackrel{(\ref{t-limit})}{=}   \ 
 { \frac{1}{(2 \pi)^2}}   \frac{1}{ 4 \tau_2^2}    \left(  \sum_{C_{\beta_1} + C_{\beta_2} = C_\beta }  I^{\alpha \dot\gamma} (D_\alpha \cdot C_{\beta_1})  \cF_{a|C_{\beta_1}} \cF_{\dot\gamma| C_{\beta_2}} 
 -    \psi \cdot \langle \langle  \pi^\ast(\pi_\ast(G_a))  \big\rangle\rangle_{C_\beta}    \right)
\,.}
\ee

Note that since $I^{\alpha \dot\gamma} (D_\alpha \cdot C_{\beta_1}) = c_{\beta_1}^{\dot\gamma}$ is nothing but the expansion coefficients of the class $C_{\beta_1}^{\rm b} = c_{\beta_1}^{\dot\gamma} \Sigma_{\dot\gamma}^{\rm b}$,
the first term can also be written more suggestively in the form given in (\ref{HAEfirsttermsugg}).

As remarked already, the gravitational descendant term  can be non-vanishing only for $(0)$-fluxes of the form $G_{\alpha_\tau} = D_\tau \cdot  \pi^\ast(D^{\rm b}_\alpha)$.
How it can be evaluated is detailed in Appendix \ref{app_desc}, and the result reads:
\be   \label{psiequation}
\begin{split}
\psi\cdot \langle \langle \pi^\ast(D^{\rm b}_\alpha) \rangle \rangle_{C_\beta} &= \frac{1}{(H^{\rm b} \cdot_{B_3} C_\beta^{\rm b})^2} (({\rm I}) + ({\rm II})) -  \frac{2}{(H^{\rm b} \cdot_{B_3} C_\beta^{\rm b})}   ({\rm III})  \,,   \cr
({\rm I}) &\ = (D^{\rm b}_\alpha  \cdot_{B_3} C_\beta^{\rm b}) \sum_{\gamma, \dot\alpha}  \langle \langle G_{\dot\alpha} \rangle \rangle_{C_\beta}   \,   I^{\dot \alpha  \gamma}    \,  ( D^{\rm b}_\gamma \cdot_{B_3} H^{\rm b} \cdot_{B_3}  H^{\rm b})    \,,  \cr
({\rm II})   &\ =    \sum_{C_{\beta_1} + C_{\beta_2} = C_\beta, C_{\beta_i} \neq 0}   (D^{\rm b}_\alpha  \cdot_{B_3} C_{\beta_1}^{\rm b})  \langle \langle G_a \rangle\rangle_{C_{\beta_1}}   I^{ab} \langle\langle G_b \rangle\rangle_{C_{{\beta_2}}}   (H^{\rm b}  \cdot_{B_3} C_{\beta_2}^{\rm b})^2 \,,\cr
({\rm III})   &\ =  \langle \langle  \pi^\ast(H^{\rm b}) \cdot  \pi^\ast(D^{\rm b}_\alpha) \rangle \rangle_{C_\beta}
    \,.
\end{split}
\ee
For a visualisation, see Fig.~\ref{fig:BCOV2}.
Above, $H^{\rm b}$ denotes an auxiliary divisor class on $B_3$ whose precise choice is irrelevant provided that  $H^{\rm b} \cdot_{B_3} C_\beta^{\rm b} \neq 0$. Moreover
recall that $I^{ab}$ is the inverse of the intersection form (\ref{Iab}), and $I^{\gamma \dot\alpha}$ is the inverse of $I_{\gamma \dot\alpha} = D^{\rm b}_\gamma \cdot_{B_3} \Sigma_{\dot\alpha}^{\rm b}$ as introduced in (\ref{intrelG-gen}).

Note that $({\rm I})$ and $({\rm III})$ are linear in partition functions, which will play an important role for realising the algebra
of derivatives that will be introduced later in (\ref{TqDq}) and (\ref{DTalgebra}) as well as symbolically represented in Fig.~\ref{fig:anomalyfluxes}.

One can repeat this derivation also for the holomorphic anomaly equation with respect to $\bar z$ and $\bar t^\alpha$. However, due to the explicit form of the Zamolodchikov metric, one finds that 
in both cases the result is suppressed by powers of the base coordinates and therefore vanishes in the asymptotic regime (\ref{t-limit}). This explains the last two equations in (\ref{HAEfirsttermsugg}).

\subsection{Example: Elliptic Fibration over $B_3 = \mathbb P^3$}

It is instructive to evaluate the holomorphic anomaly equation (\ref{HAEversion1}) for the simplest possible example,
namely for an elliptic fibration over base $B_3 = \mathbb P^3$  (later in Section~\ref{SecExS} we will consider a much more involved case). Gromov-Witten invariants for this model have been considered before in refs.~\cite{Klemm:2007in,Haghighat:2015qdq,Cota:2017aal}. We will consider a refinement of this model, namely in order to obtain an
extra $U(1)$ gauge symmetry,
we introduce an additional section $S$ with associated Shioda image $\sigma(S)$. 
 The basis (\ref{basish11-Y4}) of $H^{1,1}(Y_4)$ boils down to
  \bea
 D_\tau = S_0 + \pi^\ast(2L)   \,,  \quad  D_z = \sigma(S)   \,,  \quad D_1 = \pi^\ast(L)\,,
 \eea
 where $\bar K_{{\mathbb P}^3} = 4L$ denotes the anti-canonical class and $L$ the hyperplance class of the base $\mathbb P^3$. 
In this notation, the basis (\ref{basisH22vert-Y4}) of $H^{2,2}_{\rm vert}(Y_4)$ reduces to
\bea
 G_{1_\tau}\ \equiv\ G^{(0)}_{1_\tau} &=& D_\tau \cdot D_1\,,   \\
 G_{1_z}\ \equiv\ G^{(-1)}_{1_z}   &=& D_z \cdot  D_1   \,,   \\
  G_{\dot 1}\ \equiv\ G^{(-2)}_{\dot 1} &=& D_1 \cdot D_1  \,.
\eea
For this simple geometric background the holomorphic anomaly equation (\ref{HAEversion1}) becomes
\bea \label{FaP3int1}
\overline{\partial}_{\bar \tau} \cF_{a|C_\beta} \ \stackrel{(\ref{t-limit})}{=} \ 
{ \frac{1}{2 \pi i}} \frac{1}{ 4 \tau_2^2}    \left(
 \sum_{C_{\beta_1} + C_{\beta_2} = C_\beta }   (\pi^*(L) \cdot C_{\beta_1})  \cF_{a| C_{\beta_1}} \cF_{\dot 1| C_{\beta_2}}
 -     \delta_{a1_\tau}   \,\psi\cdot \langle \langle \pi^*(L) \rangle \rangle_{C_\beta}  \right)  \,.
\eea
We first evaluate this expression for the $(0)$-flux, $G_a = G_{1_\tau}$, 
for which the gravitational descendant term is non-zero.
Let us parametrise $C_\beta^{\rm b}= d \, (L \cdot L) \equiv C_d^{\rm b}$  on $\mathbb P^3$.
Then we find that the various terms in (\ref{psiequation}), with the choice $H^{\rm b} = L$, turn into 
\bea
({\rm I})   &=&  d \,  \langle \langle G_{\dot 1}  \big\rangle \rangle_{C_d}    \,\\
({\rm II}) &=&  \sum_{s=1}^{d-1}   (d^2 s - s^2 d)  \,    \langle \langle G_{1_\tau}  \big\rangle \rangle_{C_s}    \langle \langle G_{\dot 1}  \big\rangle \rangle_{C_{d-s}} \, \\
({\rm III})   &=&   \langle \langle G_{\dot 1}  \big\rangle \rangle_{C_d}    \,, 
\eea
which altogether yields for the gravitational descendant term:
\bea
\psi \cdot \langle \langle \pi^*( L ) \big\rangle \rangle_{C_d}   & =&   - \frac{1}{d}  \langle \langle G_{\dot 1}  \big\rangle \rangle_{C_d}  +  \sum_{s=1}^{d-1}   (s - \frac{s^2}{d})  \,    \langle \langle G_{1_\tau}  \big\rangle \rangle_{C_s}    \langle \langle G_{\dot 1 }\big\rangle \rangle_{C_{d-s}}   \\
&=& - \frac{1}{d} \cF_{\dot 1| C_{d}}  + \sum_{s=1}^{d-1}   (s - \frac{s^2}{d})       \cF_{1_\tau| C_{s}} \cF_{\dot 1| C_{d-s}}   \,.
\eea

The terms proportional to $s$ cancel against the quadratic terms in (\ref{FaP3int1}) and so the final result is
\bea \label{0fluxP3}
\overline{\partial}_{\bar \tau} \cF_{1_\tau|C_d} \ \stackrel{(\ref{t-limit})}{=} \ { \frac{1}{2 \pi i}}   \frac{1}{ 4 \tau_2^2}  \left(   \sum_{s=1}^{d-1} \frac{s^2}{d}     \cF_{1_\tau| C_{s}} \cF_{\dot 1| C_{d-s}}   + \frac{1}{d} \cF_{\dot 1| C_{d}}    \right)     \,.
\eea
By contrast, for the $(-1)$- and $(-2)$-fluxes the gravitational descendant term vanishes and the result has the simpler standard form
\bea
\overline{\partial}_{\bar \tau} \cF_{a|C_d} \ \stackrel{(\ref{t-limit})}{=} \  { \frac{1}{2 \pi i}} \frac{1}{ 4 \tau_2^2}  \left(     \sum_{s=1}^{d-1} s    \cF_{a| C_{s}} \cF_{\dot 1| C_{d-s}}    \right)     \,,   \qquad a=1_z, \dot 1 \,.
\eea
While these equations have already been observed in~\cite{Cota:2017aal} for the specific example of a smooth Weierstrass model over $\mathbb P^3$, our derivation 
via the holomorphic anomaly equation (\ref{HAEversion1})
puts them
on general grounds, making contact to \cite{Oberdieck:2017pqm}. Our derivation shows, in particular, how the linear term on the right-hand side of (\ref{0fluxP3}), which has no analogue for Calabi-Yau threefolds, originates in the flux-induced gravitational descendant invariants (which in addition  contribute also
quadratic terms to the holomorphic anomaly equation).

\section{Holomorphicity versus Modularity}\label{sec_modular}

So far we have been discussing holomorphic anomalies as they
arise from topological strings at genus zero in the formalism of BCOV~\cite{Bershadsky:1993ta,Bershadsky:1993cx}.  
For the topological string on elliptic fibrations, one can equivalently trade holomorphic against
modular anomalies, the latter being more transparent in geometry. 
Indeed, Calabi-Yau spaces which are elliptic fibrations
are well known to have distinguished modular symmetries acting on their moduli space.
Often one can exploit these symmetries to
determine infinitely many Gromow-Witten invariants via modular completion,
and therefore the exact partition functions in terms of finite input data. See especially
refs.~\cite{Alim:2012ss,Klemm:2012sx,Schimannek:2019ijf,Cota:2019cjx} for detailed expositions of the
properties of elliptic threefolds.

More specifically, our concern are the (partly anomalous)
modular properties of elliptic fourfolds with
various fluxes switched on. The modular or almost modular objects in question will be the relative
genus-zero flux-induced partition functions defined by
 \be\label{relativeZ2}
 {\cal Z}_{w,m}[G, C_\beta](\tau,z) =
  -q^{E_0}  {\cal F}_{G| C_\beta}(\tau,z) = -q^{E_0}\sum_{n, r} N_{G} (C_\beta(n,r))   \,  q^n  \, \xi^{r}   \,,
\ee
where $E_0$ has been introduced in (\ref{E0def}).

Following refs.~\cite{Lee:2019tst,Lee:2020gvu} we already pointed out that the modular properties of the partition functions,
in particular their modular weight $w$, depend on the flux background. This  is reflected by
labelling the flux  as $G=G^{(w)}$.
Depending on the flux geometry, $ {\cal Z}_{w,m}[G, C_\beta]$
can have modular weight $w\in\{-2,-1,0\}$. 
In presence of an extra $U(1)$ gauge symmetry, the partition function depends
also on $\xi \equiv e^{2\pi iz}$.
The modular symmetries get extended such as to include elliptic transformations, which express the (potentially anomalous) double periodicity in the variable $z$. Then the partition function has an extra label, the integral
index $m$ (with obvious generalization if there are several $U(1)$ symmetries\footnote{See ref.~\cite{Lee:2018spm} for a concrete exposition of such a generalisation for Calabi-Yau threefolds.}).

Note that the object ${\cal Z}_{-1,m}[G^{(-1)}, C_\beta](\tau,z)$ is not only defined, as presently,
 via the topological $A$-model on $Y_4$, but
can also be interpreted as the elliptic genus (\ref{ellgen-def1}) of a string obtained by wrapping a D3-brane on $C_\beta$ in F-theory compactified on $Y_4 \times T^2$.
In this picture, $E_0$
represents the ground state energy of the Ramond sector of the string worldsheet theory (which for heterotic strings is given by $E_0=-1$).
In Section \ref{sec:outlook} we will speculate about extending this interpretation also to the other flux backgrounds of type $(0)$ and $(-2)$.

Irrespective of their physics interpretation, our task will be to write the partition functions $ {\cal Z}_{w,m}[G, C_\beta]$ 
 in terms of suitable modular functions.
Most of these functions are well known and we will briefly review them in the  next section. We will put particular emphasis on the relation between
modular anomalies, holomorphic anomalies and the appearance of derivatives with respect to $\tau$ and $z$. 
In Section~\ref{sec_modularanom} we will then translate the holomorphic anomalies derived in Section \ref{sec_top}  into the system of modular anomalies summarised in (\ref{MAEgen}) and (\ref{E1equ}).

\subsection{The Ring of Quasi-Jacobi Forms}   \label{sec_RingofQuasis}

A key role is played by certain (quasi-)modular
 and (quasi-)Jacobi forms, which make
the modular symmetries and their anomalies manifest. We begin with a brief review
of some familiar facts  and refer to
Appendix~\ref{app_jacobi} for definitions and more details. 

An important feature is that the graded ring of holomorphic
modular forms $\cR^M=\oplus_w \cR^M_w$ is freely generated by the Eisenstein series $E_4=E_4(\tau)$ and $E_6=E_6(\tau)$  of modular weight
4 and 6, respectively. This means that any holomorphic
modular form of given weight $w$, generically denoted by
$\Phi_w^M$, can be written as a polynomial in these generators,
\be
\Phi_w^M=\Phi_w^M(E_4,E_6)\in \cR^M_w\,.
\ee
We have seen that for
 flux compactifications on fourfolds certain partition functions are related via derivatives to
others.  
This statement has been made precise in (\ref{qderivative1}), and the 
relation to the present discussion has also been anticipated in
 the last paragraph 
of Section \ref{sec_Fldeppre}.
Derivatives however map outside of $\cR^M_w$, and in particular we have
\be\label{Serre}
q\partial_q \Phi_w^M=\tilde\Phi_{w+2}^M+\frac w{12}E_2\Phi_w^M\,,
\ee
where the Eisenstein series $E_2=q\partial_q \log \et24(q)$ is not a modular, but just a quasi-modular form. That is, playing the role of a connection, it transforms with an anomalous piece:
\be\label{E2quasi}
E_2\left(\frac{a\tau+b}{c\tau+d}\right)\ =\  (c\tau+d)^2E_2(\tau)-\frac{6i}\pi c (c\tau+d)\,.
\ee
As an extra generator it extends $\cR^M$ to the ring $\cR^{QM}$ of quasi-modular forms,
\be
\Phi_w^{QM}=\Phi_w^{QM}(E_2,E_4,E_6)\in \cR^{QM}_w\,,
\ee
which maps under the action of $q\partial_q$ into itself.

A well known and important point is that $E_2 $ can be uniquely completed 
into a good modular, but only {\it almost holomorphic} form by defining
\be\label{e2hat}
\hat E_2(\tau)\ =\ E_2(\tau)-24\nu\,,\qquad \nu\equiv\frac1{8\pi {\rm Im}\tau}.
\ee
This leads to the ring of almost holomorphic modular forms with elements
\be
\hat\Phi_w=\Phi_w(\hat E_2,E_4,E_6)\in \cR^{AH}_w\,,
\ee
which we customarily denote by a hat. Demanding that partition functions be modular
leads to $\cR^{AH}$ as the physically relevant modular functions to consider (as remarked in the Introduction
and later in Section~\ref{sec:outlook},     
the non-holomorphic part generically arises from zero-modes due to degenerating geometries).
Since in these functions $E_2$ and $\nu$ always appear packaged together in terms of $\hat E_2$, taking
derivatives with respect to either one yields
the same result  (up to factors), ie.,
\bea\label{antiholder}
\partial_{E_2}\hat\Phi_w &=& -\frac1{24}\partial_{\nu}\hat\Phi_w
\\
&=&  -\frac132\pi i\,( {\rm Im}\tau)^2\partial_{\bar \tau}\hat\Phi_w\,.
\eea
In the second line we have transformed $\partial_\nu$ to the anti-holomorphic derivative with
respect to $\bar\tau$, which makes contact between the holomorphic
anomaly equations discussed in Section~\ref{sec_BCOV} and the modular anomaly equation
that will be discussed in Section~\ref{sec_modularanom}.

From (\ref{Serre}) and (\ref{antiholder}) it is clear that derivatives
with respect to $\tau$ and $E_2$ (or~$\bar\tau$) are in a sense dual to each other; 
this will play an important role later.  In fact one can define, following
\cite{Oberdieck:2017pqm},
abstract derivative operators, $T_*$ and $D_*$, whose specific representation depends
on whether they act on $\cR^{QM}$ or $\cR^{AH}$.
Explicitly, one defines the following operators acting on holomorphic quasi-modular
forms:
\bea
D_q:=q\partial_q:  \qquad\ \cR^{QM}_w &\rightarrow& \cR^{QM}_{w+2}
\\
-\frac1{24}T_q:=\partial_{E_2}:\qquad \ \cR^{QM}_w &\rightarrow& \cR^{QM}_{w-2}\,.
\eea
On the other hand, the equivalent operators acting
on almost holomorphic modular forms take the form
\bea
D_\nu:=
\nabla_{q,w} \equiv 
( q\partial_q - 2 w \nu + 2 \nu^2 \partial_\nu):
\qquad
\cR^{AH}_w &\rightarrow& \cR^{AH}_{w+2}
\\
T_\nu:= \, \partial_{\nu}\ 
=\  16\pi i\,( {\rm Im}\tau)^2\partial_{\bar \tau}:
\qquad\qquad\ \ \
 \cR^{AH}_w  &\rightarrow&  \cR^{AH}_{w-2}\,.
\eea
Evidently the representation on holomorphic forms is simpler, and this is why anomaly equations are often represented in terms of
derivatives with respect to $E_2$ rather than to $\nu$ or $\bar \tau$.

Either way, the vague statement (\ref{deristruct})  that holomorphic and anti-holomorphic  derivatives are dual
 to each other can now be sharpened by writing
\be\label{TqDq}
\big[T_q,D_q\big] \ =\ \big[ T_\nu,D_\nu\big]\ =\ -2w\, {\rm id}\,.
\ee

We now extend the previous discussion to Jacobi forms and quasi-modular generalizations thereof,
which depend on the extra elliptic variable $\xi\equiv e^{2\pi i z}$. Our presentation is guided by the expositions given by  refs.~\cite{libgober2009elliptic,oberdieck2012serre,Oberdieck_2018,Oberdieck:2017pqm},  deferring again basic definitions and details to Appendix~\ref{app_jacobi}.

The starting point is the bi-graded ring $\cR^J  = \oplus_{w,m}\cR^J_{w,m}$
of holomorphic weak Jacobi forms whose generators can be taken as\footnote{Note that  $\phi_{-1,2}^2$ is not independent.}
\be
\cR^J\ =\ \IQ\big[E_4,E_6,\phi_{-2,1},\phi_{-1,2},\phi_{0,1}\big]\,.
\ee
Here $\phi_{w,m}=\phi_{w,m}(\tau,z)$ are the standard Jacobi generators with given
modular weight $w$ and index $m$, whose definition is given in (\ref{defJacobi}). 
Any polynomial in the generators with definite weight and index, 
$\Phi^J_{w,m}\in \cR^J_{w,m}$, transforms nicely under modular (\ref{Jacmodular}) and elliptic 
(\ref{periodicity}) transformations.

As before, we will need to figure out how to express
derivatives acting  on $\cR^J$, with respect to both $\tau$ and $z$, in terms of automorphic
functions.  This will lead to the ring 
$\cR^{QJ}$ of {\it meromorphic quasi-Jacobi forms}, which is much more intricate
than the ring of quasi-modular forms, $\cR^{QM}$. 

Concretely, since the derivative $\frac1{2\pi i}\partial_z\equiv \xi\partial_\xi$ increases the modular weight by one unit, we need to
find a connection with modular weight one. The relevant object to
consider is \cite{weil1976elliptic}
\beq\label{E1def}
E_1(q,\xi)\ =\ \xi\partial_\xi \log \vartheta_1(z,\tau)\,,
\eeq
which is a prime example of a meromorphic quasi-Jacobi form.
Indeed, in analogy to $E_2$, it displays an anomalous
behavior under modular and elliptic transformations:
\bea\label{E1trans}
E_1\left(\frac{a \tau + b}{c \tau +d}, \frac{z}{c \tau +d} \right)&=&  (c\tau+d)E_1(\tau,z)+c\, z\nn   \,,
\\
E_1\left( \tau , z + \lambda \tau + \mu \right) &=&E_1(\tau,z)-\lambda    \,.
\eea
Moreover, it is meromorphic in the sense of having a pole in $1/z$. This
exhibits the fundamental need to go beyond holomophic forms. 
More details about the ring of meromorphic quasi-Jacobi forms, $\cR^{QJ}$, can be found in Appendix~\ref{app_jacobi}. 

Suffice it to mention here  what will be immediately relevant
for our purposes, namely the action of derivatives
on arbitrary weak Jacobi forms, $\Phi^J_{w,m}\in \cR^J_{w,m}$:
\bea\label{xiaction}
\xi\partial_\xi\, \Phi^J_{w,m} &=& \frac{\tilde \Phi^J_{w,m+2}}{\phi_{-1,2}} + 2m\, E_1\, \Phi^J_{w,m}\,,
\\\label{qaction}
q\partial_q \Phi^J_{w,m} &=& \frac{\tilde \Phi^J_{w,m+1}}{\phi_{-2,1}} 
+ E_1\frac{\tilde \Phi^J_{w,m+2}}{\phi_{-1,2}} 
+ \left(\frac w{12}E_2+m\, E_1^2\right) \Phi^J_{w,m}\,,
\eea
where on the right hand side  some unspecified generic weak Jacobi forms, $\tilde\Phi^J_{w,*}\in \cR^J_{w,*}$, appear.
Note, importantly, that despite of  the meromorphic
building blocks, the poles in $1/z$ and $1/z^2$ must cancel out, so that the expressions are holomorphic and the derivatives map within the subset of {\it holomorphic} quasi-Jacobi forms.

 In analogy to the familiar modular completion of $E_2$ in 
 eq.~(\ref{e2hat}), one can augment also $E_1$ by a
 mildly anholomorphic piece, 
 \be\label{e1hat1}
 \hat E_1(\tau,z)=E_1(\tau,z)+\alpha \,, \qquad 
 \alpha\equiv \frac{{\rm Im}z} {{\rm Im}\tau}\,,
 \ee
 to yield what we call an {\it almost meromorphic Jacobi form}.  Indeed, given that
  \bea
\alpha\left(\frac{a \tau + b}{c \tau +d}, \frac{z}{c \tau +d} \right) 
 &=&
 (c\tau+d) \alpha(\tau,z) -c\,z\,,
 \\
\alpha\left( \tau , z + \lambda \tau + \mu \right) 
 &=&
\alpha(\tau,z) +\lambda\,,
 \eea
we see that 
$\hat E_1$ transforms nicely under modular (\ref{Jacmodular}) and elliptic (\ref{periodicity}) transformations, namely like a Jacobi form with weight $w=1$ and index $m=0$.

The upshot is that the functions which are relevant in the present context
are {\it almost holomorphic Jacobi forms}, $\Phi^{AHJ}$. Loosely speaking, these are
polynomially generated by the meromorphic Jacobi forms
\be\label{RAHJdef}
\cR^{AHJ}\ =\ \IQ\big[\hat E_1,\hat E_2, E_4,E_6,\phi_{-2,1},\phi_{-1,2},\phi_{0,1}\big]
\big/\{\phi_{-1,2},\phi_{-2,1}\}\,,
\ee
modulo division by powers of $\phi_{-1,2}$ and $\phi_{-2,1}$ such that all poles in $z$ cancel;
this is signified by the formal divison above  
(a more precise definition is given in Appendix~\ref{app_jacobi}).
Prime examples for such
are the expressions in (\ref{xiaction}) and (\ref{qaction}) with the replacements
$E_1\to\hat E_1$, $E_2\to\hat E_2$.

Now turning to holomorphic anomaly equations, we immediately observe from
(\ref{xiaction}) and (\ref{qaction}) that when acting on weak Jacobi forms
$\Phi^J_{w,m}\in \cR^J_{w,m}$ we get:
\bea\label{simplederEi}
\partial_{E_1} \xi\partial_\xi\, \Phi^J_{w,m} &=& 2m \,\Phi^J_{w,m}  \,,\nn
\\
\partial_{E_2} q\partial_q\, \Phi^J_{w,m} &=& \frac w{12} \Phi^J_{w,m} \,,
\\
\partial_{E_1} q\partial_q\, \Phi^J_{w,m} &=& \xi\partial_\xi  \,\Phi^J_{w,m}   \,.\nn
\eea
When acting on quasi-Jacobi forms, these simple relations
do not hold any more.  Instead, invariant statements can be made by considering
commutators of derivatives, in analogy to eq.~(\ref{TqDq}).

\begin{figure}[t!]
\centering
\includegraphics[width=13cm]{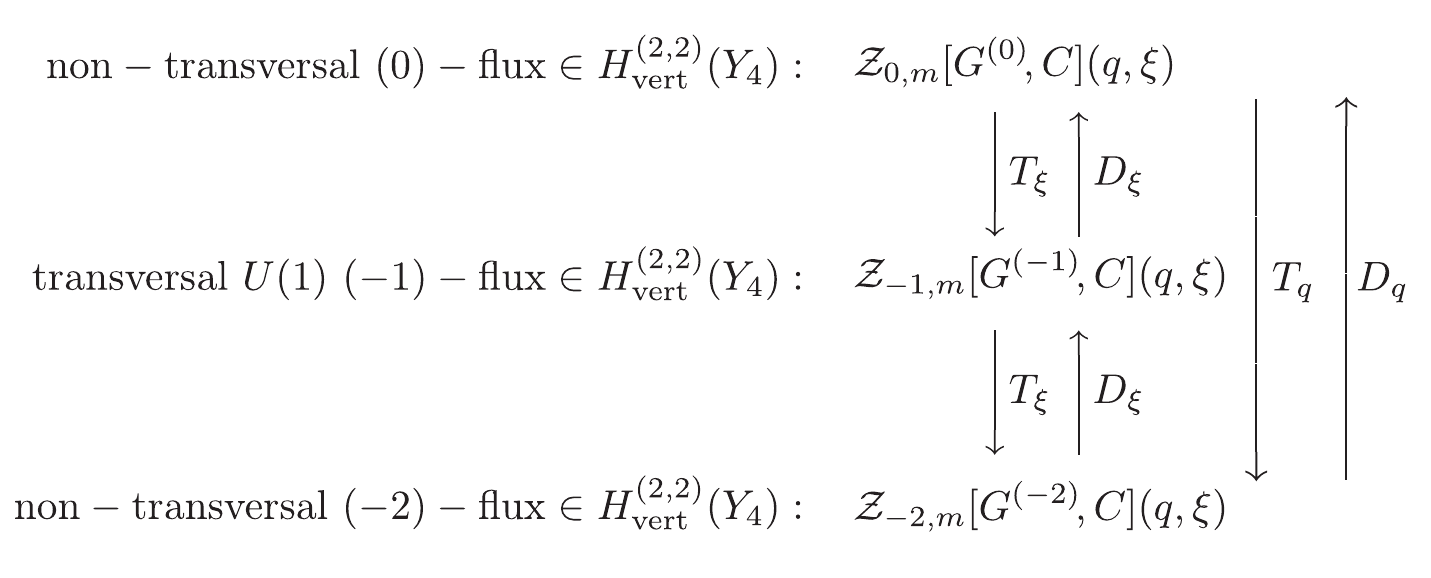}
\caption{
Shown is how the holomorphic version of the algebra of derivatives (\ref{DTalgebra}) acts between the various flux partition functions. As discussed in  \cite{Lee:2020gvu}, the partition function of modular weight $w=-1$,
${\cal Z}_{-1,m}$, coincides with the refined elliptic genus of a chiral $N=1$ supersymmetric theory in four dimensions with  $U(1)$ gauge group, while ${\cal Z}_{-2,m}$  formally corresponds, for certain geometries, to the elliptic genus of a six dimensional theory.
}
\label{fig:anomalyfluxes}
\end{figure}

Analogously to our previous discussion,  there is an isomorphism between holomorphic quasi-Jacobi forms and almost holomorphic Jacobi forms, and one can formalize the action of the various derivatives as follows \cite{Oberdieck:2017pqm}:
\bea 
\label{RQJAHJ}
\cR^{QJ} \qquad &\longleftrightarrow&\qquad  \cR^{AHJ}
\\
-\frac1{24}T_q:= \partial_{E_2}   \qquad && \qquad
 T_\nu:=\partial_\nu=16 \pi i\, {\rm Im}\tau\,( {\rm Im}\tau \, \partial_{\bar \tau}+{\rm Im}z \,\partial_{\bar z})\nn
\\
T_\xi:= \partial_{E_1}   \qquad && \qquad 
T_\alpha:=\partial_\alpha
= -2i\,{\rm Im}\tau\,\partial_{\bar z} \nn
\\
D_q:= q\partial_{q}   \qquad && \qquad  D_\nu:=
 q\partial_{q} - 2 w \nu + 2 \nu^2 \partial_\nu + \alpha \xi \partial_\xi + m \alpha^2\nn
\\
D_\xi:= \xi\partial_{\xi}   \qquad && \qquad D_\alpha:=
 \xi\partial_{\xi}  + 2m \alpha - 2 \nu \partial_\alpha\nn    \,.
\eea
The operators on the right are meant to act on functions of the form 
$ \Phi(q,\xi,\nu,\alpha)\in \cR^{AHJ}_{w,m}$.
Then one can show, in  extension of (\ref{TqDq}),  the  following commutation relations defining two isomorphic algebras:
\bea\label{DTalgebra}
\big[ T_\xi,D_q\big] &=& D_\xi\,,   \qquad  \ \ \ \big[ T_\alpha,D_\nu\big]\  =\ D_\alpha\,,  \nn
\\
\big[ T_\xi,D_\xi\big] &=& 2m\,{\rm id}\,, \qquad  \big[ T_\alpha,D_\alpha\big] \ =\ 2m\,{\rm id}\,,
\\
\big[ T_q,D_\xi\big] &=& -2T_\xi\,, \qquad  \big[ T_\nu,D_\alpha\big]\ =\ -2T_\alpha\,,\nn
\eea
with the understanding that the remaining commutators except for~\eqref{TqDq} vanish.
See Fig.~\ref{fig:anomalyfluxes} for how the algebra acts between the various flux sectors. 

While this structure directly follows from the properties of quasi-Jacobi forms and their derivatives,  our discussion explains it also purely in geometry in that the derivative structure of the partition functions, 
as expressed by the up-arrows $\uparrow\!\!D_*$ in Fig.~\ref{fig:anomalyfluxes}, 
ties together with the linear terms in the holomorphic anomaly equations, 
which underlie the down-arrows, $\downarrow\!T_*$.

\subsection{Modular and Elliptic Anomaly Equations}   \label{sec_modularanom}

The upshot of the previous section is that 
 the holomorphic anomaly quantified in (\ref{HAEfirsttermsugg}) can equivalently be expressed in terms of a modular anomaly.
The modular anomaly is encoded in the dependence of the flux-induced generating functions, $\cF_{a|{C_\beta}}$,
on the quasi-modular and quasi-Jacobi forms, $E_2$ and $E_1$. To each of these we can associate a certain type of
modular anomaly equation. More precisely,
as we will explain later in this section, we find for the generating functions of relative genus-zero Gromov-Witten invariants on fourfolds
the following two types of anomaly equations at genus zero:
\begin{whitebox}
  \bea
&&  \text{Modular Anomaly Equation:} \label{MAEgen}  \\
 && {\partial}_{E_2} \cF_{a|C_\beta} \ =\ 
 - \frac{1}{12 }  
   \big(  \sum\limits_{\substack{ C_{\beta_1} + C_{\beta_2}\\  =  C_\beta }}     \langle\langle  G_a    \rangle \rangle_{C_{\beta_1}}  \langle\langle  \pi^\ast(C_{\beta_1}) \rangle \rangle_{C_{\beta_2}}
 -  \psi \cdot   \langle \langle  \pi^\ast \pi_\ast(G_a) \rangle\rangle_{C_\beta}    \big) \nn \\
&&  \text{Elliptic Anomaly Equation:} \label{E1equ}  \\
&& \partial_{E_1} \cF_{a|{C_\beta}}  = - \langle\langle  \pi^\ast \pi_\ast( D_z \cdot G_a)   \rangle\rangle_{C_\beta}     +    \langle\langle     D_z \cdot \pi^\ast \pi_\ast(G_a) \rangle\rangle_{C_\beta}  \,.\nn
  \eea
 \end{whitebox}
 Here we introduced
\bea
    \pi^\ast \pi_\ast( D_z \cdot G_a)  &=& \left\{\begin{array}{ll}
        - \pi^\ast(b) \cdot  \pi^\ast(D_\alpha^{\rm b}) & \text{for }  G_a = D_z \cdot \pi^\ast(D^{\rm b}_\alpha) \\
        \quad 0 & \text{otherwise } 
        \end{array}\right. \\
   \pi^\ast \pi_\ast(G_a)  &=& \left\{\begin{array}{ll}
      \pi^\ast(D_\alpha^{\rm b}) & \text{for }  G_a = D_\tau \cdot \pi^\ast(D^{\rm b}_\alpha) \\
        \quad 0 & \text{otherwise, } \qquad\qquad 
        \end{array}\right.
  \eea
while $b$ denotes the height-pairing (\ref{heightpairingdef}) associated with the rational section of the fibration $Y_4$.

Hence, the elliptic anomaly equation (\ref{E1equ}) can be written even more explicitly as

 \be
 \begin{aligned}
 \label{E1equ-2}  
\partial_{E_1}   \cF_{\alpha_\tau|{C_\beta}} &= \cF_{\alpha_z|{C_\beta}}   \cr
 \partial_{E_1}   \cF_{\alpha_z|{C_\beta}} &= \cF_{G= \pi^\ast(b \cdot D^{\rm  b}_\alpha)|{C_\beta}}   \cr
  \partial_{E_1}   \cF_{\dot \gamma|{C_\beta}} &=0  \,.
 \end{aligned}
 \ee

Equations (\ref{MAEgen}) and (\ref{E1equ}) should be viewed as an explicit example (for fourfolds at genus zero)
of the abstract modular and elliptic anomaly equations that were proposed  for general elliptic $n$-folds in~\cite{Oberdieck:2017pqm}.
This work conjectured the analogue of (\ref{MAEgen}) and deduced from it (\ref{E1equ}) with the help of the commutator relations (\ref{DTalgebra}). 
One of the new points of our work is to provide a detailed derivation of (\ref{MAEgen}) via the holomorphic anomaly equations in their form (\ref{HAEfirsttermsugg}), which in turn we have deduced from the physical, conformal field theoretic BCOV formalism as applied to 
Calabi-Yau fourfolds with flux background.

Let us begin with the derivation of (\ref{MAEgen}), which is to be understood as an equation acting on some holomorphic quasi-Jacobi form.
As explained in the previous section, the dependence of such objects  on $E_2$ and $E_1$
induces by isomorphism a dependence on the anti-holomorphic variables $\bar \tau$ and $\bar z$,
namely by replacing $E_2$ by $\hat E_2$ and $E_1$ by $\hat E_1$, respectively.
The first replacement leads to an appearance of powers of $\nu =1/(8 \pi \tau_2)$  via 
 (\ref{e2hat}) and the second replacement leads to powers of $\alpha= {\rm Im}(z)/\tau_2$ via  (\ref{e1hat1}).
 
 Hence, we can find an equation for $\partial_{E_2} {\cal F}_{a|C_{\beta}}$ if we are able to isolate the dependence on $\bar \tau$ only via powers of $1/\tau_2$. At first, the expression for $T_\nu$ in (\ref{RQJAHJ}) may appear as an obstacle against doing so, because of its dependence on Im$(z)\,\partial_{\bar z}$.
 
Luckily, the limit (\ref{t-limit}) for which we have derived the holomorphic anomaly equation (\ref{HAEfirsttermsugg}) with respect to $\bar\tau$
provides a resolution of this.  That is,
up to an overall rescaling of the K\"ahler form, this limit can equivalently be characterised
by stating that the fiber moduli $\tau_2$ and ${\rm Im}(z)$ scale to zero while the base moduli stay
 finite, in such a way that ${\rm Im}(z)/\tau_2$ stays fixed.
In this limit, powers of $1/\tau_2$, associated with the replacement of $E_2$ by $\hat E_2$, are enhanced, while powers of ${\rm Im}(z)/\tau_2$ from the replacement
 of $E_1$ by $\hat E_1$ are suppressed. In other words, the holomorphic anomaly equation for $\bar\tau$ in the limit (\ref{t-limit}) automatically measures the dependence on $E_2$ (after replacing it by $\hat E_2$) without any admixture from terms associated with $E_1$.
Therefore, we can trade in the limit  (\ref{t-limit})  $\partial_{\bar \tau}$ against $\partial_{E_2}$ via the naive replacement
 \be  \label{partialE2def}
\partial_{E_2}  \quad   \stackrel{(\ref{t-limit})}{\longleftrightarrow}    \quad  \frac{\tau_2^2}{3} \frac{4 \pi^2}{2 \pi i} \partial_{\bar \tau}  \,,
\ee
rather than having to deal with the exact isomorphism   $\partial_{E_2} \leftrightarrow T_\nu$ as written in (\ref{RQJAHJ}).
This then immediately leads to (\ref{MAEgen}).

To derive the elliptic anomaly equation (\ref{E1equ}), we could similarly start from a suitable version of  the holomorphic anomaly for ${\rm Im}(z)$.
Note that even though in the asymptotic regime near the limit (\ref{t-limit}) we have $\partial_{\bar z} {\cal F}_{a|C_{\beta}} = 0$, it would be incorrect to conclude
from this that there is no modular anomaly equation with respect to $E_1$.
However, it is not even necessary to derive such a holomorphic anomaly equation for $\partial_{\bar z}$ in a suitable different limit, because, like in ref.~\cite{Oberdieck:2017pqm}, one can 
 start from the following identity stated in (\ref{DTalgebra})  
\bea   \label{Txicomm}
T_\xi = \partial_{E_1} = 12\, \big[\partial_{E_2}, \xi \partial_{\xi}\big]  =  12\, \big[\partial_{E_2}, \frac{1}{2 \pi i} \partial_{z}\big]\,,
\eea
and use (\ref{partialE2def}) to relate the action of the commutator (\ref{Txicomm}) on  $\cF_{a|C_{\beta}}$ to the commutator
\bea   \label{antiholcomm1}
\bar \partial_{\bar \tau} (\partial_z   \cF_{a|C_{\beta}})  -   \partial_z (\bar \partial_{\bar \tau} \cF_{a|C_{\beta}})    \,.
\eea
Here  we work again in the regime (\ref{t-limit}) so that the dependence on $\bar \tau$ arises solely from~$\hat E_2$.

As we will detail in Appendix \ref{Appproofz},
most of the terms in the difference cancel, except for two terms which can be traced back to special splittings of the curve $C_\beta = C_{\beta_1} + C_{\beta_2}$ where one of the two $C_{\beta_i}$ is trivial.
As result, one finds 
\bea   \label{commrelexp}
\bar \partial_{\bar \tau} (\partial_z   \cF_{a|C_\beta})  -   \partial_z (\bar \partial_{\bar \tau} \cF_{a|C_\beta})  \stackrel{(\ref{t-limit})}{=}   \frac{1}{4 \tau_2^2}    \left(  \langle\langle  \pi^\ast \pi_\ast( D_z \cdot G_a)   \rangle\rangle_{C_\beta}     -    \langle\langle     D_z \cdot \pi^\ast \pi_\ast(G_a) \rangle\rangle_{C_\beta} \right) \,.  
\eea
Transforming $ \partial_{\bar \tau} $ back to $\partial_{E_2}$ via (\ref{partialE2def}) yields the elliptic anomaly equation (\ref{E1equ}).

{ Finally, the counterparts of these equations for the higher genus invariants on elliptic fourfolds are presented in Appendix \ref{genusone}}.

\section{Evaluation of Holomorphic Anomaly Equations for Prototypical Geometries} \label{Examples}

We now apply the general formalism set up in the previous sections to a specific class of examples, where we take the base $B_3$ of the elliptic fourfold $Y_4$
to admit a rational fibration. The physical significance  is that this leads to dual heterotic and non-critical E-strings.
Aspects of this geometry have been studied \cite{Lee:2019tst,Lee:2020gvu, Klaewer:2020lfg} before,
in the context of proving the Weak Gravity Conjecture in four dimensions. We will take the viewpoint that the partition functions (\ref{relativeZ2}) correspond to elliptic genera
of suitable solitonic strings in F-theory on $Y_4$, as detailed further in Section \ref{sec:outlook}. Our interest here is in exemplifying  the details of modular and elliptic anomaly equations for 
the critical heterotic and the non-critical E-strings, for all possible vertical flux backgrounds.

\subsection{Rationally Fibered Base $B_3$}   \label{setupsecEx}

Let us denote the rational fibration of the base $B_3$ 
by
\beq\label{p}
p: B_3 \to B_2 \,.
\eeq
In F-theory compactified on the elliptic fibration over $B_3$, a D3-brane wrapping the class $C_0$ of the generic rational fiber of $B_3$ gives rise to a four-dimensional solitonic heterotic string.
We may furthermore assume hat the generic fiber $C_0$ of $B_3$ splits into two rational curves,
\beq
C_0 = C_E^1 + C_E^2 \,,
\eeq
over some divisor $\Gamma$ of $B_2$. 
Each of these curves is then associated with a four-dimensional, non-critical
E-string \cite{Lee:2019tst,Lee:2020gvu}.
More precisely, the rational fibration~\eqref{p} is obtained by blowing up a fibration $p': B_3' \to B_2$ along a single curve $\Gamma$ in $B_2$, over which no fibers of $p'$ degenerate. The exceptional divisor for this blowup will henceforth be denoted as $E$. It has the structure of a fibration 
\be\label{E-to-Gamma}
p_E: E \to \Gamma
\ee
with fiber $C_E^2$. More generally, one may also consider blowing up $B_3'$ along a set of curves $\{\Gamma_i\}$ in $B_2$. For simplicity, however, we analyse only a single blow-up for a general base twofold $B_2$, as the relevant physics of the corresponding heterotic string is already manifest in such a background. 

A K\"ahler threefold $B_3$ of this type is therefore defined by a choice of K\"ahler surface $B_2$, a divisor $\Gamma \subset B_2$ and a line bundle ${\cal L}$ on $B_2$ which defines the twisting of the
rational fibration. In particular, the fibration $p$ admits an exceptional section $S_- \in H^{1,1}(B_3)$ that has the following property
\beq   \label{S-selfint}
S_- \cdot_{B_3} S_- = - S_- \cdot_{B_3} p^*(c_1(\cL)) \,.
\eeq

In addition to specifying $B_3$, we will adopt a choice of elliptic fibration $Y_4$ over $B_3$ which has an additional rational section in order to engineer an extra $U(1)$ gauge group. This has been explained in Section~\ref{sec_top}.

Our aim is now to provide concrete expressions for the modular and elliptic anomaly equations (\ref{MAEgen}) and (\ref{E1equ}) for the
geometries specified above. To this end we first write the relative descendant invariants that appear 
in the  anomaly equation for a general  $(0)$-flux background  in terms of certain $(-2)$-flux invariants.
The latter can in turn be interpreted as the relative invariants of the embedded threefolds  $\mathbb Y_3^A$, as 
anticipated in eq.~(\ref{qderivative1}) and already observed in our
previous work \cite{Lee:2020gvu}.

Let us start by making a suitable choice of bases for the background fluxes on $Y_4$.
The cohomology group $H^{1,1}(B_3)$ is spanned as
\beq\label{H11B3}
H^{1,1}(B_3) = {\rm Span}\left< S_-,\;E,\;p^*(C_A) \right>\,,  \quad A=1, \ldots, h^{1,1}(B_2)\,. 
\eeq
Here $S_-$ is the exceptional section of the rational fibration (\ref{p}) that obeys~(\ref{S-selfint}),
 $E$ denotes the aforementioned blowup divisor described as~\eqref{E-to-Gamma}, and the divisors $C_A$ of $B_2$ form a basis of $H^{1,1}(B_2)$. 
 We can write this simply as 
 \be
 H^{1,1}(B_3) ={\rm Span}\left<D^{\rm b}_\alpha \right>\,,   \qquad    \alpha= -1,0, 1, \ldots, h^{1,1}(B_2)
 \ee
with
 \be   \label{DB2asis}
D^{\rm b}_{-1}= S_-,    \qquad D^{\rm b}_0 = E,  \qquad D^{\rm b}_{A} = p^*(C_A) \,.
\ee
We then define the following basis  of  $(0)$- and $(-1)$-fluxes:
\be
\begin{split}\label{basis0}
G_{\alpha_\tau} &= D_\tau \cdot \pi^* (D^{\rm b}_{\alpha})\,,   \\ 
G_{\alpha_z} &= D_z \cdot \pi^* (D^{\rm b}_{\alpha})\,,
\end{split}
\ee
with $D_\tau$, $D_z$ as introduced in (\ref{basish11-Y4}).

Next, in order to label the $(-2)$-fluxes, we adopt the following basis of  two-cycles:
\bea\label{H2B3}
H_2(B_3) &=&{\rm Span}\left<C_0,\; C_E^2,\; S_- \cdot p^*(C_A) \right>  \\
&=&  {\rm Span}\left<\Sigma_{\dot \alpha}^{\rm b} \right>         \,, \qquad \dot\alpha=-1,0,1, \ldots, h^{1,1}(B_2)\,. 
\eea
We can then pull back  their Poincar\'e dual elements to $Y_4$, which by abuse of notation we denote by the same symbol $\Sigma_{\dot \alpha}^{\rm b}$.  This yields a basis of $(-2)$-fluxes given by
\beq\label{basis-2}
G_{\dot\alpha} = \pi^*\Sigma_{\dot \alpha}^{\rm b} \,, \quad~ \dot\alpha= -1, 0, 1, \dots, h^{1,1}(B_2)\,,
\eeq
and leads to the intersection pairing
\beq
G_{\alpha_\tau}   \cdot G_{\dot \beta}   :=  I_{\alpha_\tau \dot \beta} = 
\begin{bmatrix}
\;1 & 0 & -\ell_B\;\\
\; 0 & -1 & 0  \\ 
\; 0 & 0 & I_{AB} 
\end{bmatrix} \,.
\eeq
Here $\ell_B$ and $I_{AB}$ denote the following intersection vector and matrix on $B_2$, 
\bea \label{ellA}
\ell_A &:=& C_A \cdot_{B_2} c_1(\cL) \,, \\ 
I_{AB} &:=& C_A \cdot_{B_2} C_B  \,,
\eea
with $A,B=1, \ldots, h^{1,1}(B_2)$. The matrix $I_{\alpha_\tau \dot \beta}$ is then inverted as
\beq\label{ginv}
I^{\dot \beta  \gamma_\tau} = 
\begin{bmatrix}
\; 1 & 0 & \ell^C \\
\; 0 & -1 & 0 \\
\; 0 & 0 & I^{BC}  
\end{bmatrix} \,,
\eeq
where $I^{BC}$ denotes the inverse matrix of $I_{AB}$, which is
 used to raise and lower the index of $\ell$.
\subsection{Modular and Elliptic Anomaly Equations for Heterotic Strings}

We are now ready to 
 evaluate the genus-zero anomaly equations for rationally fibered base geometries $B_3$, beginning with
 the modular anomaly equation (\ref{MAEgen}).

Our first task is to compute the descendant invariant $\psi \cdot \langle \langle  \pi^\ast \pi_{\ast} G_a \rangle \rangle_{C_0}$ which appears on the right-hand side of (\ref{MAEgen}).
We have already explained that  this invariant can be non-zero only for a $(0)$-flux, $G_a = G_{\alpha_\tau} = D_\tau \cdot \pi^\ast(D^{\rm b}_\alpha)$,  and in this case
\bea
\psi\cdot \langle \langle  \pi^\ast \pi_{\ast} G_{\alpha_\tau} \rangle \rangle_{C_0} =  \psi \cdot  \langle \langle  \pi^\ast(D^{\rm b}_\alpha) \rangle \rangle_{C_0}   \,.
\eea

To evaluate this further, we systematically apply~\eqref{psiequation}. The reader is walked through this computation in Appendix~\ref{app_descrational}. The end result is that 
\be
\begin{aligned}   \label{psi-invariants}
 \psi \cdot\langle \langle \pi^\ast(D^{\rm b}_{-1}) \rangle \rangle_{C_0}  &=   \sum_A \ell^A \langle \langle G_{\dot A} \rangle \rangle_{C_0}  &\equiv&  \,  \langle \langle \pi^*(S_-) \cdot \pi^\ast (p^\ast c_1(\cL)) \rangle \rangle_{C_0}       \cr
 \psi \cdot\langle \langle \pi^\ast(D^{\rm b}_{0}) \rangle \rangle_{C_0}  &=    - \sum_A \Gamma^A \langle \langle G_{\dot A} \rangle \rangle_{C_0} &\equiv& - \langle \langle \pi^* (S_-) \cdot \pi^\ast(p^\ast(\Gamma)) \rangle \rangle_{C_0}      \cr
\psi \cdot\langle \langle  \pi^\ast(D^{\rm b}_{A}) \rangle \rangle_{C_0}  &=   -2 \langle \langle G_{\dot A} \rangle \rangle_{C_0}    &\equiv& -2 \langle \langle \pi^* (S_-) \cdot \pi^\ast (D^{\rm b}_A) \rangle \rangle_{C_0}     \,.
\end{aligned}
\ee
Here we  are referring to the basis of divisors and fluxes introduced in Section~\ref{setupsecEx}, and we have expanded the blowup curve $\Gamma$ on $B_2$ as 
\bea
\Gamma = \Gamma^A C_A     \,.
\eea 
Note from the equations~\eqref{psi-invariants} that the descendant invariants relative to the rational fiber curve $C_0$ are expressible entirely in terms of $(-2)$-fluxes, and in fact solely
as linear combinations of the fluxes $G_{\dot A} = \pi^*(S_-) \cdot   \pi^\ast(D^{\rm b}_A)$   with $D^{\rm b}_A = p^\ast(C_A)$.

The relative $(-2)$-flux invariants appearing on the right-hand side of (\ref{psi-invariants}) have the following interesting geometric
interpretation: As observed in our previous work \cite{Lee:2020gvu} and remarked above,
they represent  relative Gromov-Witten invariants at genus zero pertaining to the embedded threefolds 
\bea   \label{Y3Adef1}
\mathbb Y_3^A =  \pi^\ast(D^{\rm b}_A) =  Y_4 |_{D^{\rm b}_A}   \,,
\eea
obtained as restriction of the elliptic fibration of $Y_4$ to the base divisors $D_A^{\rm b}$. Thus,
denoting  the generating functions for these threefolds invariants by  $\cF_{C_0}^{\mathbb Y_3^A}$, we have that 
\bea   \label{reinv4fold3fld}
\langle \langle G_{\dot A} \rangle \rangle_{C_0}  = \langle \langle \pi^*(S_-) \cdot \pi^\ast (D^{\rm b}_A) \rangle \rangle_{C_0}   = \cF_{C_0}^{\mathbb Y_3^A}     \,.
\eea
This relation follows from a direct application of the third line in (\ref{qderivative1}): Indeed, we can identify $D^{\rm b}_\alpha$ with $D^{\rm b}_A$, for which $ C_0 \subset D^{\rm b}_A$,
and $D^{\rm b}_\gamma$ with $S_-$, with the property $S_- \cdot_{B_3} C_0 =1$ (see also the explanation after~\eqref{(-2)-to-3}).

Having understood the structure of the gravitational descendant term, we can now turn to the quadratic term in the modular anomaly equation (\ref{MAEgen}).
In the present geometrical setup, this term contains as building blocks the invariants
 $\langle \langle \pi^\ast(C_E^1)  \rangle \rangle_{C_E^2}$  and $\langle \langle \pi^\ast(C_E^2)  \rangle \rangle_{C_E^1}$.
 
To evaluate these, we first express the two exceptional curves on $B_3$ as
\bea  \label{CE2CE1}
C_E^2 &=& E \cdot  p^\ast(\Gamma_D)    \,\\
C_E^1 &=& p^\ast(C_{A_1})    \cdot p^\ast(C_{A_2}) -    E \cdot  p^\ast(\Gamma_D)\,,
\eea
where $\Gamma_D$, $C_{A_1}$ and $C_{A_2}$ represent any divisor classes on $B_2$ with the properties
\bea
\Gamma_D   \cdot_{B_2} \Gamma = 1   \,, \qquad C_{A_1}   \cdot_{B_2}C_{A_2}   = 1 \,.
\eea
Application of the third line in (\ref{qderivative1})
then produces
\be
\begin{aligned}   \label{CE1CE2inv}
\langle \langle \pi^\ast(C_E^1)  \rangle \rangle_{C_E^2} &= \langle \langle  \pi^\ast(  p^\ast(C_{A_1})    \cdot p^\ast(C_{A_2})) - \pi^\ast (E) \cdot  \pi^\ast(p^\ast(\Gamma_D)) \rangle \rangle_{C_E^2} =     \cF_{C_E^2}^{\pi^\ast (p^\ast{\Gamma_D})}  =:  \cF_{C_E^2}    \,,   \cr
\langle \langle \pi^\ast(C_E^2)  \rangle \rangle_{C_E^1} &=  \langle \langle  \pi^*(E) \cdot  \pi^\ast(p^\ast(\Gamma_D)) \rangle \rangle_{C_E^1}        =    \cF_{C_E^1}^{\pi^\ast (p^\ast{\Gamma_D})}    =:    \cF_{C_E^1}\,,
\end{aligned}
\ee
where we used the fact that  $\pi^\ast(p^\ast(\Gamma_D))$ contains the curve classes $C_E^i$, together with the intersection numbers $\pi_*(C_E^i) \cdot_{B_3} D^{\rm b}_\alpha =0$ and $E \cdot_{B_3} \pi_*(C_E^2) = -1$, $E \cdot_{B_3} \pi_*(C_E^1) = 1$.
As a result we obtain the generating functionals for the invariants relative to ${C_E^1}$ or ${C_E^2}$ inside the threefold $\pi^\ast (p^\ast{\Gamma_D})$.
As it turns out, these invariants do not depend on the specific choice of $\Gamma_D$ as long as $\Gamma_D   \cdot_{B_2} \Gamma = 1$. This is reflected in our notation by writing $\cF_{C_E^1}$ and $\cF_{C_E^2}$.
In fact, these generating functions are proportional to the elliptic genera of the non-critical E-strings obtained by wrapping D3-branes on $C_E^1$ or
$C_E^2$, respectively.  They only depend on the structure of the elliptic fibration $Y_4$, the details of which govern the refinement with respect to the $U(1)$ fugacity \cite{Lee:2020gvu}.
Concretely,
\bea
\cF_{C_E^i} =  q^{\frac{1}{2}} \frac{E_{4,m_i}(q,\xi)}{\eta^{12}}    \,,  \qquad m_i = \frac{1}{2}  C_E^i \cdot_{B_3} b\,,
\eea
where  $b$ is the height-pairing associated with the $U(1)$ gauge symmetry, and the index $m_i$ determines the
Kac-Moody level of the latter's affine extension.

For the modular anomaly equation (\ref{MAEgen}) we therefore conclude that
\bea   \label{HAEforC0}
{\partial}_{E_2} \cF_{a|C_0} \ &=&\ 
 - \frac{1}{12 }  \left(  \cF_{a|C^1_{E}}   \cF_{C_E^2}   +   \cF_{a|C^2_{E} }  \cF_{C_E^1}   
 -     \psi\cdot \langle \langle  \pi^\ast(\pi_\ast(G_a))  \big\rangle\rangle_{C_0}    \right)   \,,
\eea
where the descendant term is as given in (\ref{psi-invariants}).

For certain fluxes, the quadratic term in the expression can be even further simplified by applying (\ref{qderivative1}).
Specifically, the premise that $C_E^1$ and $C_E^2$ are both contained in one of the divisor classes forming the flux is satisfied for 
$(0)$-fluxes of the form $G_{A_\tau} = D_\tau \cdot \pi^\ast(D^{\rm b}_A)$,
 as well as for their $(-1)$-flux counterparts of the form $G_{A_z} = D_z \cdot \pi^\ast(D^{\rm b}_A)$.
For such backgrounds, (\ref{HAEforC0})   becomes
\bea
{\partial}_{E_2} \cF_{A_\tau|C_0} &=&  - \frac{1}{12 }  \left(  (C_{A} \cdot_{B_2} \Gamma) ( ((q \partial_q + E_0) \cF_{C^1_{E}})   \cF_{C_E^2}   +  ((q \partial_q + E_0) \cF_{C^2_{E}})   \cF_{C_E^1} )
+  2 \cF_{C_0}^{\mathbb Y_3^A}    \right) \,,  \nonumber  \\
{\partial}_{E_2} \cF_{A_z|C_0} &=&  - \frac{1}{12 }  \left(   (C_{A} \cdot_{B_2}\Gamma) ( (\xi \partial_\xi \cF_{C^1_{E}})   \cF_{C_E^2}   +  (\xi \partial_\xi \cF_{C^2_{E}})   \cF_{C_E^1} )
  \right)  \label{Azresult}   \,,
\eea
where $E_0 = -1/2$ for the E-string.
The linear term in the first line follows from (\ref{psi-invariants}).
On the other hand, for the $(-2)$-fluxes we find
\be
\begin{aligned}   \label{partialE20fluxesS}
{\partial}_{E_2} \cF_{\dot {-1}|C_0} &= {\partial}_{E_2} \langle\langle  \pi^\ast(p^\ast(C_{A_1})    \cdot p^\ast(C_{A_2}))   \rangle  \rangle_{C_0}   &=& \ \, 0   \, ,  \cr
{\partial}_{E_2} \cF_{\dot 0|C_0} &= {\partial}_{E_2} \langle\langle \pi^\ast(  E \cdot  p^\ast(\Gamma_D))    \rangle  \rangle_{C_0}    &=& \,  - \frac{1}{12} (  \cF_{C_E^1} \cF_{C_E^2} - \cF_{C_E^2} \cF_{C_E^1}) \ = \ 0 \,,  \cr
{\partial}_{E_2} \cF_{\dot {A}|C_0} &= {\partial}_{E_2} \langle\langle   \pi^\ast(S_- \cdot  p^\ast(C_A) )   \rangle  \rangle_{C_0} & =& \, - \frac{1}{12}   (C_{A} \cdot_{B_2}\Gamma) \cF_{C_E^1}  \cF_{C_E^2}    \,.
\end{aligned}
\ee

Before verifying these equations for an explicit example, let us  evaluate also
 the elliptic anomaly equation, eq.~(\ref{E1equ-2}).
Again, for $(0)$- and $(-1)$-fluxes of the form $G_{A_\tau}$ and $G_{A_z}$, respectively, 
the right-hand side of this equation is tailor-made for applying the general relation (\ref{qderivative1}). This allows us to express the result in terms of the Gromov-Witten invariants relative to $C_0$
within  the embedded threefold $\mathbb Y_3^A =D_A \subset Y_4$; see the discussion around (\ref{Y3Adef1}).
This  yields the elliptic anomaly equations in the following concrete form:
\bea
\partial_{E_1}   \cF_{A_\tau|{C_0}} &=&  \xi \partial_\xi  \,  \cF_{C_0}^{\mathbb Y_3^A }   \\
\partial_{E_1}   \cF_{A_z|{C_0}} &=&    (b \cdot \pi_*(C_0))\,  \cF_{C_0}^{\mathbb Y_3^A}   \,.   
\eea

\subsection{Example: $B_3=dP_2 \times \mathbb P^1_{\l'}$}  \label{SecExS}

We now 
 apply the formulae derived in the previous section
 to a specific example 
 of a rationally fibered base $B_3$, c.f.,~\eqref{p}. 
The Calabi-Yau fourfold $Y_4$ is elliptically fibered over the K\"ahler threefold
\be   \label{Dp2xP1}
B_3=dP_2 \times \mathbb P^1_{\l'}  \,.
\ee
This example has already been analysed in~\cite{Lee:2020gvu}, to which we refer for an in-depth description of the background geometry (see esp.~Appendix B therein for details). 
As a new result, we will first present the exact partition functions for all types of flux backgrounds, including those for the  $(0)$-fluxes which had not been provided in~\cite{Lee:2020gvu}. We will then use 
these findings to test and exemplify the modular and elliptic anomaly equations we derived above.

Note that we can view the del Pezzo surface $dP_2$ 
in (\ref{Dp2xP1}) as a single blowup of a Hirzebruch surface $\mathbb F_1$. Let us denote its base by $\IP^1_h$. Thus $B_3$ admits a natural
rational fibration \eqref{p} over $B_2 =\IP^1_h \times \mathbb P^1_{\l'}$, with generic fiber $C_0$.
The divisor classes of $B_3$ can be expressed as linear combinations of 
\beq
p^\ast(C_1)\,, \;\; p^\ast(C_2)\,, \;\;  S_-\,, \;\; E \,,
\eeq 
where $C_1 \simeq \IP^1_{\ell'}$, $C_2 \simeq \IP^1_h$, and $S_-$ is the section satisfying (\ref{S-selfint}) with 
\bea
c_1({\cal L})   = C_1 \,.
\eea
Since the rational fiber of the del Pezzo surface $dP_2$ splits into a union of rational curves, $C_0 = C_E^1 + C_E^2$, over a point on $C_2 \simeq \IP^1_h$, we identify the exceptional divisor
 $E$ within $B_3$ as  $E \simeq C_E^2 \times C_1$. In particular, this means that 
 \bea
 \Gamma \simeq C_1   \,.
 \eea
We will also make use of the following intersection numbers in $B_3$:\footnote{While we have so far been carefully distinguishing the curve class in $B_3$ and the corresponding base curve class in $Y_4$, we will  for simplicity of presentation denote in this section  both classes by a common symbol.}
\be
C_0 \cdot_{B_3} S_-=C_E^1\cdot_{B_3} S_-=1,\qquad C_E^1\cdot_{B_3} E = 1,\qquad C_E^2\cdot_{B_3} E = -1 \,,
\ee
where $C_0$ is the generic rational fiber of the $dP_2$, extended to that of $B_3$.
The intersection numbers between these rational curves $C_0$, $C_E^1$, $C_E^2$ and the pull-back divisors $p^*(C_1)$, $p^*(C_2)$ are all 0.
Useful relations include furthermore
\begin{align}   \label{intrelB3S}
 &p^\ast(C_1) \cdot_{B_3} p^\ast(C_1) = 0 \,, \quad 
p^\ast(C_2) \cdot_{B_3} p^\ast(C_2) = 0  \,,   \quad  S_-   \cdot_{B_3} S_- = - S_- \cdot p^\ast(C_1) \,,   \\
&S_- \cdot_{B_3} E = 0  \,,    \quad  E \cdot_{B_3} E = - S_- \cdot_{B_3} p^\ast(C_1)\,, \quad 
E  \cdot_{B_3} p^\ast(C_1) = 0 \,.
\end{align}
With their help we can express the basis of curve classes $\Sigma_{\dot i}^{\rm b}$ as follows:
\be
\begin{aligned}   \label{SigmadotS}
\Sigma_{\dot{(-1)}}^{\rm b} &= C_0 =  p^\ast(C_1) \cdot_{B_3} p^\ast (C_2)   \cr
\Sigma_{\dot 0}^{\rm b} &= C_E^2 = p^\ast(C_2) \cdot_{B_3} E  \cr
\Sigma_{\dot 1}^{\rm b} &= S_- \cdot_{B_3} p^\ast(C_1) \cr
\Sigma_{\dot 2}^{\rm b} &= S_- \cdot_{B_3} p^\ast(C_2)  \,.
\end{aligned}
\ee
The elliptic fibration over $B_3$ is designed, as in \cite{Lee:2020gvu}, such as to realise an independent rational section with height-pairing
\be   \label{bmodelS}
b=2\overline{K}_{B_3}=6 p^\ast(C_1)+4S_-+4 p^\ast(C_2)-2E \,.
\ee

In summary, our choice of flux basis is listed in Table \ref{tab:Fluxbasis}, together
with the modular weight of the corresponding partiton functions.\newline

 \begin{table}[htp]
\begin{center}
\begin{tabular}{c||c|c}
mod.~weight $w$ & notation   & basis flux class $G_a\in H^{2,2}_{\rm vert}(Y_4)$\\
\hline\hline
\multirow{4}{*}{$-2$} & $G_{\dot{(-1)}}$ & $\pi^\ast(p^\ast(C_1)) \cdot \pi^\ast(p^\ast(C_2))$\\
 & $G_{\dot 0}$ & $\pi^\ast(p^\ast(C_2)) \cdot \pi^\ast(E)$ \\
 & $G_{\dot 1}$ &  $\pi^\ast(S_-) \cdot \pi^\ast(p^\ast(C_1))$ \\
 & $G_{\dot 2}$ &  $\pi^\ast(S_-) \cdot \pi^\ast(p^\ast(C_2))$ \\  
 \hline
  \multirow{4}{*}{$-1$} & $G_{(-1)_z}$ & $D_z \cdot \pi^\ast(S_-)$ \\
 & $G_{0_z}$ & $D_z \cdot \pi^\ast(E)$ \\
 & $G_{1_z}$ &  $D_z \cdot \pi^\ast(p^\ast(C_1))$ \\
 & $G_{2_z}$ &  $D_z \cdot \pi^\ast(p^\ast(C_2))$ \\
 \hline
\multirow{4}{*}{$0$} & $G_{(-1)_\tau}$ & $D_\tau \cdot \pi^\ast(S_-)$ \\
 & $G_{0_\tau}$ & $D_\tau \cdot \pi^\ast(E)$ \\
 & $G_{1_\tau}$ &  $D_\tau \cdot \pi^\ast(p^\ast(C_1))$ \\
 & $G_{2_\tau}$ &  $D_\tau \cdot \pi^\ast(p^\ast(C_2))$ \\
 \hline
\end{tabular}
\end{center}
\caption{Basis of $H^{2,2}_{\rm vert}(Y_4)$ for our example geometry, 
which is given by an elliptic fibration over $B_3=dP_2 \times \mathbb P^1_{\l'}$.
Indicated is also the modular weight of the associated partiton functions, ${\cal Z}_{w,*}[G_*,C_*]$.}
\label{tab:Fluxbasis}
\end{table}

The partition functions (\ref{relativeZ2}) can now be obtained explicitly by employing standard methods of
mirror symmetry, starting from
the toric data of the fourfold and flux geometry. These were already written down in ref. 
\cite{Lee:2020gvu}, to which we refer for details. As explained there, this procedure results in
a finite number of Gromov-Witten invariants, which can be used to determine
the exact partition functions via modular completion in terms of suitable Jacobi forms.
In order to concisely write these down, it is convenient to define
the following modular and quasi-modular Jacobi forms\footnote{Compared to \cite{Lee:2020gvu}, we have exchanged the labeling for $Z_{-2,2}^{{1}}(q,\xi)$ and $Z_{-2,2}^{{2}}(q,\xi)$, i.e. $Z^1_{\rm here} = Z^2_{\rm there}$ and $Z^2_{\rm here} = Z^1_{\rm there}$. Similarly, $\mathbb Y_{3,\rm here}^1 = \mathbb Y_{3,\rm there}^2$ and $\mathbb Y_{3,\rm here}^2 = \mathbb Y_{3,\rm there}^1$ for the embedded threefolds defined in (\ref{embthreefoldsS}).}
\bea
\label{Zwmdefs}
Z_{-2,2}^{{1}}(q,\xi) &=&  \frac1{12\et24}(14E_4E_{6,2}+10 E_{4,2}E_6),\nn \\
Z_{-2,2}^{{2}} (q,\xi) &=&  Z_{-2,2}^{{1}}+\frac1{12\et24}E_{4,1}(E_2E_{4,1}- E_{6,1}),\\
Z_{-1,2}^0 (q,\xi) &=& 84\, \phi_{-1,2},\nn\\
Z_{0,2}^0(q,\xi)  &=& \frac{-137 E_4^2E_{4,2}+120 E_4 E_{4,1}^2-169 E_6 E_{6,2}+ 4 E_2(37 E_{4,1} E_{6,1}+ 8 E_{6,2}E_4)+6 E_2^2E_{4,1}^2}{2\cdot 12^2\,\eta^{24}},\nn
\eea
where the subscripts indicate the respective weight and index; we will suppress the
arguments $(q,\xi)$ in the following. The Jacobi forms appearing on the right-hand side of these equations have been defined in Appendix \ref{app_jacobi}.
In terms of these building blocks, we find for the heterotic partition functions (defined generally in (\ref{relativeZ2})) in the background of the $(-2)$-fluxes:
\bea
\label{Zm2defs}
{\cal Z}_{-2,2}[G_{\dot{(-1)}},C_0]&=&  0\nn \\ 
{\cal Z}_{-2,2}[G_{\dot0 },C_0]&=&  0 \\ 
{\cal Z}_{-2,2}[G_{\dot 1},C_0]&=& Z_{-2,2}^{{1}}   \equiv {\cal Z}_{-2,2}^{\mathbb Y_3^1}[C_0]   \nn \\  
{\cal Z}_{-2,2}[G_{\dot 2},C_0]&=& Z_{-2,2}^{{2}}  \equiv {\cal Z}_{-2,2}^{\mathbb Y_3^2}[C_0]   \nn \,. 
\eea
The notation ${\cal Z}_{-2,2}^{\mathbb Y_3^A}[C_0]$ indicates that the latter two quantities coincide with the heterotic elliptic genera arising from compactifications on
\be   \label{embthreefoldsS}
\mathbb Y_3^1 = Y_4 |_{p^\ast(C_1)}\,, \qquad  \mathbb Y_3^2 = Y_4 |_{p^\ast(C_2)}   \,.
\ee 
For the $(-1)$-fluxes we have:
\bea
\label{Zm1defs}
{\cal Z}_{-1,2}[G_{{(-1)_z}},C_0]&=&  \xi \partial_\xi(\frac12Z_{-2,2}^{{1}}+  Z_{-2,2}^{{2}})+Z_{-1,2}^0  \nn \\ 
{\cal Z}_{-1,2}[G_{0 _z},C_0]&=&   \xi \partial_\xi(\frac{1}{2}Z_{-2,2}^{{1}})\\ 
{\cal Z}_{-1,2}[G_{1_z},C_0]&=& \xi \partial_\xi(Z_{-2,2}^{{1}}) \nn\\  
{\cal Z}_{-1,2}[G_{2_z},C_0]&=&\xi \partial_\xi(Z_{-2,2}^{{2}}) 
 \,, \nn
\eea
while we get for the $(0)$-fluxes:
\bea
\label{Z0defs}
{\cal Z}_{0,2}[G_{{(-1)_\tau}},C_0]&=&  q\partial_q(\frac{1}{2}Z_{-2,2}^{{1}}+Z_{-2,2}^{{2}})+\xi\partial_\xi(\frac{1}{4}Z_{-1,2}^0)+Z^0_{0,2}  \nn \\ 
{\cal Z}_{0,2}[G_{0_\tau},C_0]&=&  q\partial_q(\frac{1}{2}Z_{-2,2}^{{1}}) \\ 
{\cal Z}_{0,2}[G_{1_\tau},C_0]&=& q\partial_q(Z_{-2,2}^{{1}})\nn\\  
{\cal Z}_{0,2}[G_{2_\tau},C_0]&=& q\partial_q(Z_{-2,2}^{{2}}) \,.  \nn 
\eea
Here we clearly see the derivative relationships between the various flux sectors,  which
correspond to the symbolic up-arrows in Fig.~\ref{fig:anomalyfluxes}. At the same time it is evident that
not all $w=-1,0$ partition functions are derivatives of the $w=-2$ ones. This ties in with our previous results \cite{Lee:2019tst,Lee:2020gvu}. 

The  derivative structure we observe
perfectly reflects the general properties of the prepotentials deduced towards the end of Section \ref{sec_Fldeppre}, as follows:
Consider first the modular weight $w=-2$ partition functions (\ref{Zm2defs}) for the heterotic string.
The partition functions for fluxes $G_{\dot 1}$ and $G_{\dot 2}$ coincide with the partition functions associated with a six-dimensional 
heterotic string obtained by wrapping a D3-brane on the curve $C_0$ inside the embedded threefolds $\mathbb Y_3^1$ and $\mathbb Y_3^2$:
\bea
{\cal Z}_{-2,2}[G_{\dot 1},C_0]&=& Z_{-2,2}^{{1}} = {\cal Z}_{-2,2}^{\mathbb Y_3^1} [C_0]   \\  
{\cal Z}_{-2,2}[G_{\dot 2},C_0]&=& Z_{-2,2}^{{2}} = {\cal Z}_{-2,2}^{\mathbb Y_3^2} [C_0] \,. 
\eea
This was predicted by the third equation in (\ref{qderivative1}), with $D_\gamma^{\rm b} = S_-$ and $D_\alpha^{\rm b} = p^\ast(C_A)$.
Similarly, the vanishing of ${\cal Z}_{-2,2}[G_{\dot {(-1)}},C_0]$ and ${\cal Z}_{-2,2}[G_{\dot 0},C_0]$
is explained by the same formula if we now identify $D_\gamma^{\rm b}$ with $p^\ast(C_A)$ or $E$  and   take into account that $p^\ast(C_A) \cdot_{B_3} C_0 =0$ and $E  \cdot_{B_3} C_0 =0$.

As for the weight $w=-1$ and $w=0$ partition functions, given in (\ref{Zm1defs}) and (\ref{Z0defs}),
the expressions for $G_{{A}_\tau}$ and $G_{{A}_z}$, $A=1,2$, match the predictions of the first two equations in (\ref{qderivative1}).
Indeed, the curve $C_0$ is fully contained in the divisors $D_\alpha^{\rm b} = p^\ast(C_A)$, in the sense defined in (\ref{cond1}). This explains the characteristic derivative structure for the partition functions.
Interestingly, for flux $G_{{0}_\tau} = D_\tau \cdot \pi^\ast(E)$ and $G_{{0}_z} = D_z \cdot \pi^\ast(E)$ this conclusion is {\it a priori} not justified. 
However, in this case the divisor $D_\alpha^{\rm b} = E$ contains a component of $C_0$ - namely the exceptional curve $C_E^2$ -, which is the fiber of $E \simeq C_E^2 \times C_1$. Due to the symmetric relationship of $C_E^1$ and $C_E^2$
the partition function is just one-half of the partition function in the flux backgrounds $G_{{1}_\tau}$ and $G_{{1}_z}$.

However, for more general fluxes, in particular for  $G_{{{(-1)}}_\tau}$ and $G_{{{(-1)}}_z}$, we clearly see that the partition functions are not total derivatives and contain in addition fully modular, resp.~quasi-modular, contributions via $Z_{0,2}^0(q,\xi)$ and $Z_{-1,2}^0 (q,\xi)$.
 
Similarly we can compute the partition functions associated with the two four-dimensional E-strings obtained by wrapping  D3-branes along the exceptional fibral curves $C_E^1$ and $C_E^2$ in $B_3$.
The complete expressions are listed in Appendix \ref{App-Estrings}.
To understand their structure, recall that $C_E^1$ and $C_E^2$ lie in the fiber of $B_3$ over the curve $\Gamma \simeq C_1$. 
 Since on the base  $B_2$ of $B_3$ we have $C_1 \cdot_{B_2} C_2= 1$ (while $C_1 \cdot_{B_2} C_1= 0 = C_2 \cdot_{B_2} C_2$),
the embedded threefold $\mathbb Y_3^2$ defined in (\ref{embthreefoldsS}) contains both curves $C_E^1$ and $C_E^2$. In fact, this threefold is Calabi-Yau, and the generating functions
for the relative genus-zero invariants for $C_E^1$ and $C_E^2$ within $\mathbb Y_3^2$, called ${\cal F}_{C_E^1}$ and ${\cal F}_{C_E^2}$ in   (\ref{CE1CE2inv}),  are related to the elliptic genera for the corresponding  six-dimensional E-strings  \cite{Lee:2020gvu} as follows:
\bea   \label{FvsZCEi}
{\cal F}_{C_E^i} = - q^{\frac{1}{2}}  {\mathcal Z}_{C_E^i} &\equiv& - q^{\frac{1}{2}}    {\mathcal Z}^{\mathbb Y_3^2}[C_E^i]   \,.
\eea
These can in turn be written as partition functions on $Y_4$ in the background of suitable fluxes. This presentation is not unique, but a canonical choice is the following:
\be
\begin{aligned}   \label{ZCEirel}
{\mathcal Z}_{C_E^1} &\equiv {\mathcal Z}^{\mathbb Y_3^2}[{C_E^1}]  =    {\mathcal Z}_{-2,1}[G_{\dot 0}, C_E^1]    \cr
{\mathcal Z}_{C_E^2} &\equiv {\mathcal Z}^{\mathbb Y_3^2}[{C_E^2}]  =  - {\mathcal Z}_{-2,1}[G_{\dot 0}, C_E^2]    \,.
\end{aligned}
\ee 
Here we used that 
$G_{\dot 0} =   \pi^\ast(E) \wedge \pi^\ast(p^\ast(C_2))$ 
 along with the fact that 
 $E \cdot_{B_3} C_E^1 = 1$ and  $E \cdot_{B_3} C_E^2 = -1$, while $p^\ast(C_2)$ contains both $C_E^i$, as discussed above.
 Hence (\ref{ZCEirel}) follows from the general property (\ref{qderivative1}).

Note that there are various relations between the flux partition functions, for example since  $S_- \cdot_{B_3} C_E^1 = 1$, we can equivalently write
 ${\mathcal Z}_{-2,1}[G_{\dot 0}, C_E^1]  = {\mathcal Z}_{-2,1}[G_{\dot 2}, C_E^1]$   (where we expressed $G_{\dot 2} =  \pi^\ast(S_-) \wedge \pi^\ast(p^\ast(C_2))$), while 
 ${\mathcal Z}_{-2,1}[G_{\dot 2}, C_E^2] =0$ (since $S_- \cdot_{B_3} C_E^2 = 0$). 
 In this way all expressions for the $(-2)$-fluxes in (\ref{-2FluxesE}) follow from the structure given in (\ref{qderivative1}).
 The same is true for the derivative structure of the background of fluxes $G_{A_z}$ and $G_{A_\tau}$ for $A=1,2$ in (\ref{-1FluxesE}) and 
  (\ref{0FluxesE}).
 The structure of the remaining $(-1)$- and $(0)$-fluxes is not captured by  (\ref{qderivative1}), and correspondingly we observe more complicated expressions for the corresponding partition functions in (\ref{-1FluxesE}) and 
  (\ref{0FluxesE}).   \\

Having determined the partition functions of the heterotic and the E-strings 
in the various flux sectors exactly,
we are now ready to discuss in detail the modular and elliptic anomaly equations
they satisfy.

We begin with the  partition functions for the dual heterotic string.
To evaluate the anomaly equations, we first express the partition functions listed in (\ref{Zm2defs}), (\ref{Zm1defs}) and (\ref{Z0defs})
in terms of 
the generators of the ring of quasi-Jacobi forms, as introduced in Section \ref{sec_RingofQuasis}.
Some details have been collected in Appendix \ref{app_PartfunC0}.
This allows us to read off the derivatives with respect to $E_1$ and $E_2$, which correspond to the down-arrows in Fig.~\ref{fig:anomalyfluxes}. The resulting anomaly equations can be summarised as follows:

For the $(-2)$-fluxes we find
\bea
\label{Zm2pE2}
\partial_{E_2} {\cal Z}_{-2,2}[G_{\dot{(-1)}},C_0]&=&  0\nn \\ 
\partial_{E_2} {\cal Z}_{-2,2}[G_{\dot0 },C_0]&=&  0 \\ 
\partial_{E_2} {\cal Z}_{-2,2}[G_{\dot 1},C_0]&=&  0 \nn \\  
\partial_{E_2} {\cal Z}_{-2,2}[G_{\dot 2},C_0]&=&  - \frac{1}{12}   {\cal Z}_{-2,1}[G_{\dot 2},C^1_E]    {\cal Z}_{-2,1}[G_{\dot0 },C^2_E]  \nn \,,  
\eea
while manifestly
\bea\label{elliptic-example-(-2)}
\partial_{E_1}   {\cal Z}_{-2,2}[G_{\dot{\alpha}},C_0] =  0    \qquad \forall\,  \alpha    \,.
\eea
The partition functions, or elliptic genera, for the $(-1)$-fluxes obey 
\bea
\label{Zm1pE2}
\partial_{E_2} {\cal Z}_{-1,2}[G_{{(-1)_z}},C_0]&=& -\frac{1}{12}   {\cal Z}_{-1,1}[G_{{(-1)_z}},C^1_E] \,
   {\cal Z}_{-2,1}[G_{{\dot 0}},C^2_E]  \nn \\ 
\partial_{E_2}  {\cal Z}_{-1,2}[G_{0 _z},C_0]&=&  0\\ 
\partial_{E_2}  {\cal Z}_{-1,2}[G_{1_z},C_0]&=&  0 \nn\\  
\partial_{E_2}  {\cal Z}_{-1,2}[G_{2_z},C_0]&=& 
 \frac{1}{12}  \Big (   {\cal Z}_{-2,1}[G_{{\dot 0}},C_E^1]  {\cal Z}_{-1,1}[G_{{2}_z},C_E^2] -
                            {\cal Z}_{-1,1}[G_{{2}_z},C_E^1]  {\cal Z}_{-2,1}[G_{{\dot 0}},C_E^2]   \Big )       
                            \nn
\eea
as well as  
\bea
\label{Zm1pE1}
\partial_{E_1} {\cal Z}_{-1,2}[G_{{(-1)_z}},C_0]&=& 2 {\cal Z}_{-2,2}[G_{{\dot 1}},C_0]   +   4  {\cal Z}_{-2,2}[G_{{\dot 2}},C_0]     \nn \\ 
\partial_{E_1}  {\cal Z}_{-1,2}[G_{0 _z},C_0]&=&     2 {\cal Z}_{-2,2}[G_{{\dot 1}},C_0]   \\ 
\partial_{E_1}  {\cal Z}_{-1,2}[G_{1_z},C_0]&=&   4 {\cal Z}_{-2,2}[G_{{\dot 1}},C_0]  \nn\\  
\partial_{E_1}  {\cal Z}_{-1,2}[G_{2_z},C_0]&=&   4  {\cal Z}_{-2,2}[G_{{\dot 2}},C_0] 
 \,. \nn
\eea
Finally, the weight $w=0$ partition functions satisfy the following anomaly equations:
\bea
\label{Z0pE2}
\partial_{E_2} {\cal Z}_{0,2}[G_{{(-1)_\tau}},C_0]&=&   \frac{1}{12} {\cal Z}_{-2,2}[G_{\dot 1},C_0]  - \frac{1}{12}   {\cal Z}_{0,1}[G_{{(-1)_\tau}},C^1_E]  {\cal Z}_{-2,1}[G_{\dot0 },C^2_E],    \nn \\ 
\partial_{E_2} {\cal Z}_{0,2}[G_{0_\tau},C_0]&=&   - \frac{1}{12}  {\cal Z}_{-2,2}[G_{\dot 1},C_0] ,
\\
\partial_{E_2} {\cal Z}_{0,2}[G_{1_\tau},C_0]&=&  -\frac{1}{6}   {\cal Z}_{-2,2}[G_{\dot 1},C_0] ,
  \nn\\
\partial_{E_2} {\cal Z}_{0,2}[G_{2_\tau},C_0]&=&   -\frac{1}{6}   {\cal Z}_{-2,2}[G_{\dot 2},C_0]   + \frac{1}{6}    {\cal Z}_{-2,1}[G_{\dot 0},C^1_E]    {\cal Z}_{0,1}[G_{2_\tau},C_E^2]       \,  \nn 
\eea
as well as 
\bea\label{elliptic-example-(0)}
\partial_{E_1} {\cal Z}_{0,2}[G_{{\alpha_\tau}},C_0]&=&    {\cal Z}_{-1,2}[G_{{\alpha_z}},C_0]    \qquad \forall  \, \alpha \,.
\eea

These results are in complete agreement with the modular and elliptic anomaly equations, (\ref{MAEgen}) and (\ref{E1equ}) as derived above, when applied to the specific geometry of our example.
For the modular anomaly equation (\ref{MAEgen})  we can start from the form given in (\ref{HAEforC0}).
As for the quadratic terms, the discussion around (\ref{FvsZCEi}) and  (\ref{ZCEirel}) implies that in the anomaly equation for the partition function the role of ${\cal F}_{C_E^1}$ will be played by $\mathcal{Z}_{-2,1}[G_{\dot 0},C_E^1]$ and that of ${\cal F}_{C_E^2}$ will be played by $- \mathcal{Z}_{-2,1}[G_{\dot 0},C_E^2]$ (modulo the sign changes from going from the prepotentials to the partition functions).

The weight $w=0$ partition functions are the only ones for which the {\it modular} anomaly equation receives an extra contribution from the gravitational descendant terms.
For rationally fibered base manifolds the latter have been computed in (\ref{psi-invariants}), which in our case ($\Gamma = C_1$ and $c_1({\cal L}) = C_1$) reduce to
\be
\begin{aligned}   \label{psi-invariantsS}
\psi \cdot\langle \langle \pi^\ast(D^{\rm b}_{-1}) \rangle \rangle_{C_0}  &= \langle \langle \pi^\ast(S_-) \cdot \pi^\ast (p^\ast (C_1)) \rangle \rangle_{C_0}  = \langle \langle G_{\dot 1} \rangle \rangle_{C_0} = {\cal F}^{\mathbb Y_3^1}_{C_0}  \cr
 \psi \cdot\langle \langle \pi^\ast(D^{\rm b}_{0}) \rangle \rangle_{C_0}  &= - \langle \langle  \pi^\ast(S_-) \cdot \pi^\ast(p^\ast(C_1)) \rangle \rangle_{C_0}  = -  \langle \langle G_{\dot 1} \rangle \rangle_{C_0}  = - {\cal F}^{\mathbb Y_3^1}_{C_0}  \cr
\psi \cdot\langle \langle  \pi^\ast(D^{\rm b}_{A}) \rangle \rangle_{C_0}  &=  -2 \langle \langle  \pi^\ast(S_-) \cdot \pi^\ast (D^{\rm b}_A) \rangle \rangle_{C_0}  = -2  \langle \langle G_{\dot A} \rangle \rangle_{C_0}    = -2 {\cal F}^{\mathbb Y_3^A}_{C_0}     \,, 
\end{aligned}
\ee
where~\eqref{reinv4fold3fld} was used for the last equality in each line. 

Putting everything together we indeed confirm that the anomaly equations in~(\ref{Z0pE2}) follow from~\eqref{HAEforC0}; in particular, for
$\partial_{E_2} {\cal Z}_{0,2}[G_{A_\tau},C_0]$, $A=1,2$, these equations can equivalently be derived directly from the first line in
 (\ref{Azresult}). The relations (\ref{Zm1pE2}) and (\ref{Zm2pE2}) are likewise consistent with (\ref{HAEforC0}), and where applicable, agree with the expressions (\ref{Azresult}) and  (\ref{partialE20fluxesS}), respectively.

 As for the {\it elliptic} anomaly equation (\ref{E1equ}), we start from (\ref{E1equ-2}) and observe immediately that the equations for the weight $w=-2$ and $w=0$ partition functions are satisfied by~\eqref{elliptic-example-(-2)} and~\eqref{elliptic-example-(0)}, respectively.
 To also understand the form of the equations listed in (\ref{Zm1pE1}), we must express the fluxes $\pi^\ast(b) \cdot D_\alpha$, which appear in the middle equation of~\eqref{E1equ-2}, in terms of the basis elements for the 
$(0)$-fluxes. Using
(\ref{intrelB3S}), (\ref{SigmadotS}) and (\ref{bmodelS})
this gives
\be   \label{bdotDmodelS}
\begin{aligned}
b \cdot_{B_3} S_- &= 2 \Sigma_{\dot 1}^{\rm b} + 4 \Sigma_{\dot 2}^{\rm b} \cr
b \cdot_{B_3}  E &= 4 \Sigma_{\dot 0}^{\rm b} + 2 \Sigma_{\dot 1}^{\rm b} \cr
b \cdot_{B_3}  p^\ast(C_1) &= 4 \Sigma_{\dot{(-1)}}^{\rm b} +  4 \Sigma_{\dot{1}}^{\rm b}   \cr
b \cdot_{B_3} p^\ast(C_2) &= 6 \Sigma_{(\dot{-1})}^{\rm b}       +  4 \Sigma_{\dot 2}^{\rm b} -  2  \Sigma_{\dot 0}^{\rm b}    \,.
\end{aligned}
\ee
Since ${\cal Z}_{-2,2}[G_{{\dot{(-1)}}},C_0]=0$ and ${\cal Z}_{-2,2}[G_{{\dot{0}}},C_0]=0$, this indeed explains
the anomaly equations in (\ref{Zm1pE1}).

Finally, we now consider the anomaly equations for the E-strings.
It is clear that their modular anomalies can only arise from the gravitational descendant terms, because
the curves $C_E^i$ do not further split as sum of holomorphic curve classes.
The gravitational descendant terms with respect to the E-string curves $C_E^i$ can be computed in a similar way as spelled out for $C_0$ in Appendix \ref{app_descrational}.
The result of this computation is 
\be 
\begin{aligned}
\psi  \cdot \langle \langle  \pi^\ast p^\ast(C_2)     \rangle \rangle_{C_E^1} &= - 2 \langle \langle G_{\dot 0}   \rangle \rangle_{C_E^1}       \cr   
\psi  \cdot \langle \langle  \pi^\ast p^\ast(C_2)     \rangle \rangle_{C_E^2} &=  + 2  \langle \langle G_{\dot 0}   \rangle \rangle_{C_E^2}  \,, \cr  
\end{aligned}
\ee
while
\be 
\begin{aligned}
\psi  \cdot \langle \langle  \pi^\ast p^\ast(C_1)     \rangle \rangle_{C_E^i} &= 0   \cr
\psi  \cdot \langle \langle  \pi^\ast S_-     \rangle \rangle_{C_E^i} &= 0    \cr   
\psi  \cdot \langle \langle  \pi^\ast E     \rangle \rangle_{C_E^i} &= 0 \,.   \cr   
\end{aligned}
\ee
This perfectly explains the structure of the following modular anomaly equations 
\be 
\begin{aligned}
\partial_{E_2} {\cal Z}_{0,1}[G_{2_\tau},C_E^1] &=& - \frac{1}{6}  {\cal Z}_{-2,1}[G_{\dot 0},C_E^1]    \cr
\partial_{E_2} {\cal Z}_{0,1}[G_{2_\tau},C_E^2] &=&   \frac{1}{6}  {\cal Z}_{-2,1}[G_{\dot 0},C_E^2]\,,
\end{aligned}
\ee
which can be checked to be satisfied by the explicit expressions given in Appendix~\ref{App-Estrings}
(all other equations vanish identically).
Analogously, the predicted form of the elliptic anomaly equations
\be 
\begin{aligned}
\partial_{E_1} {\cal Z}_{-2,1}[G_{\dot \alpha},C_E^i] &= 0     \,   \cr 
\partial_{E_1} {\cal Z}_{0,1}[G_{\alpha_\tau},C_E^i] &=  {\cal Z}_{-1,1}[G_{\alpha_z},C_E^i]   
\end{aligned}
\ee
as well as
\be 
\begin{aligned}
\partial_{E_1} {\cal Z}_{-1,1}[G_{(-1)_z},C_E^1] &=   4  {\cal Z}_{-2,1}[G_{\dot 2},C_E^1]    \cr
\partial_{E_1} {\cal Z}_{-1,1}[G_{(-1)_z},C_E^2] &=   0  \cr
\partial_{E_1} {\cal Z}_{-1,1}[G_{0_z},C_E^i] &=  4  {\cal Z}_{-2,1}[G_{\dot 0},C_E^i]      \,   \qquad i=1,2  \cr
\partial_{E_1} {\cal Z}_{-1,1}[G_{1_z},C_E^1] &=  4  {\cal Z}_{-2,1}[G_{\dot 2},C_E^1] - 2 {\cal Z}_{-2,1}[G_{\dot 0},C_E^1]     \cr
\partial_{E_1} {\cal Z}_{-1,1}[G_{1_z},C_E^2] &= - 2 {\cal Z}_{-2,1}[G_{\dot 0},C_E^2]    
\end{aligned}
\ee
is perfectly matched by the explicit expressions we find for the partition functions from mirror symmetry.
This is in agreement with (\ref{bdotDmodelS}) if one takes into account that many of these partition functions vanish, see (\ref{-2FluxesE}).

\section{Physics Discussion}\label{sec:outlook}

In this article we have studied the generating functions $\cF$ for relative genus-zero Gromov-Witten invariants on elliptic Calabi-Yau fourfolds $Y_4$ with fluxes.
A main result is the derivation of their modular and elliptic anomaly equations, (\ref{MAEgen}) and (\ref{E1equ}), starting
from the $tt^\ast$ formalism introduced by BCOV \cite{Bershadsky:1993ta,Bershadsky:1993cx}.

These equations can be interpreted from various different angles. From the point of view of Gromov-Witten theory, the anomaly equations (\ref{MAEgen}) and (\ref{E1equ}) for fourfolds, as well as their generalisations
to arbitrary elliptic $n$-folds, had been conjectured in \cite{Oberdieck:2017pqm}. 
Some of their properties can be understood in a purely geometric way, such as the appearance of derivatives for special classes of flux backgrounds, as explained around eq.~(\ref{qderivative1}).

The derivation of the anomaly equations via $tt^\ast$ geometry, as detailed in the present paper, makes use of the interpretation of the generating functions $\cal F$
in the topological $A$-model as prepotentials of two-dimensional flux compactifications of Type IIA string theory on 
fourfolds. 
Up to a prefactor, the prepotentials coincide with the partition functions $\mathcal Z$ defined in (\ref{relativeZ2}).
The latter have distinguished modular behavior, i.e., they are given by quasi-modular extensions of Jacobi forms or their generalisations, which are called quasi-Jacobi forms.

A third interpretation is in terms of elliptic genera of certain chiral $N=1$ supersymmetric strings in four dimensions.
It uses the duality between Type IIA string theory compactified to two dimensions on some elliptic fourfold $Y_4$,
 and F-theory on $Y_4\times T^2$. The strings in question arise from D3-branes wrapped on some curve, $C_\beta$, on the base $B_3$ of the elliptic fibration~$Y_4$.
However, as pointed out in \cite{Lee:2020gvu}, such an interpretation of relative prepotentials 
as four-dimensional elliptic genera
is {\it a priori} possible only for certain flux backgrounds, namely those which can be uplifted
 from two to four dimensions while preserving Poincar\'e invariance.

More precisely, for an elliptic Calabi-Yau fourfold one can label the possible vertical flux backgrounds by types $(0)$, $(-1)$ or $(-2)$, as in (\ref{basisH22vert-Y4}). These refer to the modular weight of the respective partition functions.
Of these, only the $(-1)$-fluxes describe gauge backgrounds in fully Poincar\'e invariant compactifications of F-theory to four dimensions.
Nevertheless from the worldsheet perspective of the strings, all flux sectors should appear on a similar footing, even though the $(-2)$- and $(0)$-fluxes break Poincar\'e invariance when uplifted from two to four dimensions.  
In other words, we expect 
all partition functions 
to admit an index-like interpretation in four dimensions. This is also
suggested by the fact
that they are related by the anomaly equations. 

In the sequel we develop this  more physical, though somewhat tentative interpretation further.
Our aim is to shed more light on the derivative
relationships between flux partition functions, to better understand  
the role of the embedded threefolds $\mathbb{Y}_3^A$, introduced in \cite{Lee:2020gvu} and
encountered here in eq.~(\ref{Y3adef}), and to elucidate the 
physics behind the appearance of
the linear terms in the anomaly equations (\ref{MAEgen}) and (\ref{E1equ}). 
After all, all these features are intertwined and ought to reflect a common physical origin.

\subsubsection*{Flux Backgrounds as Defects}

In order to get a handle on a possible worldsheet interpretation,
we start from the M-theory formulation of our geometry, where we deal with 
four-form flux on a spacetime of the form
\be
\mathbb{C}\times S_a^1\times Y_4\,.
\ee
This is dual to Type IIB string theory on
 \be\label{IIBframe}
 \mathbb{C}\times S_a^1\times S_b^1\times B_3 \,,
 \ee
 where $B_3$ denotes the base of the elliptic fibration, $Y_4$.
In this duality frame, the strings whose elliptic genus is computed by the topological $A$-model prepotential arise from D3 branes wrapped on two-cycles, $C_\beta \in H_2(B_3)$.

One may think of the four-form flux as being sourced by M5-brane domain walls,  as explained for instance~in Section~3 of \cite{Denef:2008wq}; the M5-branes are extended along a two-dimensional subspace $\mathbb{C}$ of three-dimensional spacetime $\mathbb{C}\times S^1_a$, and wrap the four-cycle in the Calabi-Yau fourfold that is Poincar\'e dual to the four-form flux. In the dual Type IIB formulation, the M5-branes map to different objects depending on whether the original flux is of type $(-2)$, $(-1)$, or $(0)$. 

The situation is easiest understood for 
$(-2)$-fluxes of the form $G_{\dot{\alpha}}=\pi^*(\Sigma_{\dot \alpha}^{\rm b})$, where $\Sigma_{\dot \alpha}^{\rm b}$ is a curve on the base $B_3$.
Such a flux is sourced by an M5-brane along $\mathbb C \times \pi^*(\Sigma_{\dot \alpha}^{\rm b})$. Note that this M5-brane in particular wraps the full elliptic fiber.
Locally, we can identify one of the one-cycles in the fiber with the M-theory circle. Dualising to the Type IIA frame we obtain a D4-brane that is locally wrapped on $\mathbb C \times \Sigma_{\dot \alpha}^{\rm b}$ times the remaining 1-cycle in the fiber.
T-duality along the latter then takes us to Type IIB theory on  $\mathbb{C}\times S_a^1\times S_b^1\times B_3$ with a D3-brane wrapped on  $\mathbb C \times \Sigma_{\dot \alpha}^{\rm b}$.
From the perspective of the four-dimensional spacetime,
this  ``flux'' D3 brane represents a defect. Indeed even in the limit of infinite radii for $S_a^1\times S_b^1$, four-dimensional Poincar\'e invariance is broken.  See the left-hand side of Fig.~\ref{fig:fluxgeo} for a visualization.

\begin{figure}[t!]
\centering
\includegraphics[width=14cm]{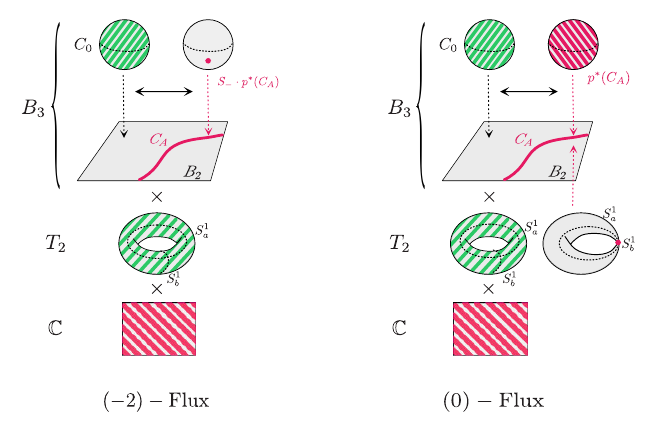}
\caption{
Shown is the interplay of the string and flux geometry for the rational fibrations $B_3 \to B_2$ 
which we consider as an example, referring to the geometry (\ref{IIBframe}) in the Type IIB duality frame.
The green hatching shows the wrapping locus of the D3-brane that leads to
a heterotic string which is further compactified on $S^1_a\times S^1_b$ to two dimensions.
The red hatching shows the loci of the ``flux'' branes that encode the background flux.\\
The left side corresponds to a $(-2)$-flux which is described by a D3-brane on ${\mathbb C}
\times (S_-\cdot p^*(C_A))\equiv {\mathbb C}\times\Sigma_{\dot A}^{\rm b}$. When uplifting to four dimensions by making
the circles large, this turns into a defect in four dimensions.\\
The right side corresponds to a $(0)$-flux of the form $G_{\alpha_\tau}=D_\tau\cdot\pi^*p^*(C_A)\equiv D_\tau\cdot D_A$,
which is described by a KK monopole defect (red hatched locus), as explained in the text.
We will argue below that the linear term in the holomorphic anomaly of the elliptic genus arises, formally, from the branch where the red and green hatched loci intersect. 
}
\label{fig:fluxgeo}
\end{figure}

The brane sources against which we can trade the remaining types of $(-1)$- or $(0)$-fluxes are more complicated \cite{Denef:2008wq}.  
For a $(0)$-flux $G_{\alpha_\tau} = D_\tau \cdot \pi^\ast(D^{\rm b}_\alpha)$, one obtains a Kaluza-Klein (KK) monopole along $\mathbb{C}\times D^{\rm b}_\alpha$, with $S^1_b$ being fibered nontrivially over $S^1_a$ times the two-cycle dual to the four-cycle $D^{\rm b}_\alpha$ in $B_3$.  See the right-hand side of Fig.~\ref{fig:fluxgeo} for a visualization.

On the other hand, for a $(-1)$-flux $G_{\alpha_z}= D_z \cdot \pi^\ast(D^{\rm b}_\alpha)$, one finds a domain wall realised by a Type IIB five-brane along $\mathbb{C}\times S^1_b $ times a three-chain in $B_3$ ending on seven-branes. This is in agreement with the interpretation of this flux background as an internal gauge flux, for which the four-dimensional theory is Poincar\'e invariant  (after decompactifying $S^1_a \times S^1_b$). This reflects, as emphasized above, that only the $(-1)$-flux partition functions lift nicely to four-dimensional elliptic genera without any defects  \cite{Lee:2020gvu}.
For the  $(0)$- or $(-2)$-flux backgrounds, by contrast, we instead propose an interpretation as elliptic genera of strings in the presence of KK-monopoles or string-like defects in four dimensions.

As a first, crude test of this picture, we observe that while for the $(-2)$-fluxes the geometry is symmetric under exchange of $S_a^1$ and $S_b^1$, for
the other flux types the brane configurations distinguish between the two circles. This serves
as an intuitive explanation of the result~\eqref{E1equ-2} that in general only the $(-2)$-fluxes give rise to good  (quasi-)modular partition functions, while the modular properties in the presence of generic  
$(-1)$- and $(0)$-fluxes are much more intricate. 
 
 \subsubsection*{Localisation of Partition Functions on Defects}

 Next we would like to understand, from this perspective,
  why for suitable flux backgrounds the partition functions on the fourfold $Y_4$ are given \cite{Lee:2020gvu} 
  in terms of the 
 prepotentials (or derivatives thereof) associated with certain embedded threefolds, ${\mathbb Y}_3^A$. This statement was formalized by eqs.~(\ref{qderivative1}) which we derived from relative Gromov-Witten theory.
 
For concreteness,  let us focus on geometries where the base space, $B_3$, is by itself a rational fibration,      
of the form as detailed in Section~\ref{setupsecEx}.
Staying in the Type IIB picture, we consider a D3-brane wrapped on the rational fiber $C_0$ of the fibration $p:B_3\to B_2$, corresponding to a heterotic string in four dimensions. As above,
we furthermore consider a $(-2)$-flux background dual to a second, ``flux'' D3 brane wrapped on $\mathbb{C}$ times a representative of the two-cycle $\Sigma^{\rm b}_{\dot A} = S_- \cdot_{B_3} p^*(C_A)$; see again Fig.~\ref{fig:fluxgeo}.
For such special $(-2)$-flux we know from (\ref{qderivative1}) that the prepotential encodes the Gromov-Witten invariants of the threefold ${\mathbb Y}_3^A = Y_4 |_{p^\ast (C_A)}$.
To understand this from a stringy worldsheet perspective, note that 
as the rational fiber can be moved over $B_2$, the moduli space of the heterotic string will include a component which is fibered over $C_A$. 
According to our initial remarks, the relative prepotential with respect to the curve $C_0$ should be proportional to the elliptic genus
\be  \label{ellgen(-2)-a}
\Tr_{{\cal H}_{-2}} (-1)^{F_R} {F_R}\, q^{H_L} \overline{q}^{H_R} \xi^{Q}   \,.
\ee
Here ${\cal H}_{-2}$ denotes the Hilbert space of string excitations for the solitonic heterotic string probing the Type IIB spacetime $\mathbb C \times S^1_a \times S^1_b \times B_3$ in the background of said D3-brane defect. 
Locally, at a generic point of the moduli space, the worldsheet CFT does not sense the presence of the flux background on $Y_4$, or equivalently,
of the D3-brane defect along $\mathbb C \times \Sigma^{\rm b}_{\dot A}$. Hence away from the defect,
the elliptic genus (\ref{ellgen(-2)-a})
of the string should give the same answer as for a Type IIB background with no flux at all, i.e., it should vanish.
In the duality frame of the heterotic string, this means in particular  that the spectrum of its excitations is non-chiral, except possibly for contributions localised at the defect dual to the ``flux'' D3-brane along 
$\mathbb C \times \Sigma^{\rm b}_{\dot A}$. 

Restricting to this locus is equivalent to constraining the D3-brane to the four-cycle 
$p^\ast(C_A)$ on $B_3$.  As remarked above, the elliptic fibration over $p^\ast(C_A)$ defines an embedded threefold,
 ${\mathbb Y}_3^A$.
This ties in with our observation \cite{Lee:2020gvu} that when $\mathbb Y_3^A$ happens to be Calabi-Yau, the elliptic genus (\ref{ellgen(-2)-a}) of the four-dimensional string reproduces the elliptic genus
of a string on  the Type IIB background $\mathbb C^2 \times S^1_a \times S^1_b \times p^\ast(C_A)$, albeit without any further defects.  Thus, if we denote the Hilbert space of the worldsheet theory in this background as  ${\cal H}_{{\mathbb Y}_3^A}$,  we expect that
\be\label{Y3ellgen}
\Tr_{{\cal H}_{-2}} (-1)^{F_R} {F_R}\, q^{H_L} \overline{q}^{H_R} \xi^{Q}    =   \Tr_{{\cal H}_{{\mathbb Y}_3^A}} (-1)^{F_R} F_R^2 \,q^{H_L} \overline{q}^{H_R} \xi^{Q} \ \equiv - q^{-1} \cF_{C_0}^{\mathbb Y_3^A}\,.
\ee
Note that the fluctuations of the string in the directions normal to $p^\ast(C_A)$ are encoded in the extra factor of $\mathbb C$, but 
since the worldsheet theory on $\mathbb C^2 \times S^1_a \times S^1_b \times p^\ast(C_A)$  has $N=(0,4)$ rather than $N=(0,2)$ supersymmetry, there must appear an extra factor of 
$F_R$ on the right-hand side in order to saturate the extra zero modes and give a non-zero result. This
fits together with the observation  \cite{Lee:2020gvu}  
that formally the right-hand side of (\ref{Y3ellgen}) looks like the
elliptic genus of a six-dimensional theory, in particular it has the proper
modular weight, $w=-2$.

These preliminary considerations may serve as a precursor for the deeper understanding 
of the other types of flux backgrounds, the $(-1)$- and $(0)$-fluxes:
In particular, it would be interesting to explain
 the derivative contributions to the $(-1)$- and $(0)$-flux partition functions, as also encoded in (\ref{qderivative1}),  in an analogous manner. 
 While the localisation of the partition function to the threefolds $\mathbb Y_3^A$
  follows along the same lines as for $(-2)$-fluxes discussed above,
 it is more challenging to explain the physical rationale behind the derivatives.

\subsubsection*{Holomorphic Anomalies from a Geometric Perspective}

Rather than exhaustively solving this problem here, let us adopt a worldsheet perspective to identify
a possible physical mechanism that underlies the holomorphic anomalies under consideration.
This purported mechanism is complementary in that it ought to supply the non-holomorphic completions of 
$\hat E_2(\tau)= E_2(\tau)-\frac{3}{\pi{\rm Im}\tau}$ and 
$\hat E_1(\tau,z)= E_1(\tau,z)+\frac{{\rm Im}z}{{\rm Im}\tau}$
that are needed to restore invariance under modular and elliptic transformations. 
We have seen that the appearance of the holomorphic quasi-modular and quasi-Jacobi forms, $E_2(\tau)$ and $E_1(\tau,z)$,  is a consequence of certain partition functions being derivatives of others. As we have just argued, this derivative structure is tied to the localisation of the elliptic genera on certain defects.
This calls for a more direct explanation of the holomorphic anomalies from the perspective of such defects. 

For illustration, let us focus on
the prototypical anomaly equation for $(0)$-type flux, which has the schematic form
\be\label{crudeHAE}
\bar\partial_{\bar\tau}\cF_{G^{(0)}|C_\beta}\ =\ \sum_{C_{\beta_1}+C_{\beta_2}=C_\beta}
\cF_{G^{(0)}|C_{\beta_1}}\cF_{G^{(-2)}|C_{\beta_2}}+ \cF_{G^{(-2)}|C_{\beta}}\,.
\ee
We have exhibited that there is in general a mixture of
both quadratic and linear terms.  
The latter originates in the gravitational descendant term in (\ref{MAEgen}) and vanishes for $(-1)$ and $(-2)$ flux backgrounds.

The quadratic terms correspond to the split of a reducible
curve $C_\beta$ into two irreducible components and are familiar from
the modular anomaly equations at genus zero on threefolds \cite{Klemm:2012sx,Alim:2012ss}. In six dimensions,
for the special case where a D3-brane on $C_\beta=C_0$ 
describes a heterotic string,  they have a physical interpretation in terms of a heterotic string splitting into two non-critical E-strings \cite{Minahan:1998vr,Haghighat:2014pva}.  
 On this component of moduli space, new zero modes appear. The contribution to the modular anomaly is then proportional to the elliptic genus of the system localized on the component of moduli space 
where such zero modes emerge, i.e., to the product of the elliptic genera of the two strings into which the original bound state marginally decomposes.

More specifically, it is well known that the elliptic genus can receive non-holomorphic contributions  if the spectrum of the worldsheet theory contains a continuum of states (see e.g., \cite{Troost:2010ud}). In this case, the cancellation of right-moving bosonic and fermionic modes in the index, which would be responsible for holomorphicity of the elliptic genus when the spectrum is discrete, can fail. 
A~continuous spectrum points to a non-compact sigma model target space, and the holomorphic anomaly localises on its boundary \cite{Gaiotto:2019gef,Dabholkar:2020fde}. 
For example, this phenomenon occurs if the worldsheet theory contains St\"uckelberg-type compensator fields whose shift symmetry is gauged in such a way as to cancel one-loop gauge anomalies on the worldsheet \cite{Murthy:2013mya,Harvey:2014nha}. 

The relation to the quadratic term on the right-hand side of (\ref{crudeHAE}) is most clear if
we specialise to a heterotic string as before and note that 
the split into two E-strings occurs at the position of an NS5-brane.
At the same time, NS5-branes in the background of a heterotic string
provide precisely the ingredients described above: As explained in \cite{Blaszczyk:2011ib,Quigley:2011pv}, in such backgrounds the GLSM underlying the heterotic worldsheet theory suffers from a 1-loop anomaly, which is cancelled by a two-dimensional version of the Green-Schwarz mechanism involving a  St\"uckelberg-type compensator field. Even though the technical details differ, this puts the holomorphic anomaly associated with the split of 
the heterotic string into two E-strings into the general context of the holomorphic anomalies observed for various GLSMs with 1-loop gauge anomalies on the worldsheet \cite{Murthy:2013mya,Harvey:2014nha}.

It is then suggestive that an analogous mechanism should be at work behind the linear term in (\ref{crudeHAE}). Let us again
focus on the heterotic string that arises from a D3-brane wrapped on $C_0$.
The KK-monopole that describes the $(0)$-flux background, $G_{\alpha_\tau} = D_\tau \cdot D_\alpha$, on the Type IIB side, dualises to a heterotic NS5-brane that wraps some divisor in the dual heterotic threefold and extends along the subspace $\mathbb C$ in four dimensions. 
By analogy, we may expect that the presence of this NS5-brane induces a holomorphic anomaly which should be localised on the NS5-brane, acting as the boundary of the target space.
This time, however, the anomaly is not tied to a split of the heterotic string into two constituent strings.

This can be most easily seen when the divisor $D_\alpha$ defining the $(0)$-flux is of the form $\pi^\ast(p^\ast C_A)$, where $C_A$ is a curve on $B_2$; see Fig.~\ref{fig:fluxgeo}.
If the anomaly follows the same logic as before,
it is proportional 
 to the elliptic genus of the component of moduli space 
 where the string meets the NS5-brane.
This component corresponds to the moduli of $C_0$ inside the vertical divisor $\pi^\ast(p^\ast C_A)$, and 
is given by the embedded threefold, $\mathbb Y_3^A$.
In other words, the linear term in the anomaly should be  proportional to the elliptic genus $\cF_{C_0}^{\mathbb Y_3^A}$, as given on the  right-hand side of (\ref{Y3ellgen}).

This heuristic picture  perfectly matches our quantitative evaluation of the linear piece of (\ref{crudeHAE}), for the class of geometries under consideration. Recall that it is given in Gromov-Witten theory by the gravitational descendant invariant as computed in the third line of (\ref{psi-invariants}). Indeed this reproduces via~\eqref{reinv4fold3fld} the elliptic genus 
along the divisor $\pi^\ast(p^\ast C_A)$, precisely as expected.
Similar reasoning goes through also for those $(0)$-fluxes whose divisors $D_\alpha$ are exceptional divisors on $B_3$, see the second line of (\ref{psi-invariants}).
The remaining case, where $D_\alpha$ intersects $C_0$ topologically, is however more involved. From  the first line of (\ref{psi-invariants}) we observe that the linear term of the anomaly equation is now proportional to the elliptic genus along the pullback of the self-intersection
of such a divisor on $B_2$ to $B_3$. While it is tempting to speculate that this may have to do with a certain localisation of zero modes, a more quantitative analysis to support this is beyond the scope of this work. 

Suffice it to mention in closing that the structure of the linear term
seems analogous to the holomorphic anomaly discussed in \cite{Gaiotto:2019gef,Dabholkar:2020fde}.
There one considers sigma-models on non-compact target spaces, $\mathcal X$, with boundaries 
$\mathcal Y$. The elliptic genus in turn suffers from a holomorphic anomaly that localizes on $\mathcal Y$,
similar to what we find for the linear term in (\ref{crudeHAE}) given by $\cF_{C_0}^{\mathbb Y_3^A}$.
One difference is that our analysis involves only mildly non-holomorphic modular and Jacobi forms and not complicated
mock modular forms as in those works, but this may be due to the fact that we consider the limit 
(\ref{antitlimit}) in which anti-holomorphic $\bar q$-series vanish. Another is that we actually deal with anomalies induced by background fluxes, and we chose to represent the latter by ``flux''-branes assuming that these capture the correct physics. While this seems to make sense for (co-)homological aspects, it is not clear to what extent the suggestive arguments we made above apply to actual flux backgrounds.

To summarize,  in this paper we have promoted the familiar, fruitful interplay between topological string theory, enumerative geometry, holomorphic anomalies  and the worldsheet interpretation of 
critical and non-critical strings to the realm of $N=1$ supersymmetric  theories in four dimensions.  
The novel features we encountered include the enumerative geometry of relative Gromov-Witten invariants on fourfolds with fluxes, and linear terms in holomorphic anomaly equations. These reflect derivative relationships between partition functions and arise from gravitational descendant invariants 
in Gromov-Witten theory or from degenerating flux geometries.

\subsection*{Acknowledgements}
We thank Bumsig Kim and Jeongseok Oh for useful discussions. 
The work of SJL is supported by IBS under the project code, IBS-R018-D1.
The work of T.W. is supported in part by Deutsche Forschungsgemeinschaft under Germany's Excellence Strategy EXC 2121 Quantum Universe 390833306.


\appendix

\section{Gravitational Descendant Invariants}   \label{app_desc}

In this appendix we evaluate the gravitational descendant invariants at genus zero, which contribute to the linear terms in
the modular anomaly equation (\ref{MAEgen}), or its non-holomorphic cousin (\ref{HAEversion1}).
As stated in (\ref{psiequation}), one can rewrite the genus-zero descendant invariants in terms of non-descendant Gromov-Witten invariants via the 
so-called Dubrovin method \cite{Hori:2003ic}. In Appendix \ref{App_A1} we will give a detailed derivation of (\ref{psiequation}) and then evaluate it in Appendix \ref{app_descrational} 
 for the case where the base curve is the rational fiber of a $\mathbb P^1$ fibration $B_3$.

\subsection{General Derivation}   \label{App_A1}

Our goal is to compute general genus-zero descendant invariants of the form 
\beq 
\psi\cdot \langle D \rangle_{C_\beta}\,,
\eeq
with one point fixed, for $D \in H^2(Y_4)$ and  $C_\beta \in H_2(Y_4)$. 
In expressions of this form, the `dot' denotes the intersection form of $\psi$ with $\langle D \rangle_{C_\beta}$ along (the virtual cycle in) the moduli space of stable maps at genus zero with one point fixed.
The idea is to first relate this invariant to the 3-point invariant,
\beq\label{psi-HHD}
\psi \cdot  \langle  H,H, D \rangle_{C_\beta}\,,
\eeq
where $H \in H^2(Y_4)$ is some suitably chosen auxiliary divisor class. Its  purpose is, morally,
  to mimick a stable degeneration and while its precise choice does not matter provided that $H \cdot C_\beta \neq 0$. Then one
invokes the boundary lemma to eliminate the $\psi$-class from~\eqref{psi-HHD}. Here, the $\psi$-class in~\eqref{psi-HHD}, as well as in any of the ensuing descendant invariants, is always understood to act on the rightmost marked point. 

Concretely, repeated use of the divisor equation (\ref{DivEqu}) for descendant invariants yields 
\bea
\psi \cdot\langle  H,  D \rangle_{C_\beta} &=&  (H \cdot C_\beta) \,\psi\cdot \langle   D \rangle_{C_\beta} +  \langle  H \cdot D \rangle_{C_\beta} \\
 \psi \cdot \langle  H, H,D \rangle_{C_\beta} &=&(H \cdot C_\beta) \,\psi\cdot  \langle H, D \rangle_{C_\beta} + \langle  H, H \cdot D \rangle_{C_\beta}  \,,
\eea
which allows  to solve for  $\psi\cdot \langle  D \rangle_{C_\beta}$ in terms of $\psi \cdot \langle  H,H,  D \rangle_{C_\beta}$ as long as $H \cdot C_\beta \neq 0$.
In the second step this three-point descendant invariant
is expressed with the help of the boundary lemma  \cite{Hori:2003ic} as 
\bea \label{HHpsi-1}
\psi \cdot  \langle  H,H, D \rangle_{C_\beta} =   \sum_{C_{\beta_1} + C_{\beta_2} =C_\beta}    \langle  D, A_i  \rangle_{C_{\beta_1}}   g^{ij}   \langle A_j, H, H \rangle_{C_{\beta_2}}  \,,
\eea
where the sum is over all splittings of the curve class $C_\beta = C_{\beta_1} + C_{\beta_2}$, and $\{A_i\}$ represents a basis of $H^*(Y_4)$ with intersection form $g_{ij} = \int_{Y_4} A_i \wedge A_j$ and inverse $g^{ij}$.
Since for a Calabi-Yau fourfold $Y_4$ the moduli space of stable maps at genus zero with $n$ points fixed has complex virtual dimension $1+n$, the only non-trivial contributions on the right-hand side can come from $A_i \in H^{2,2}(Y_4)$.

The sum over curve class splittings $C_\beta = C_{\beta_1} + C_{\beta_2}$ in (\ref{HHpsi-1}) includes, as special cases, also the splittings corresponding to $C_{\beta_1} = 0$, $C_{\beta_2} = C_\beta$ and $C_{\beta_1} = C_\beta$, $C_{\beta_2} = 0$.
Such ``trivial'' splittings do not contribute to the familiar quadratic terms in the BCOV equations, but in the present context
of fourfolds they can contribute to the gravitational descendant invariant.
They are easily dealt with because the Gromov-Witten invariants for homologically trivial curves reduce to simple ``classical'' intersection integrals  over $Y_4$.  The relevant expressions in our situation, where $A_i \in H^{2,2}(Y_4)$ and $D \in H^{2}(Y_4)$, are
\bea
\langle  A_j, H, H \rangle_{0}    &=& \int_{Y_4} A_j \wedge H \wedge H \ \equiv \ (A_j\cdot H \cdot H) \,, \\
\langle  D, A_i \rangle_{0}    &=&  \int_{Y_4} D \wedge A_i   = 0    \,.
\eea
The remaining invariants that appear on the right of (\ref{HHpsi-1}) can be reduced to one-point invariants with the help of the divisor equation, which for primary invariants (i.e. those not including any $\psi$ classes) reads
\bea
\langle  D_1, D_2, \ldots, D_n, A_i  \rangle_{C_\beta} =  (D_1 \cdot C_\beta)  (D_2 \cdot C_\beta)  \ldots   (D_n \cdot C_\beta)   \langle A_i \rangle_{C_\beta}   \,, \quad  \quad  D_i \in H^2(Y_4)   \,.
\eea
This leads to
\bea   \label{HHpsD}
\psi \cdot\langle  H,H,  D \rangle_{C_\beta} &=& ({\rm A}) + ({\rm B})   \,,   \\
({\rm A}) &=& (D \cdot {C_\beta})  \sum_{i,j}  \langle A_i \rangle_{C_\beta}   g^{ij}  (\int_{Y_4} A_j \wedge H \wedge H)    \,,   \\
({\rm B})   &=&     \sum_{C_{\beta_1} + C_{\beta_2} = C_\beta, \,C_{\beta_i} \neq 0}    (D \cdot C_{\beta_1})  \langle A_i \rangle_{C_{\beta_1}}   g^{ij} \langle A_j \rangle_{C_{\beta_2}}   (H \cdot C_{\beta_2})^2   \,.
\eea
Putting everything together, we then arrive at the following expression of the gravitational descendant term:
\bea
 \psi \cdot  \langle D \rangle_{C_\beta}  = \frac{1}{(H \cdot C_\beta)^2} (({\rm A}) + ({\rm B})) -  \frac{2}{H \cdot C_\beta}  (C)   \,,
\eea
where $(C)=\langle  H \cdot D \rangle_{C_\beta}$.

So far we have been considering general Gromov-Witten invariants on Calabi-Yau fourfolds. If we consider {\it relative}
 genus-zero invariants $ \psi \cdot\langle \langle \pi^\ast(D^{\rm b}) \rangle \rangle$ as appearing in our anomaly equations, 
the same logic goes through provided we pick the divisor $H=\pi^\ast(H^{\rm b})$ suitably.
Explicitly,   
\bea\label{psi-app}
 \psi \cdot \langle \langle\pi^\ast(D^{\rm b}) \rangle \rangle_{C_\beta}  &=& \frac{1}{(H^{\rm b}\cdot C_\beta^{\rm b})^2} (({\rm I}) + ({\rm II})) -  \frac{2}{H^{\rm b}\cdot C_\beta^{\rm b}}  ({\rm III})  \,,   \\ \label{psi-app-1}
({\rm I}) &=& (D^{\rm b} \cdot C_\beta^{\rm b})  \sum_{i,j}  \langle \langle A_i  \rangle \rangle_{C_\beta}   g^{ij}  (\int_{Y_4} A_j \wedge H\wedge H )  \,,  \\ \label{psi-app-2}
({\rm II})   &=&     \sum_{C_{\beta_1} + C_{\beta_2} = C_\beta,\, C_{\beta_i} \neq 0}    (D^{\rm b} \cdot C_{\beta_1}^{\rm b})  \langle\langle A_i \rangle\rangle_{C_{\beta_1}}   g^{ij} \langle\langle A_j \rangle\rangle_{C_{\beta_2}}   (H^{\rm b}\cdot C_{\beta_2}^{\rm b})^2 ,  \\ \label{except}
({\rm III})   &=&     \langle \langle  H\cdot  \pi^*(D^{\rm b}) \rangle \rangle_{C_\beta} 
  \,.
\eea
Here all intersection products (except for the one in~\eqref{except}) are evaluated directly on the base $B_3$, and the base curve classes are also distinguished by the superscript, e.g.  $C^{\rm b}_\beta := \pi_* C_\beta$.

\subsection{Application to Rationally Fibered $B_3$}   \label{app_descrational}

We now evaluate the equations~\eqref{psi-app},~\eqref{psi-app-1} and~\eqref{psi-app-2} for the geometries considered in Section~\ref{setupsecEx}, for which $B_3$ is rationally fibered. Since the invariants are linear in the divisor $D^{\rm b}$, it suffices to evaluate them  for the basis elements $D^{\rm b} = D^{\rm b}_\alpha$ separately.  Recall that these encode the linear pieces in the anomaly equations related to the $(0)$-fluxes $G_{\alpha_\tau}$. To this end, we will evaluate the equations for the auxiliary base divisor $H^{\rm b}$ of the form, 
\beq
H^{\rm b}= 2 S_- + a^A \,p^* (C_A) + a\,E \,.
\eeq
Here the parameters $a^A$ and $a$ are to be chosen appropriately, depending on the divisor $D^{\rm b} \equiv D^{\rm b}_\alpha$ 
that we consider. 

Firstly, when choosing $\alpha = A$ for $A=1, \ldots, h^{1,1}(B_2)$ (so that $ D^{\rm b} = D_A^{\rm b} = p^*(C_A)$), one immediately sees that 
\beq
D^{\rm b} \cdot C_0 = 0 \,,\qquad D^{\rm b} \cdot C_E^1 = 0 = D^{\rm b} \cdot C_E^2 \,,
\eeq
which leads to the trivial vanishings $({\rm I}) = ({\rm II})=0$, as per~\eqref{psi-app-1} and~\eqref{psi-app-2}. Therefore, the descendant invariant simplifies to:
\bea
\psi \cdot \langle \langle  \pi^\ast(D^{\rm b}) \rangle \rangle_{C_0} &=&  - \langle \langle \pi^*(2S_- + a^B \,p^* (C_B) + a\, E) \cdot \pi^*(D_{A}^{\rm b}) \rangle \rangle_{C_0}\nn \\
&=& -2 \langle \langle \pi^*(S_-) \cdot \pi^*(D^{\rm b}_A) \rangle \rangle_{C_0}  \\
&=& -2 \langle \langle G_{\dot A} \rangle \rangle_{C_0} \,. \nn
\eea
In the second step we have used the vanishing $E \cdot p^*(C_A) = 0$ as well as the independence\footnote{In fact this consistency requirement demands that $\langle \langle \pi^*p^*(C_B) \,\cdot\, \pi^*p^*(C_A) \rangle \rangle_{C_0}$ should vanish.} of the invariants $ \psi \cdot\langle \langle \pi^\ast(D^{\rm b}_{A}) \rangle \rangle_{C_0} $ on the parameters $a^B$ for $H^{\rm b}$. That is,  the descendant invariant has turned into nothing but the partition function for the $(-2)$-flux $G_{\dot A}$.

This partition function coincides  with the relative invariant on the induced fibration $\mathbb Y_3^A := Y_4|_{p^*(C_A)}$, that is:
\beq\label{4-to-3}
 \langle \langle G_{\dot A} \rangle \rangle_{C_0} =  \langle \langle  ~\rangle \rangle_{C_0}^{\mathbb Y_3^A} \,,
\eeq
where the embedded threefold,  $\mathbb Y_3^A$, may or may not be a Calabi-Yau manifold by itself.
More specifically,  the connection between the Gromov-Witten theories on $Y_4$ and $\mathbb Y_3^A$ arises as follows:
 \beq\label{(-2)-to-3}
\langle \langle \pi^*(S_-) \cdot \pi^*(D^{\rm b}_A) \rangle \rangle_{C_0}^{Y_4} = \langle \langle \pi^*(S_-)  \rangle \rangle_{C_0}^{\mathbb Y_3^A}  =\langle \langle~\rangle\rangle_{C_0}^{\mathbb Y_3^A} \,,
\eeq
where, in the first step, the reduction formula~\eqref{ReductionFormula} has been applied, based on the fact that $C_0$ is fully contained in the divisor $D_A^{\rm b}$ of $B_3$. In the second step the divisor equation has been used to remove the marked point. 
We therefore conclude that for the choice $\alpha = A$ the gravitational descendant term evaluates to the 
following relative invariant
of the embedded threefold,~$\mathbb Y_3^A$:
\beq
\psi \cdot\langle \langle  \pi^\ast(D^{\rm b}_{A}) \rangle \rangle_{C_0} = -2 \langle \langle G_{\dot A} \rangle \rangle_{C_0} = -2  \langle \langle  ~\rangle \rangle_{C_0}^{\mathbb Y_3^A} \,. 
\eeq

Secondly, for the choice $\alpha =0$ (so that $D^{\rm b} = E$), one sees again that
\beq
D^{\rm b} \cdot_{B_3} C_0 = 0 \,,
\eeq
and hence  $({\rm I}) = 0$ via~\eqref{psi-app-1}. On the other hand, $D^{\rm b}$ intersects non-trivially with the split components:
\beq
D^{\rm b} \cdot_{B_3} C_E^1 = 1 = - D^{\rm b} \cdot_{B_3} C_E^2 \,.
\eeq
If we want the term $({\rm II})$ to manifestly vanish as well, we can choose  $a=-1$ and thus specialise the base divisor $H^{\rm b}$ to
\beq
H^{\rm b}= 2 S_- + a^A \, p^* (C_A) - E \,.
\eeq
The descendant invariant is then computed as
\bea
 \psi \cdot \langle \langle\pi^\ast(D^{\rm b}) \rangle \rangle_{C_0} &=&  - \langle \langle \pi^*(2S_- + a^A \,p^*(C_A) - E) \cdot \pi^*(E) \rangle \rangle_{C_0}\nn \\
&=& -  \langle \langle \pi^*(S_-) \cdot \pi^*p^*(\Gamma) \rangle \rangle_{C_0} \,, \\
&\equiv:&-  \langle \langle G_\Gamma \rangle \rangle_{C_0} \,,\nn
\eea
where in the second step we have used the vanishing of $S_- \cdot E$ and $p^*(C_A) \cdot E$, as well as the relation $E \cdot E = -S_- \cdot p^*(\Gamma)$. The last step is simply a definiton:
\beq
G_{\Gamma}:=\pi^*(S_-) \cdot \pi^*p^*(\Gamma)\,.
\eeq
Since $\mathbb Y_3^\Gamma :=Y_4|_{p^*(\Gamma)}$ fully contains $C_0$, we can follow steps analogous to the ones used in~\eqref{(-2)-to-3}, and thus obtain  
\beq
 \psi \cdot\langle \langle \pi^\ast(E) \rangle \rangle_{C_0} = - \langle \langle G_\Gamma \rangle \rangle_{C_0} = -  \langle \langle  ~\rangle \rangle_{C_0}^{\mathbb Y_3^\Gamma} \,.
\eeq

Finally, for the choice $\alpha = -1$ (so that $D^{\rm b} = S_-$), we take for the base divisor $H^{\rm b}$:
\beq
H^{\rm b}= 2 S_- \,,
\eeq
by turning off all the parameters $a^A$ and $a$. With this choice, 
we immediately infer the vanishing of $({\rm II})$ in~\eqref{psi-app-2}  due to
\bea
D^{\rm b} \cdot  (C_E^1, C_E^2) &=& (1,0) \,,\\
H^{\rm b}\cdot (C_E^1, C_E^2) &=& (2, 0) \,.\nn
\eea
However, the term $({\rm I})$ will in general lead to a non-trivial expression. Specifically, from~\eqref{psi-app-1} we have:
\bea
\frac{1}{(H\cdot C_0)^2}({\rm I}) &=&    \langle \langle A_a  \rangle \rangle_{C_0} \, I^{ab} \, (\int_{Y_4} A_b \wedge \pi^\ast(S_-) \wedge \pi^\ast(S_-))  \nn\\ \label{2nd}
&=&   \langle \langle G_{\dot \alpha} \rangle \rangle_{C_0} \, I^{\dot \alpha \beta_\tau} \, (\int_{Y_4} G_{\beta_\tau} \wedge \pi^\ast(S_-) \wedge \pi^\ast(S_-)) \,,
\eea 
where the sum over cohomology classes has been reduced to one over $(-2)$-/$(0)$-fluxes upon ignoring vanishing terms. Note that the integrals that appear in~\eqref{2nd} can be simplified to
\bea
\int_{Y_4} G_{(-1)_{\tau}} \wedge \pi^\ast(S_-) \wedge \pi^\ast(S_-) &=& \int_{B_3} S_- \cdot p^*(c_1(\cL)) \cdot p^* (c_1(\cL)) = c_1(\cL) \cdot_{B_2} c_1(\cL) \,,\nn\\
\int_{Y_4} G_{0_\tau} \wedge \pi^\ast(S_-) \wedge \pi^\ast(S_-) &=&  -\int_{B_3} E \cdot S_- \cdot p^* (c_1(\cL)) = 0 \,, \\
\int_{Y_4} G_{A_\tau} \wedge \pi^\ast(S_-) \wedge \pi^\ast(S_-) &=& -\int_{B_3} p^*(C_A) \cdot S_- \cdot p^* (c_1(\cL)) = - \ell_A \,,
\nn\eea
where, in the last equation, the definition~\eqref{ellA} has been used for $\ell_A$. 
Plugging the inverse intersection form~\eqref{ginv} into~\eqref{2nd}, we thus obtain
\beq
\frac{1}{(H\cdot C_0)^2}({\rm I}) =  \langle \langle G_{\dot \alpha} \rangle \rangle_{C_0}  \Lambda^{\dot \alpha} \,,
\eeq
where $\Lambda^{\dot \alpha}$ are the entries in the inverse matrix that mix the $(-2)$-fluxes.
Explicitly they are given by
\bea
\Lambda^{-1} &=& c_1(\cL)\cdot_{B_2} c_1(\cL) - \ell_A\ell^A  = 0 \,,\nn \\
\Lambda^0 &=& 0 \,,\\
\Lambda^A &=& -\ell^A \,. \nn
\eea
Upon  evaluating the second term  $({\rm III})$  in~\eqref{psi-app} in a similar manner, we eventually obtain
\beq
 \psi \cdot \langle \langle\pi^\ast(S_-) \rangle \rangle_{C_0} =\sum_A \ell^A \langle \langle G_{\dot A} \rangle \rangle_{C_0} = \sum_A \ell^A  \langle \langle  ~\rangle \rangle_{C_0}^{\mathbb Y_3^A} \,.
\eeq
Summarizing, for the specific geometries under consideration (namely rationally fibered bases~$B_3$),
the term $({\rm II})$ can be arranged to trivially vanish for judicious choices for the auxiliary divisor $H$, in which case the gravitational descendant invariant $\psi \cdot \langle \langle  \pi^\ast(D^{\rm b}) \rangle \rangle_{C_0}$ manifestly reduces to expressions 
that are purely linear in partition functions (note, as mentioned, that
the final result does actually not depend on the choice of $H$,
as long as $H \cdot C_\beta \neq 0$).  For more general geometries, however, there will be additional quadratic pieces. 

\section{Details on the Derivation of the Modular and Elliptic Anomaly Equations}\label{details_AnomalyEqs}

Here we present some technicalities concerning our derivation of the holomorphic anomaly equation in the form (\ref{HAEfirsttermsugg})
and of the elliptic anomaly equation (\ref{E1equ}).

\subsection{Proof of Equation (\ref{Cbaraufinal})}  \label{AppproofC}

We show that the overall coupling 
 \bea
 \overline C_{\bar \tau}^{j b} =  \overline{\cF}_{\bar c;\bar{\tau} \bar{k}}   \,  e^{2K}    \, G^{j \bar k} \, G^{b \bar c} \,,
\eea
which appears in the holomorphic anomaly equation (\ref{HAEFaC-1}) for $\bar i = \bar \tau$, reduces  in the limit (\ref{t-limit})
to the expression
\bea    \label{Cbaraufinal-App}
\overline C_{\bar \tau}^{j b}    \stackrel{(\ref{t-limit})}{=}  { \frac{1}{(2 \pi)^2}}\frac{1}{4 \tau^2_2} \,  \delta^j_\alpha  \delta^b_{\dot\beta}  \,    I^{\alpha \dot\beta}       \,.
\eea
To see this, note first that 
 the only block submatrix of the inverse Zamolodchikov metric that survives the limit   (\ref{t-limit})  involves the indices $i,j = \alpha, \beta$, 
\bea   \label{inverseGibarj}
 G^{i\bar j} &\stackrel{(\ref{t-limit})}{=}& G^{\alpha \bar\beta} \,   \delta^i_\alpha  \,   \delta^{\bar j}_{\bar \beta} \,,
\eea
where we recall that $D_{j=\alpha} = \pi^\ast(D_\alpha^{\rm b})$ with $\alpha = 1, \ldots, h^{1,1}(B_3)$.
This can be shown by direct inspection of (\ref{Gdef1}), using (\ref{Kaehlerpot}) and the intersection form on $Y_4$.

As a result the coupling $\overline C_{\bar \tau}^{jb}$ is non-zero only if the index $j$ refers to a pullback divisor  $\pi^\ast(D^{\rm b}_\alpha)$. More precisely, 
  \be   \label{Cbartau1}
 \overline C_{\bar \tau}^{j b} \stackrel{(\ref{t-limit})}{=}      \delta^j_\alpha  \,   \overline{\cF}_{\bar c;\bar{\tau} \bar{\gamma}}   \,  e^{2K}    \, G^{\alpha \bar \gamma} \, G^{b \bar c}     = \delta^j_\alpha   \,  I_{\bar\gamma_\tau \bar c}  \, e^{2K}    \, G^{\alpha \bar \gamma} \, G^{b \bar c}     \,.
  \ee
In the last term we used the notation introduced in
(\ref{basisH22vert-Y4}) to rewrite the three-point function
in terms of the topological pairing (\ref{intrelG-gen}) as
\bea   \label{tripleantihola}
\overline{\cF}_{{\bar c};\bar{\tau}\bar{\gamma}}   = (D_\tau \cdot  \pi^\ast(D_\gamma^{\rm b})  \cdot G_c)^\ast =  (G_{\gamma_\tau}  \cdot  G_c)^\ast = I_{\bar\gamma_\tau   \bar c}       \,.
\eea
In (\ref{Cbartau1}) it is understood that we sum over index $\bar \gamma_\tau$, which is identified with the index $\bar\gamma$.
This notation will be kept also in subsequent equations with a similar structure.

In the next step we 
replace $G^{\alpha \bar \gamma}$ on the right-hand side of  (\ref{Cbartau1}) by the inverse metric for the $(2,2)$ fields. To this end, consider the (0)-flux $G_{\alpha_\tau} = D_\tau \cdot \pi^\ast(D_\alpha^{\rm b})$. 
Due to its factorised structure, the pairing $G_{\alpha_\tau   \bar\gamma_\tau}$ can be written as
\bea \label{Gfactored}
G_{\alpha_\tau   \bar\gamma_\tau} = e^{K}  \langle \bar\gamma_\tau | \alpha_\tau\rangle = e^{2K}   ( \langle \bar\tau | \tau \rangle \langle \bar\gamma | \alpha \rangle  +  \langle \bar\tau | \alpha \rangle \langle \bar\gamma | \tau \rangle ) =  G_{\tau \bar\tau}   G_{\alpha \bar\gamma} +     G_{\tau \bar\gamma}   G_{\alpha \bar\tau}.
\eea
In the scaling limit (\ref{t-limit}), one finds that
\bea   \label{Gbartautau}
G_{\bar \tau j} \stackrel{(\ref{t-limit})}{=} { \frac{1}{(2 \pi)^2}}\frac{1}{4 \tau^2_2}   \delta_{j  \tau}  \,,
\eea
again modulo  irrelevant contributions that are relatively  suppressed by additional powers of the base K\"ahler moduli $v^\alpha$.
Hence only the first term in (\ref{Gfactored}) survives, i.e. 
\bea
G_{\alpha_\tau   \bar\gamma_\tau}   \stackrel{(\ref{t-limit})}{=}  G_{\tau \bar\tau}  G_{\alpha \bar\gamma}  \stackrel{(\ref{t-limit})}{=}   { \frac{1}{(2 \pi)^2}} \frac{1}{4 \tau^2_2} G_{\alpha \bar\gamma}        \,.
\eea
By similar reasoning we have more generally
\be
G_{\alpha_\tau   \bar d}  \stackrel{(\ref{t-limit})}{=}  G_{\alpha_\tau   \bar\gamma_\tau}  \delta_{\bar d}^{\bar \gamma_\tau}   \stackrel{(\ref{t-limit})}{=}    { \frac{1}{(2 \pi)^2}}\frac{1}{4 \tau^2_2} G_{\alpha \bar\gamma} \,  \delta_{\bar d}^{\bar \gamma_\tau}    
\ee
and for the inverse matrix
\bea
G^{\alpha_\tau   \bar d}  \stackrel{(\ref{t-limit})}{=}   { {(2 \pi)^2}} 4 \tau^2_2   \,     G^{\alpha   \bar\gamma}    \,    \delta^{\bar d }_{\bar\gamma_\tau} \,.
\eea

With this input, we can trade $G^{\alpha   \bar\gamma}$ against the inverse metric $G^{\alpha_\tau   \bar d}$  in (\ref{Cbartau1}), which becomes
\bea   \label{Cbartau2}
\overline C_{\bar \tau}^{j b}   \stackrel{(\ref{t-limit})}{=}   { \frac{1}{(2 \pi)^2}} \frac{1}{4 \tau^2_2} \,  \delta^j_\alpha   \,   I_{\bar d \bar c}  \, e^{2K}    \, G^{\alpha_\tau \bar d} \, G^{b \bar c}  =   { \frac{1}{(2 \pi)^2}} \frac{1}{4 \tau^2_2} \,  \delta^j_\alpha  I^{\alpha_\tau b}  \,. 
\eea
In the last step we made use of the general identity
\bea
I_{\bar d \bar c} \,  (e^{K} G^{a \bar d})  \, (e^K G^{b \bar c})  =    I^{a b}   \, ,
\eea
which follows from the definitions   (\ref{Gibarjdef}) and (\ref{toppairingdef})  
of the quantities in the underyling topological field theory.
To see this, consider instead the inverted equation,
\bea   \label{IGGgen}
I^{\bar a \bar b} (e^{-K} G_{c \bar a}) (e^{-K} G_{d \bar b})  =     \langle \bar a| c  \rangle   I^{\bar a \bar b}   \langle \bar b |  d  \rangle   = \langle d | \bar b  \rangle    I^{\bar b \bar a} \langle \bar a| c  \rangle = \langle d | \unit | c  \rangle  = I_{dc}  = I_{cd}\,,
\eea 
where we used the reality of the metric, $G_{d \bar b} = G_{b \bar d}$.

Note that the pairing $I^{ab}$ ties together fluxes whose combined associated modular weight totals $-2$.
This is a consequence of the factorization of the four-point function
(\ref{fourptfact}) and the fact that the latter can be assigned \cite{Haghighat:2015qdq,Cota:2017aal} a modular weight $w=-2$.
In the present context this implies that
$I^{\alpha_\tau b}$ in  (\ref{Cbartau2}) can be non-zero only if $b$ refers to the a $(-2)$-flux index $\dot \beta$. In this case we can evaluate the pairing entirely on the base $B_3$,
\bea
I^{\alpha_\tau b} = \delta^b_{\dot\beta}    \,   I^{\alpha_\tau \dot\beta} =  \delta^b_{\dot\beta}  \,    I^{\alpha \dot\beta}        \,,   
\eea
where $ I^{\alpha \dot\beta} $ is the inverse of the intersection pairing $I_{\alpha \dot\beta} = D_\alpha^{\rm b} \cdot_{B_3} \Sigma_{\dot\beta}^{\rm b}$.
This concludes our derivation of (\ref{Cbaraufinal-App}).

\subsection{Proof of Equation (\ref{commrelexp})} \label{Appproofz}

In this appendix we derive the expression (\ref{commrelexp}) for the commutator (\ref{antiholcomm1}). This derivation leads to the elliptic anomaly equation
as discussed in Section \ref{sec_modularanom}.

To compute the first term of (\ref{antiholcomm1}), we go back to the original expression (\ref{BCOV4}), but  focus only on the index $i_1$ with
 $t^{i_1} = z$.    
Upon reproducing the normalisation factors analogous to~\eqref{Fanorm}, we are lead to
\be   \label{bartauzF}
\begin{split}
-\frac{1}{(2 \pi i)^2}\bar \partial_{\bar \tau} (\partial_z   \cF_{a|C_{\beta}})  &\stackrel{(\ref{t-limit})}{=} {\overline C_{\bar \tau}}^{\alpha \dot\gamma}  (\!\sum\limits_{\substack{ C_{\beta_1} + C_{\beta_2}\\  =  C_\beta }}
  \cF_{a; \alpha z |C_{\beta_1}}  \cF_{\dot\gamma |C_{\beta_2}} +  \cF_{a; \alpha |C_{\beta_1}}  \cF_{\dot\gamma;z |C_{\beta_2}} )    \cr
 &- G_{\bar \tau z }   \cF_{a|C_{\beta}}    -  {\overline C_{\bar \tau}}^{\alpha \dot\gamma}  I_{\dot\gamma a}   \cF_{z \alpha{|C_{\beta}}}   \,,
\end{split}
\ee
where we furthermore used our result from Section~\ref{FromBCOV} concerning the specific form~\eqref{Cbaraufinal} of the overall coupling, i.e.,
 ${\overline C_{\bar \tau}^{j b} = \delta^j_\alpha \delta^b_{\dot\gamma} \overline C_{\bar \tau}}^{\alpha \dot\gamma}$.
From this we subtract 
\be \label{zbartauF}
\begin{split}
-\frac{1}{(2\pi i)^2}\partial_z (\bar \partial_{\bar \tau} \cF_{a|C_{\beta}}) \stackrel{(\ref{t-limit})}{=} {\overline C_{\bar \tau}}^{\alpha \dot\beta}(\partial_z \sum\limits_{\substack{ C_{\beta_1} + C_{\beta_2}\\  =  C_\beta }}
  \cF_{a; \alpha |C_{\beta_1}}  \cF_{\dot\beta |C_{\beta_2}}
-   I_{a\dot\beta}  \partial_z   (\psi \cdot \langle \langle  \pi^\ast(D_\alpha) \rangle\rangle_{C_\beta}) )  \,,
\end{split}
\ee
where the asymptotic equality follows from~\eqref{HAEFaC-1}.

In performing the subtraction, all but three terms cancel.
The first obvious candidate contribution is the term $- G_{\bar \tau z }   \cF_a$. However, in the limit (\ref{t-limit})  in which we are working, the Zamolodchikov metric $G_{\bar \tau z } \sim \frac{{\rm Im}(z)}{\tau_2^2} \frac{1}{{\rm Im}(t)}$ contributes a term which vanishes as ${\rm Im}(t) \to \infty$.

The second candidate arises from the quadratic piece associated with the sum over all splittings, $ C_{\beta_1} + C_{\beta_2}  =  C_\beta$. A priori, this sum includes, as special cases, the pairs $(C_{\beta_1}, C_{\beta_2}) =(C_\beta,0)$ and 
$(C_{\beta_1}, C_{\beta_2}) =(0,C_\beta)$.  Indeed, (\ref{bartauzF}) receives a contribution from the splitting $(C_{\beta_1}, C_{\beta_2}) =(0,C_\beta)$ of the form
\bea
{\overline C_{\bar \tau}}^{\alpha \dot\gamma} \cF_{a; \alpha z |C_{\beta_1}=0}  \cF_{\dot\gamma |C_{\beta_2}=C_{\beta}}  = {\overline C_{\bar \tau}}^{\alpha \dot\gamma}  (G_a \cdot D_\alpha \cdot D_z) \cF_{\dot\gamma |C_{\beta}} =  {\overline C_{\bar \tau}}^{\alpha \dot\gamma} C_{a \alpha z}  \cF_{\dot\gamma |C_{\beta}}   \,.
\eea
The corresponding split does not, however, contribute in (\ref{zbartauF}) because prior to taking the $z$-derivative, both factors involve only two insertions each and hence the term with $C_{\beta_1}=0$ (or $C_{\beta_2}=0$) vanishes.

The only other term which does not cancel originates in the descendant invariant  $-  {\overline C_{\bar \tau}}^{\alpha \dot\gamma}  I_{\dot\gamma a}   \cF_{z \alpha} $  in  (\ref{bartauzF}). 
As implied by (\ref{HHpsi-1}), its evaluation likewise involves a sum of all possible splits of the curve class $C_\beta$. 
Due to the presence of the additional divisor $D_z$ in  $\cF_{z\alpha}$, (\ref{HHpsi-1}) contains a term 
\be
 \langle  D_\alpha, D_z,  G_c  \rangle_{C_{\beta_1}=0}   I^{cd}   \langle G_d, H, H \rangle_{C_{\beta_2}=C_\beta}   \,,
\ee
which has no analogue in (\ref{zbartauF}).
This gives rise to a term 
\be
-  {\overline C_{\bar \tau}}^{ \alpha  \dot\gamma}  I_{\dot\gamma a}   (G_{\alpha_{z}} \cdot G_c) I^{cd} \cF_{d|C_\beta} = -  {\overline C_{\bar \tau}}^{ \alpha \dot\gamma}  I_{a \dot\gamma}   \langle \langle  G_{\alpha_{z}}  \rangle \rangle_{C_\beta}    \,,
\ee
where we  expressed $G_{\alpha_{z}} = D_z \cdot  \pi^\ast(D_\alpha)$ and used the fact that $I^{cd}$ is the inverse of the intersection pairing $I_{ab} = G_a \cdot G_b$.

All in all we therefore find 
\be \begin{split}   \label{ElHAE1}
-\frac{1}{(2\pi i)^2} \left( \bar \partial_{\bar \tau} (\partial_z   \cF_{a|C_{\beta}})  -  \partial_z (\bar \partial_{\bar \tau} \cF_{a|C_{\beta}}) \right)  &\stackrel{(\ref{t-limit})}{=} {\overline C_{\bar \tau}}^{\alpha \dot\gamma} C_{a \alpha z}  \cF_{\dot\gamma |C_{\beta}}  -   {\overline C_{\bar \tau}}^{ \alpha \dot\gamma}  I_{\dot\gamma a}   \langle \langle  G_{\alpha_z}  \rangle \rangle_{C_\beta}   \cr
&=  \frac{1}{(2\pi)^2}\frac{1}{4 \tau_2^2}   I^{\alpha \dot\gamma} (C_{a \alpha z}  \cF_{\dot\gamma |C_{\beta}} - I_{\dot\gamma a}   \langle \langle  G_{\alpha_{z}}  \rangle \rangle_{C_\beta} ) \,,
\end{split}
\ee
where~\eqref{Cbaraufinal} was used after the second equality. 
To evaluate this further, note that the topological intersection number in the first term,
\be
 C_{a \alpha z} = G_a \cdot D_z \cdot \pi^\ast(D_\alpha^{\rm b}) = G_a \cdot G_{\alpha_{z}}  \,,
\ee
is non-vanishing only if $G_a$ refers to a $(-1)$-flux, $G_a = G_{\rho_z} =  D_z \cdot \pi^\ast(D_\rho^{\rm b}) $,
for some base divisor $D_\rho^{\rm b}$. In other words
\bea
C_{a \alpha z}=  \begin{cases}   - b \cdot_{B_3} D_\rho^{\rm b}   \cdot_{B_3} D_\alpha^{\rm b}  \,,     &   \text{if} \quad  G_a =  D_z \cdot \pi^\ast(D_\rho^{\rm b})     \\
 0 &    \text{otherwise} \,,
\end{cases}
\eea
where $b$ is the height pairing associated with the extra section.
Contracting this with $ I^{\alpha \dot\gamma}$
identifies the $(-2)$-flux $G_{\dot\gamma}$ appearing in  $\cF_{\dot\gamma |C_{\beta}}$ as 
$G_{\dot\gamma} =- \pi^\ast(b) \cdot  \pi^\ast(D_\rho^{\rm b}) = \pi^\ast  \pi_\ast( \sigma \cdot G_a)$. 
Note that $ \pi_\ast( \sigma \cdot G_a)=0$ whenever $G_a$ refers to a $(0)$-flux or $(-2)$-flux. Hence in using the above compact notation it is automatically encoded that 
this contribution is present only when $G_a$ is a $(-1)$-flux.

Similarly, the intersection form $I_{\dot\gamma a}$ in the second term in (\ref{ElHAE1}) is non-zero only if $G_a$ refers to a $(0)$-flux, $G_{\rho_\tau} = D_\tau \cdot \pi^\ast(D_\rho^{\rm b})$. 
Contraction with $I^{\alpha\dot\gamma }$ then requires that $\alpha = \rho$. The flux $G_{\alpha_z}$ appearing in the gravitational descendant invariant can therefore be compactly be expressed as
$D_z \cdot \pi^\ast(D_\alpha^{\rm b}) = D_z \cdot \pi^\ast \pi_\ast(G_a)$, because $\pi^*(G_a)=0$ unless $G_a$ is a $(0)$-flux.
Thus altogether we have derived the relation
\bea
\boxed{
\bar \partial_{\bar \tau} (\partial_z   \cF_{a|C_\beta})  -   \partial_z (\bar \partial_{\bar \tau} \cF_{a|C_\beta})  \stackrel{(\ref{t-limit})}{=}   \frac{1}{4 \tau_2^2}    \left(  \langle\langle  \pi^\ast \pi_\ast( D_z \cdot G_a)   \rangle\rangle_{C_\beta}     -    \langle\langle     D_z \cdot \pi^\ast \pi_\ast(G_a) \rangle\rangle_{C_\beta} \right) \,.  }  
\eea

 \section{Jacobi and Quasi-Jacobi Forms} \label{app_jacobi}

There exists extensive literature about Jacobi forms, so we can be brief.  See for example, besides the classic books~\cite{weil1976elliptic,EichlerZagier},  also the works in physics  
  \cite{Kawai:1998md,Dabholkar:2012nd,gritsenko2018graded}.  
  We just mention here some aspects that are important for the present work. In essence, Jacobi forms are holomorphic functions of two variables, $\Phi(\tau,z):  \IH\times\IC\rightarrow \IC$, which are  characterized by their transformation properties under the modular group and ``elliptic'' (double periodic shift) symmetries:
\bea\label{Jacmodular}
\Phi_{w,m}  \left(\frac{a \tau + b}{c \tau +d}, \frac{z}{c \tau +d} \right) &=& (c \tau+d)^w e^{2\pi  i  \frac{m c}{c\tau +d}  z^2}    \Phi_{w,m}(\tau,z)\ \ \,{\rm for}    
 \left(\begin{matrix} 
      a & b \\
      c & d \\
   \end{matrix}\right) \in SL(2,\IZ),
\\
\Phi_{w,m}\left( \tau , z + \lambda \tau + \mu \right) &=& e^{-2 \pi i   m (\lambda^2 \tau  + 2  \lambda  z )  }   \Phi_{w, m} (\tau,  z)\,,
\quad \lambda, \mu \in \mathbb Z \,.\label{periodicity}
\eea
The labels indicate modular weight $w\in \IZ$ and index $m\in\IZ_{\geq0}$.  
Moreover, Jacobi forms possess a Fourier expansion 
\be
\Phi_{w,m}\ =\ \sum_{n\geq0}\sum_{r^2\leq4mn}c(n,r)\,e^{2 \pi i(n\tau+ rz)}\,,
\ee
and as such are natural building blocks \cite{Kawai:1993jk,Gritsenko:1999fk,Gritsenko:1999nm} of elliptic genera or partition functions that are refined by an extra $U(1)$ current. 

\vskip2mm
A Jacobi form  $\Phi_{w, m}(\tau, { {z}})$ is called 
\begin{itemize}
\item a holomorphic Jacobi form if $c(n,r) = 0$ unless $4 {m} n \geq r^2$,
\item a Jacobi cusp form  if $c(n,r) = 0$ unless $4 {m} n > r^2$,
\item a weak Jacobi form if $c(n,r) = 0$ unless $ n \geq 0$ \,.
\end{itemize}

Jacobi forms form a bi-graded ring which we denote by
\be
\cR^J\ = \oplus_{w,m}\cR^J_{w,m}\,,
\ee
which is polynomially generated by
\be\label{RJdef}
\cR^J\\ =\ \IQ\big[E_4,E_6,\phi_{0,1},\phi_{-2,1},\phi_{-1,2}\big],
\ee
modulo the relation
$
\phi_{-1,2}^2 =\frac1{432} {\phi_{-2,1}}\left( \phi_{0,1}^3-3 E_4  \phi_{-2,1}^2\phi_{0,1}+2 E_6  \phi_{-2,1}^3\right).
$
Above, $E_4$ and $E_6$ are the familiar, simplest examples of Eisenstein series which in general are defined by
($B_{2k}$ denotes the Bernoulli numbers):
\bea\label{app:E2kdef}
E_{2k}(\tau)&=&2^{-1}\zeta(2k)^{-1}\sum_{(m,n)\not=(0,0)}(m+n\tau)^{-2k}
\\
&=& 1-\frac{4k}{B_{2k}}\sum_{k,r\geq1}^\infty r^{2k-1}q^{kr}\,.\nn
\eea
Moreover the Jacobi generators can be written in terms of theta-functions as follows:
\bea\label{defJacobi}
&& \phi_{-2,1}(\tau, z) = -\frac{\vartheta_1(\tau,z)^2}{\eta^6(\tau)} =
\zh^2+\frac1{12}E_2\zh^4+...\,,\label{phim21}\nn
\\
&& \phi_{-1,2}(\tau, z) = \frac{i\vartheta_1(\tau,2z)}{\eta^3(\tau)} =2\zh+\frac13E_2\zh^3+...\,,
\\
&& \phi_{0,1}(\tau, z) = 4 \left(  \frac{\vartheta_2(\tau,z)^2}{\vartheta_2(\tau,0)^2}  + \frac{\vartheta_3(\tau,z)^2}{\vartheta_3(\tau,0)^2}  + \frac{\vartheta_4(\tau,z)^2}{\vartheta_4(\tau,0)^2}  \right)=12+E_2\zh^2+....\,,\nn
\eea
where $\zh\equiv2\pi i z$.
Special cases of Jacobi forms are refined versions of the Eisenstein series, which map back to the Eisenstein series upon setting $z\rightarrow0$. In the present work we will encounter
\begin{align}
\label{Eisenjacobi}
E_{4,1}& =\frac1{12}\left({E_4\phi_{0,1}-E_6\phi_{-2,1}}\right)\,,\nn
\\
E_{4,2}& =\frac1{12^2}\left({E_4\phi_{0,1}^2-2E_6\phi_{0,1}\phi_{-2,1}+E_4^2\phi_{-2,1}^2}\right)\,,
\\
E_{6,1}& =\frac1{12}\left({E_6\phi_{0,1}-E_4^2\phi_{-2,1}}\right),\nn
\\
E_{6,2}& =\frac1{12^2}\left({E_6\phi_{0,1}^2\!-\!2E_4^2\phi_{0,1}\phi_{-2,1}\!+\!E_4E_6\phi_{-2,1}^2}\right)\,.\nn
\end{align}

In our context of flux backgounds and holomorphic anomaly equations,
derivatives $\frac1{2\pi i}\partial_\tau\equiv q\partial_q$ and 
 $\frac1{2\pi i}\partial_z\equiv \xi\partial_\xi$  acting on $\cR^{J}$ play an important role. However these map outside of $\cR^{J}$, and
this is why we need to extend the space of functions to {\it quasi-Jacobi} forms, as well as  their almost holomorphic variants.
 
Resting on earlier ideas, see for example \cite{KanekoZagier,Gritsenko:1999nm,libgober2009elliptic}, quasi-Jacobi forms have been more recently discussed in \cite{oberdieck2012serre,Oberdieck_2018,Oberdieck:2017pqm},  whose approach we briefly summarize; we refer to these references for a more rigorous treatment.

The important point is that in order to capture derivatives, one needs to go to meromorphic (in $z$) Jacobi forms.  This is already evident from the expressions (\ref{xiaction}) and (\ref{qaction}) given in the text. 
A systematic approach can be given as follows \cite{oberdieck2012serre}. First, define the twisted Eisenstein series
\be
J_n(\tau,z):=\delta_{1,n}\frac\xi{1-\xi}+ B_n-n \sum_{k,r\geq1}r^{n-1}(\xi^k+(-1)^n\xi^{-k})\,q^{kr}\,, \ \ n\geq1,
\ee
which are different refinements of the Eisenstein series than the 
Eisenstein-Jacobi series  in (\ref{Eisenjacobi}); rather they coincide up to normalization with the expansions given in \cite{Gaberdiel:2009vs}.
Upon specialization to $z=0$ we have: $J_{2k}(0,\tau)=B_{2k}E_{2k}(\tau)$, $J_{2k+1}(0,\tau)=0$ ($k\geq1$).
The first instance with $n=1$ coincides with
the meromorphic quasi-Jacobi form of weight $w=1$ and index $m=0$
that we have introduced in Section~\ref{sec_modular}:
\be\label{E1expan}
J_1(\tau,z) \ \equiv\ E_1(\tau,z) \ =\ \frac1{2\pi i}\partial_z \log \theta_1(z,\tau)=\frac 1\zh+\frac 1{12}E_2 \zh+...\,.
\ee
It obeys the anomalous transformation laws  given in eq.~(\ref{E1trans}).
On the other hand,
the general $J_n$ do not transform nicely, which is why one introduces the 
better-behaved objects,
 \be
 K_n(\tau,z):= \sum_{k=0}^n(-1)^{n+k}\left( n\atop k\right)J_k J_1^{n-k}\,,   \ \ n\geq2\,.
 \ee
 These transform under modular (\ref{Jacmodular}) and elliptic (\ref{periodicity}) transformations
 as Jacobi forms with weight $w=n$ and index $m=0$.  The price to pay is that  $J_n$
 are meromorphic with poles up to order~$1/z^n$. 
 Upon a change of basis, they can also be witten in terms of the  $(n-2)$-th derivatives of the Weierstrass function.
  
 The point is now that the ring generated by the meromorphic  {\it quasi-Jacobi forms}
 of index $m=0$,
 \be
\cR^{QJ}_{*,0}\ =\ \IQ\big[E_1,E_2, K_n\big]\,,
\ee
is closed under taking arbitrary derivatives with respect to both $z$ and $\tau$.
For example,
\bea
\xi\partial_\xi E_1&=& K_2+\frac1{12}E_2\,,    \qquad  (\xi\partial_\xi)^2 E_1=  \xi\partial_\xi K_2 = K_3\,,
\nn
\\
q \partial_q E_1 &=& \frac12K_3+ E_1 K_2+\frac1{12}E_1E_2
\,,    \qquad  \xi\partial_\xi  K_3= \frac56K_4-\frac72 K_2^2\,. 
\nn
\eea

Therefore, given some (quasi-)Jacobi form $\Phi^{QJ}_{w,m}$ with given weight and index, one can 
determine the action of arbitrary derivatives on it by first transforming to a meromorphic quasi-Jacobi form 
of index $m=0$,  by first ``dividing out the index'', ie.,
\be
\Phi^{QJ}_{w,m} \ \longrightarrow \Phi^{QJ}_{w,m}/\phi_{-2,1}^m\ \in \cR^{MQJ}_{w+2m,0}\,.
 \ee
 Then one can act with arbitrary
 derivatives with respect to both $z$ and $\tau$, which, as said above,
 stays within $\cR^{QJ}_{*,0}$.
 After doing so, one can map back to a quasi-Jacobi form of the desired weight and index
 by multiplication with $\phi_{-2,1}^m$.
One may then express the result in terms 
of the standard Jacobi generators, using relations such as
\bea
K_2 &=& -\frac1{12}\frac{\phi_{0,1}}{\phi_{-2,1}}\,,\ \ \
K_3= \frac{\phi_{-1,2}}{\phi_{-2,1}^2}\,, \ \ \
K_4=20 E_4-3 K_2^2\,,
\\
K_5&=&-2K_2K_3 \,, \ \ \
K_6=9K_2^3+K_3^2-\frac1{56}E_6\,, 
\qquad{\rm etc.}\nn
\eea
Thus the result lies in the following ring of generators, modulo appropriate
divisions by powers of $\phi_{-1,2}$ and $ \phi_{-2,1}$:
\be\label{RQJdef}
\cR^{QJ}\ =\ \IQ\big[E_1, E_2, E_4,E_6,\phi_{-2,1},\phi_{-1,2},\phi_{0,1}\big]
\big/\big\{\phi_{-1,2},\phi_{-2,1}\big\}\,.
\ee
In our context the poles in powers of $1/z$ cancel so that the final result, while quasi-Jacobi, is holomorphic in $z$ after all.
This happens in particular for the  $z$ and $\tau$ derivatives of a general weak Jacobi form, which were given in eqs.~(\ref{xiaction}) and (\ref{qaction}).

For holomorphic anomaly equations also
mildly anholomorphic variants of $\Phi^{QJ}$ are important, which transform
as standard Jacobi forms under modular (\ref{Jacmodular}) and
elliptic (\ref{periodicity}) transformations. By definition, any such
{\it almost holomorphic (or almost meromorphic) Jacobi form} $\Phi^{AHJ}\in \cR^{AHJ}$ has the expansion
\be\label{almostholo}
\Phi^{AHJ}(\tau,z)\ =\ \sum_{i,j\geq0}^{d_\nu,d_\alpha} \Phi^{(i,j)}(\tau,z)\,\nu^i\alpha^j
\,,\qquad\ \ \ \nu\equiv\frac1{8\pi {\rm Im}\tau}\,,\ \ \alpha=\frac{{\rm Im}z}{{\rm Im}\tau}\,,
\ee
where the sum runs over finitely many terms and the $\Phi^{(i,j)}(\tau,z)$ are holo-  resp.~meromorphic and appropriately convergent. The maximal powers are called the depths of the almost holomorphic Jacobi form.

The expansion (\ref{almostholo}) is actually what defines quasi-Jacobi forms 
in the first place:
If, which is what we assume, the non-holomorphic function $\Phi^{AHJ}(\tau,z)$ obeys the transformation laws of a Jacobi form as given in (\ref{Jacmodular}) and 
(\ref{periodicity}), then by definition the holomorphic or meromorphic first term in the expansion is
a quasi-Jacobi form, $\Phi^{(0,0)}\in \cR^{QJ}$. From this
point of view, the remainder of the sum then provides its modular completion.\footnote{We do not consider more general, mock modular Jacobi forms \cite{s2008mock,Dabholkar:2012nd} 
whose modular completion has a much more complicated structure,
because these do not appear in our computations where we take $\bar {t}^i   \rightarrow \infty$.
}

In our context, quasi-Jacobi forms are produced by derivatives
and can be expressed in terms of the generators in (\ref{RQJdef}) in a simple way.
Their modularly completed, almost holomorphic versions are simply obtained
by substituting $E_1\rightarrow \hat E_1=E_1+\alpha$ 
and $E_2\rightarrow \hat E_2=E_2-24\nu$ for the generators in $\cR^{QJ}$.
This is what we indicated in eq.~(\ref{RAHJdef}) in the main text.

\section{Explicit Flux Partition Functions for $B_3 = dP_2 \times \mathbb P^1_{l'}$}\label{expertf}

Here we collect explicit expressions for the partition functions in the various flux sectors
for our example,  both for the emerging heterotic as well as for the non-critical E-strings. To facilitate translation to geometry, 
we remind the reader of the  basis of fluxes as in given Table~\ref{tab:Fluxbasis}.

\subsection{Heterotic String from Curve $C=C_0$}   \label{app_PartfunC0}

Recall from Section ~\ref{SecExS} the  definition of the following building blocks:
\bea
\label{Zwmdefs-A}
Z_{-2,2}^{{1}}(q,\xi) &=&  \frac1{12\et24}(14E_4E_{6,2}+10 E_{4,2}E_6),\nn \\
Z_{-2,2}^{{2}} (q,\xi) &=&  Z_{-2,2}^{{1}}+\frac1{12\et24}E_{4,1}(E_2E_{4,1}- E_{6,1}),\\
Z_{-1,2}^0 (q,\xi) &=& 84\, \phi_{-1,2},\nn\\
Z_{0,2}^0(q,\xi)  &=& \frac{-137 E_4^2E_{4,2}+120 E_4 E_{4,1}^2-169 E_6 E_{6,2}+ 4 E_2(37 E_{4,1} E_{6,1}+ 8 E_{6,2}E_4)+6 E_2^2E_{4,1}^2}{2\cdot 12^2\,\eta^{24}},\nn
\eea
in terms of which partition functions for the $(-2)$-fluxes read:
\bea
\label{Zm2defs-A}
{\cal Z}_{-2,2}[G_{\dot{(-1)}},C_0]&=&  0\nn \\ 
{\cal Z}_{-2,2}[G_{\dot0 },C_0]&=&  0 \\ 
{\cal Z}_{-2,2}[G_{\dot 1},C_0]&=& Z_{-2,2}^{{1}}\nn \\  
{\cal Z}_{-2,2}[G_{\dot 2},C_0]&=& Z_{-2,2}^{{2}}\nn \,.  
\eea
For the $(-1)$-fluxes we have
\bea
\label{Zm1defs-A}
{\cal Z}_{-1,2}[G_{{(-1)_z}},C_0]&=&  \xi \partial_\xi(\frac12Z_{-2,2}^{{1}}+  Z_{-2,2}^{{2}})+Z_{-1,2}^0  \nn \\ 
{\cal Z}_{-1,2}[G_{0 _z},C_0]&=&   \xi \partial_\xi(\frac{1}{2}Z_{-2,2}^{{1}})\\ 
{\cal Z}_{-1,2}[G_{1_z},C_0]&=& \xi \partial_\xi(Z_{-2,2}^{{1}}) \nn\\  
{\cal Z}_{-1,2}[G_{2_z},C_0]&=&\xi \partial_\xi(Z_{-2,2}^{{2}}) 
 \,,  \nn
\eea
and the $(0)$-fluxes lead to
\bea
\label{Z0defs-A}
{\cal Z}_{0,2}[G_{{(-1)_\tau}},C_0]&=&  q\partial_q(\frac{1}{2}Z_{-2,2}^{{1}}+Z_{-2,2}^{{2}})+\xi\partial_\xi(\frac{1}{4}Z_{-1,2}^0)+Z^0_{0,2}  \nn \\ 
{\cal Z}_{0,2}[G_{0_\tau},C_0]&=&  q\partial_q(\frac{1}{2}Z_{-2,2}^{{1}}) \\ 
{\cal Z}_{0,2}[G_{1_\tau},C_0]&=& q\partial_q(Z_{-2,2}^{{1}})\nn\\  
{\cal Z}_{0,2}[G_{2_\tau},C_0]&=& q\partial_q(Z_{-2,2}^{{2}}) \,.  \nn 
\eea
As pointed out before, not all weight $w=-1,0$ partition functions are given by derivatives.

We now rewrite the partition functions in terms of quasi-Jacobi forms, which will then allow us
to determine the anomaly equations by taking derivatives with respect to $E_1$ and $E_2$.

For the $(-2)$-fluxes this is already accomplished
by eqs.~(\ref{Zm2defs-A}) and
(\ref{Zwmdefs-A}). The noteworthy feature is that  
$Z_{-2,2}^{{2}}$ is only quasimodular and differs from $Z_{-2,2}^{{1}}$ by a piece proportional to
$(E_2 E_{4,1}-E_{6,1})$. This expresses that the flux sectors 
$G_{\dot 1}$ and $G_{\dot 2}$ differ by what corresponds,
in heterotic language, to a non-perturbative transition where a small instanton is traded against
a heterotic NS5-brane.
We will see this feature propagating to the other flux sectors, 
$G_{1_*}$ and $G_{2_*}$, as well, and in order to emphasize this,
we will separate out terms of this form below. In this way we can distinguish contributions to the
modular anomaly arising from this transition from contributions to $E_2$ that arise from $q$-derivatives.

Concretely, for the $(-1)$-flux sectors we can write the partition functions alternatively
 in terms of meromorphic quasi-Jacobi forms as follows:
\bea
\label{Zm1jac}
{\cal Z}_{-1,2}[G_{{(-1)_z}},C_0]&=& 
4 E_1(\frac12 Z_{-2,2}^{{1}}+Z_{-2,2}^{{2}})- 6\oet24 \frac{\phi_{-1,2}}{\phi_{-2,1}}{E_{4,1}E_6} +\frac{48}7  Z_{-1,2}^0\\   
&&\ \ -\frac1{6\et24} \frac{\phi_{-1,2}}{\phi_{-2,1}}E_{4,1} (E_2E_{4}-E_{6}) \nn\\
 \nn \\ 
{\cal Z}_{-1,2}[G_{0 _z},C_0]&=&  
\frac12 {\cal Z}_{-1,2}[G_{1_z},C_0]\\   
{\cal Z}_{-1,2}[G_{1_z},C_0]&=& \ 4 E_1 Z_{-2,2}^{{1}}
 -4\oet24 \frac{\phi_{-1,2}}{\phi_{-2,1}}{E_{4,1}E_6}  +4 Z_{-1,2}^0\nn\\  
{\cal Z}_{-1,2}[G_{2_z},C_0]&=& \
{\cal Z}_{-1,2}[G_{1_z},C_0]+\frac13 \frac{1}{\eta^{24}} E_1 E_{4,1}(E_2 E_{4,1}-E_{6,1})
-\frac17 Z_{-1,2}^0\nn\\
&&\ \ -\frac1{6\et24}\frac{\phi_{-1,2}}{\phi_{-2,1}}E_{4,1} (E_2E_{4}-E_{6})\,.
\eea
Note that these partition functions are actually holomorphic in $z$, as any poles in $z$ cancel out. The same is true for the modular weight $w=0$ partition functions:
\bea\label{14-11FluxH}
{\cal Z}_{0,2}[G_{{(-1)_\tau}},C_0]   &=& 
-{\cal Z}_{0,2}[G_{{1_\tau}},C_0] 
 -\Big(2E_1^2+\frac14E_2\Big)\Big( Z_{-2,2}^{{2}} +\frac32 Z_{-2,2}^{{1}}   \Big)\nn\\
&& + E_1\Big(   Z_{-1,2}^0  +\mathcal{Z}_{-1,2}[G_{2_z},C^0]  +\frac32 \mathcal{Z}_{-1,2}[G_{1_z},C^0]   \Big)
 \nn\\
&&+ \frac1{12^4\et24}E_2\Big(141 E_4 E_6 \phi_{0,1}^2+145 E_4^2 E_6  \phi_{-2,1}^2 
+2E_4^2 \frac{\phi_{0,1}^3}{\phi_{-2,1}}
\\
&& -9(19E_4^3+13 E_6^2)\phi_{-2,1}\phi_{0,1}\Big)
+\frac1{48\et24}E_2^2E_{4,1}^2
+\frac{118}{12^4\et24}\frac{\phi_{0,1}^3}{\phi_{-2,1}}E_4E_6
\nn\\
&&-\frac2{ 12^4\et24}\Big(\!(187E_4^3+155E_6^2)\phi_{0,1}^2\!+(68E_4^4+156E_4E_6^2) \phi_{-2,1}^2\!
-507 E_4^2E_6 \phi_{-2,1}\phi_{0,1}\!\Big)\nn
\eea

\bea\label{-13FluxH}
{\cal Z}_{0,2}[G_{{0_\tau}},C_0]   &=&  \frac12 {\cal Z}_{0,2}[G_{{1_\tau}},C_0]
\qquad\qquad\qquad\qquad\qquad
\qquad\qquad\qquad\qquad\qquad\qquad\qquad
\eea

\bea\label{11FluxH}
{\cal Z}_{0,2}[G_{{1_\tau}},C_0] 
&=&-\big(2E_1^2+\frac16E_2\big)Z_{-2,2}^{{1}}+ E_1 {\cal Z}_{-1,2}[G_{1_z},C_0]+\frac1{432\et24}\frac{ \phi_{0,1}^3}{\phi_{-2,1}}{E_4E_6}
\\
&&  +\frac1{1296\et24}\left(21E_4^2E_6 \phi_{0,1}\phi_{-2,1}
- (2E_4^4+7 E_4E_6^2)\phi_{-2,1}^2
-(9E_4^3+6 E_6^2)\phi_{0,1}^2\right)\nn
\eea

\bea\label{012Flux}
{\cal Z}_{0,2}[G_{{2_\tau}},C_0]
&=& {\cal Z}_{0,2}[G_{{1_\tau}},C_0] -\frac1{12\et24}\Big(\big(\frac14E_2-2 E_1^2\big)E_{4,1}+\frac{E_{4,2}}{\phi_{-2,1}}\Big)(E_2 E_{4,1}-E_{6,1})
\nn\\
&&-\frac1{12\et24}\frac{\phi_{-1,2}}{\phi_{-2,1}}E_1\big((E_2E_4-E_6)E_{4,1}+(E_2E_{4,1}-E_{6,1})E_4\big)
\\
&&-\frac1{48\et24}(E_2E_{4,1}-E_{6,1})E_{6,1}+\frac1{4\et24\phi_{-2,1} }(E_2E_{4,2}-E_{6,2})E_{4,1}\nn\,.\qquad\qquad\ \ \
\eea
As advertised, we see that ${\cal Z}_{0,2}[G_{{1_\tau}},C_0]$ and 
${\cal Z}_{0,2}[G_{{2_\tau}},C_0]$ differ by terms reflecting an instanton/NS5-brane transition.

\subsection {{E-strings from Curves $C=C_E^{1,2}$}}   \label{App-Estrings}

We now repeat the same exercise for the non-critical E-strings which arise from the curves~$C_E^{1,2}$. Let us first define the following modular and quasi-modular Jacobi forms:
\bea
Z_{-2,1}^0 &=& -\frac{E_{4,1}}{\eta^{12}},\\
Z_{0,1}^0 &=&  \frac{E_{6,1}-E_2 E_{4,1}}{3\,\eta^{12}}.
\eea
The second one signifies a small instanton transition as before.
For the weight $w=-2$ partition functions we find:
\be
\label{-2FluxesE}
\begin{split}
\mathcal{Z}_{-2,1}[G_{\dot{(-1)}},C_E^1]&=  0,
\\
\mathcal{Z}_{-2,1}[G_{\dot{(-1)}},C_E^2]&=  0,
\\
\mathcal{Z}_{-2,1}[G_{\dot 0},C_E^1]&= Z_{-2,1}^0,
\\
\mathcal{Z}_{-2,1}[G_{\dot 0},C_E^2]&= - Z_{-2,1}^0,
\\
\mathcal{Z}_{-2,1}[G_{\dot 1},C_E^1]&=   0,
\\
\mathcal{Z}_{-2,1}[G_{\dot 1},C_E^2]&=   0,
\\
\mathcal{Z}_{-2,1}[G_{\dot 2},C_E^1] &= Z_{-2,1}^0,
\\
\mathcal{Z}_{-2,1}[G_{\dot 2},C_E^2] &= 0\,,\\
\end{split}
\ee
while for the weight $w=-1$ partition functions we have:
\be
\label{-1FluxesE}
\begin{split}
\mathcal{Z}_{-1,1}[G_{(-1)_z},C_E^1] &=2\xi\partial_\xi \mathcal{Z}_{-1,1}[G_{\dot 2},C_E^1],\\
\mathcal{Z}_{-1,1}[G_{(-1)_z},C_E^2]&=0,
\\
\mathcal{Z}_{-1,1}[G_{0_z},C_E^1]&=2\xi\partial_\xi \mathcal{Z}_{-1,1}[G_{\dot 0},C_E^1],
\\
\mathcal{Z}_{-1,1}[G_{0_z},C_E^2]&=2\xi\partial_\xi \mathcal{Z}_{-1,1}[G_{\dot 0},C_E^2],
\\
 \mathcal{Z}_{-1,1}[G_{1_z},C_E^1]&=0,\\
 \mathcal{Z}_{-1,1}[G_{1_z},C_E^2]&=0,
\\
\mathcal{Z}_{-1,1}[G_{2_z},C_E^1]&=\xi\partial_\xi \mathcal{Z}_{-1,1}[G_{\dot 0},C_E^1],
\\
\mathcal{Z}_{-1,1}[G_{2_z},C_E^2] &=-\xi\partial_\xi \mathcal{Z}_{-1,1}[G_{\dot 0},C_E^2].
\end{split}
\ee
Finally, for the weight $w=0$ partition functions we get:
\be
\label{0FluxesE}
\begin{split}
\mathcal{Z}_{0,1}[G_{(-1)_\tau},C_E^1] &=2q\partial_q \mathcal{Z}_{-2,1}[G_{\dot 0},C_E^1]+Z^0_{0,1},\\
\mathcal{Z}_{0,1}[G_{(-1)_\tau},C_E^2]&= 0,
\\
\mathcal{Z}_{0,1}[G_{0_\tau},C_E^1]&= 2q\partial_q \mathcal{Z}_{-2,1}[G_{\dot 0},C_E^1]+Z^0_{0,1},
\\
\mathcal{Z}_{0,1}[G_{0_\tau},C_E^2]&= 2q\partial_q \mathcal{Z}_{-2,1}[G_{\dot 0},C_E^2]-Z^0_{0,1},
\\
 \mathcal{Z}_{0,1}[G_{1_\tau},C_E^1]&=0,
 \\
 \mathcal{Z}_{0,1}[G_{1_\tau},C_E^2]&=  0,
\\
\mathcal{Z}_{0,1}[G_{2_\tau},C_E^1]&= q\partial_q \mathcal{Z}_{-2,1}[G_{\dot{0}},C_E^1],
\\
\mathcal{Z}_{0,1}[G_{2_\tau},C_E^2]&= -q\partial_q \mathcal{Z}_{-2,1}[G_{\dot{0}},C_E^2].
\end{split}
\ee
These expressions can be easily rephrased in terms of $E_1, E_2$ by making use of the following identities:
\bea
\xi\partial_\xi  Z^0_{-2,1} &=&  \frac{\phi_{-1,2}}{\phi_{-2,1}}\frac{E_4}{\eta^{12}}+2E_1 Z^0_{-2,1},\\
q\partial_q Z^0_{-2,1} &=& \Big(E_1^2-\frac{1}{6}E_2\Big)Z^0_{-2,1}+ \frac{\phi_{-1,2}}{\phi_{-2,1}} E_1\frac{E_4}{\eta^{12}} -\frac{1}{\phi_{-2,1}}\frac{E_{4,2}}{\eta^{12}}+\frac{1}{12}\frac{E_{6,1}}{\eta^{12}}.
\eea
Note that again these expressions are holomorphic quasi-Jacobi forms, due to cancellations between the poles of the individual terms.  This can be easily checked by making use of the expansions (\ref{phim21}) and (\ref{E1expan}).

\section{Modular Anomaly Equation for Genus-one Prepotentials}\label{genusone}

In the main part of this article we have focused on the holomorphic anomaly equations for the genus-zero prepotentials as these are related to the elliptic genera (\ref{ellgen-def1}) of certain four-dimensional strings.
For completeness we now present their higher genus analogue on elliptic fourfolds.

From the expression (\ref{virtdimexp}) for the virtual dimension of the moduli space of stable maps on Calabi-Yau fourfolds it is clear 
that the only non-zero invariants can arise 
at genus $g=0$ or $g=1$ \cite{Klemm:2007in,cao2019stable,Cao_2020}. At genus one, the virtual dimension vanishes already without any reference to an incidence relation associated with a background flux (so the situation is similar as for threefolds).
Therefore, in this case one can consider the Gromov-Witten invariants counting stable holomorphic maps 
$f: \Sigma_{g=1,k}  \to C$
 with $k$ points fixed, subject to the condition that their image on $C$ lies on certain divisors $D_i \in H^{1,1}(Y_4)$, for $i=1, \ldots, k$.
These invariants can then be transformed via the divisor equation to the invariants at genus $g=1$ with no points fixed. 
Our aim is to obtain a holomorphic anomaly equation for the generating functions of theses building blocks.

The starting point for our derivation of the holomorphic anomaly equation is the BCOV expression for the correlation functions at genus $g=1$ \cite{Bershadsky:1993ta,Bershadsky:1993cx}.
Let us again first consider the situation for Calabi-Yau threefolds.
Compared to the genus-zero expression (\ref{BCOV}), there appears an additional term of the form 
\be \label{newtermCY3}
\frac{1}{2} {\overline C_{\bar i}}^{jk}   \cF^{(0)}_{jk i_1 i_2 i_3 i_4 |C} \,.
\ee
It describes the factorization of the genus one curve into a sphere with two punctures; these connect via ${\overline C_{\bar i}}^{jk}$ to the anti-holomorphic operator. Note that in this appendix we will indicate the genus $g$ of the maps by  the superscript of the generating function.

In analogy to what was explained in Section \ref{FromBCOV} for genus $g=0$ invariants, we can interpret the BCOV equation in terms of the generating functions for the relative $g=1$ Gromov-Witten invariants on an elliptic Calabi-Yau fourfold $Y_4$. It takes the form:
{\be \label{HAEFaC-1-genus1}
{ - \frac{1}{2 \pi i }} \overline{\partial}_{\bar i} \cF^{(1)}_{C_\beta} \ =\  
{\overline C_{\bar i}}^{jb}
\left( 
\cF^{(0)}_{b; j |C_{\beta}}    + 
\cF^{(0)}_{b  |C_{\beta}}  \cF^{(1)}_{j | C=0} + 
 \sum\limits_{\substack{ C_{\beta_1} + C_{\beta_2}  =  C_\beta  \\ C_{\beta_i} \neq 0}}
  \cF^{(0)}_{b |C_{\beta_1}}  \cF^{(1)}_{j |C_{\beta_2}}
\right)\,.
\ee  
}
The first term on the right is the qualitatively new term as compared to the genus-zero expression (\ref{HAEFaC-1}), and is 
the analogue of (\ref{newtermCY3}) for Calabi-Yau fourfolds.
It arises from the degenerations of the holomorphic map $f: \Sigma_{g=1}  \to C_\beta$ for which the genus-one Riemann surface $\Sigma_{g=1}$
develops two nodes and the anti-holomorphic operator $\bar \phi_{\bar i}$ is located on the rational curve formed by these two pinchings.
Note that we sum over all insertions of divisor classes $D_i \in H^{1,1}(Y_4)$ and four-fluxes $G_a \in H^{2,2}_{\rm vert}(Y_4)$.
The second and third term are the genus-one analogues of the quadratic degenerations appearing in  (\ref{HAEFaC-1}), with the difference that one must sum not only over all splittings $C_\beta =C_{\beta_1} + C_{\beta_2}$, but also distribute the genus of the maps as $g = g_1 + g_2$. In the present case this gives two contributions: 
For $C_{\beta_i} \neq 0$ one obtains, as the only non-vanishing contributions, the third term in (\ref{HAEFaC-1-genus1}). In addition, there is a splitting where either $C_{\beta_1} =0$ or 
$C_{\beta_2} =0$, which potentially leads to a classical contribution. While
at  $g=0$ we have  $ \cF^{(0)}_{j | C=0}=0$, we get for genus $g=1$ a non-trivial classical contribution given by
\cite{Cox:2000vi,Hori:2003ic,Klemm:2007in}
\bea  \label{genusoneinvtrivialclass}
 \cF^{(1)}_{j | C=0} = - \frac{1}{24} \int_{Y_4}  c_3(Y_4) \cdot D_j \,.
\eea
This leads to the second term on the right-hand side of eq.~(\ref{HAEFaC-1-genus1}). 
Finally, at genus $g=1$ there is no gravitational descendant term since the invariants on the left do not depend on a background flux class; hence the analogue of the third term on the righthand side
of (\ref{HAEFaC-1}) vanishes.

Equ.~(\ref{HAEFaC-1-genus1}) can be further evaluated in the limit (\ref{t-limit}), in which the overall factor of ${\overline C_{\bar i}}^{jb}$ takes the simple form (\ref{Cbaraufinal}).
The divisor index $j$ now refers to a pullback divisor $\pi^\ast(D^{\rm b}_\alpha)$. As a result, the genus-one invariants (\ref{genusoneinvtrivialclass}) which appear in the anomaly equation are of the form
\bea
 \cF^{(1)}_{\alpha | C=0} = - \frac{1}{24}  \pi_{\ast} [c_3(Y_4)] \cdot_{B_3} D_\alpha^{\rm b} \,.
\eea
Furthermore, the flux indices $b$ which are summed over in (\ref{HAEFaC-1-genus1}) refer only to the $(-2)$-fluxes~$G_{\dot\alpha}$.
All in all one finds for
 the holomorphic anomaly equation
\bea
2 \pi i (4 \tau_2^2)   \overline{\partial}_{\bar \tau} \cF^{(1)}_{C_\beta} &\stackrel{(\ref{t-limit})}{=}&   \cF^{(0)}_{\dot\alpha  |C_{\beta}}   I^{\dot\alpha \alpha} \left(D^{\rm b}_\alpha  \cdot_{B_3} (C_\beta - \frac{1}{24} \pi_{\ast} [c_3(Y_4)])\right)    \\
&&  +  \sum\limits_{\substack{ C_{\beta_1} + C_{\beta_2}  =  C_\beta  \\ C_{\beta_i} \neq 0}}      I^{\dot\alpha \alpha}      (D^{\rm b}_\alpha  \cdot_{B_3} C_{\beta_2})  \cF^{(0)}_{\dot\alpha |C_{\beta_1}}  \cF^{(1)}_{C_{\beta_2}}  \nn    \\
&=& \langle \langle \pi^\ast(C_\beta) \rangle \rangle_{g=0,C_\beta}   - \frac{1}{24}\langle \langle \pi^\ast(\pi_\ast[c_3(Y_4)])  \rangle \rangle_{g=0,C_\beta}   \\
&&+    \sum\limits_{\substack{ C_{\beta_1} + C_{\beta_2}  =  C_\beta  \\ C_{\beta_i} \neq 0}}     \langle \langle \pi^\ast(C_{\beta_2})    \rangle \rangle_{g=0,C_{\beta_1}}   \, 
 \cF^{(1)}_{C_{\beta_2}}     \,.\nn
\eea

This equation corresponds to the following modular anomaly 
\bea   \label{E2genus1}
\partial_{E_2} \cF^{(1)}_{C_\beta}  &=& - \frac{1}{12}   \left(   \langle \langle \pi^\ast(C_\beta) \rangle \rangle_{g=0,C_\beta}   - \frac{1}{24}\langle \langle \pi^\ast(\pi_\ast[c_3(Y_4)])  \rangle \rangle_{g=0,C_\beta} \right.   \\ \nn
&& \phantom{xxx}  +    \sum\limits_{\substack{ C_{\beta_1} + C_{\beta_2}  =  C_\beta  \\ C_{\beta_i} \neq 0}}     \left.  \langle \langle \pi^\ast(C_{\beta_2})    \rangle \rangle_{g=0,C_{\beta_1}}   \, 
 \cF^{(1)}_{C_{\beta_2}}      \right)   \,.\nn
\eea

On the other hand, there is no non-trivial elliptic anomaly equation because the genus-one invariants do not depend on the flux, i.e.
\bea
\partial_{E_1} \cF^{(1)}_{C_\beta} = 0 \,.
\eea

Note that the explicit form of the second term on the right-hand side of (\ref{E2genus1}) depends on the specific type of elliptic fibration $Y_4$. For example, for a smooth Weierstrass model
(which in particular does not allow for additional rational sections) one finds that 
\bea
- \frac{1}{24} \pi_\ast[c_3(Y_4)] =  + \frac{5}{2} (\bar K_{B_3})^2    \,.
\eea
Using this relation, the modular anomaly equation (\ref{E2genus1}) agrees with
equation at genus-one given in \cite{Cota:2017aal} for such fibrations.

\bibliography{papers}
\bibliographystyle{JHEP}

\end{document}